%% file: main.tex
\documentclass[]{aa}
\usepackage[capitalise]{cleveref} 
\usepackage{morefloats}
\usepackage{amsmath}
\usepackage{color}
\usepackage{textcomp}
\usepackage{commath}
\usepackage{array}
\usepackage{siunitx}
\usepackage{subcaption}

\bibpunct{(}{)}{;}{a}{}{,} 

\newcolumntype{C}[1]{>{\centering\arraybackslash}m{#1}}

\begin{document}

\title{The effect of metallicity on the atmospheres of exoplanets with fully coupled 3D hydrodynamics, equilibrium chemistry, and radiative transfer}

\titlerunning{The effect of metallicity on the atmosphere of GJ~1214b}
\authorrunning{Drummond et al.}

\author{B. Drummond\inst{\ref{inst1}}
\and N. J. Mayne\inst{\ref{inst1}}
\and I. Baraffe\inst{\ref{inst1},\ref{inst2}}
\and P. Tremblin\inst{\ref{inst3}}
\and J. Manners\inst{\ref{inst1},\ref{inst4}}
\and D. S. Amundsen\inst{\ref{inst5},\ref{inst6}}
\and J. Goyal\inst{\ref{inst1}}
\and D. Acreman\inst{\ref{inst1}}
}

\institute{Astrophysics Group, University of Exeter, EX4 4QL, Exeter, UK\label{inst1} \\
\email{b.drummond@exeter.ac.uk}
\and Univ Lyon, Ens de Lyon, Univ Lyon1, CNRS, CRAL, UMR5574, F-69007, Lyon, France\label{inst2}
\and Maison de la simulation, CEA, CNRS, Univ. Paris-Sud, UVSQ, Université Paris-Saclay, 91191 Gif-Sur-Yvette, France\label{inst3}
\and Met Office, Exeter, EX1 3PB, UK\label{inst4}
\and Department of Applied Physics and Applied Mathematics, Columbia University, New York, NY 10025, USA\label{inst5}
\and NASA Goddard Institute for Space Studies, New York, NY 10025, USA\label{inst6}}

\date{Received /
	Accepted }


\abstract {
In this work we have performed a series of simulations of the atmosphere of GJ~1214b assuming different metallicities using the Met Office Unified Model (UM). The UM is a general circulation model (GCM) that solves the deep, non-hydrostatic equations of motion and uses a flexible and accurate radiative transfer scheme, based on the two-stream and correlated-$k$ approximations, to calculate the heating rates. In this work we consistently couple a well-tested Gibbs energy minimisation scheme to solve for the chemical equilibrium abundances locally in each grid cell for a general set of elemental abundances, further improving the flexibility and accuracy of the model. As the metallicity of the atmosphere is increased we find significant changes in the dynamical and thermal structure, with subsequent implications for the simulated phase curve. The trends that we find are qualitatively consistent with previous works, though with quantitative differences. We investigate in detail the effect of increasing the metallicity by splitting the mechanism into constituents, involving the mean molecular weight, the heat capacity and the opacities. We find the opacity effect to be the dominant mechanism in altering the circulation and thermal structure. This result highlights the importance of accurately computing the opacities and radiative transfer in 3D GCMs.
}

\maketitle

\input{intro}

\input{model_description}

\input{results}

\input{discussion}

\input{conclusions}

\bibliographystyle{aa} 

\end{document}

%% file: intro.tex

\section{Introduction}

The precise elemental composition of exoplanetary atmospheres is a highly unconstrained quantity and likely depends on several factors. These factors include the composition of the initial material from which the planetary system formed, the local conditions in the proto-planetary disc at the time of planetary formation, subsequent migration through the disc and later evolution due to deposition of material by meteorites or evaporation of the atmosphere \citep[see][for a review]{MadAM16}.

The abundances of the individual elements (H, O, C, N, K, Li etc) fundamentally determine the overall chemical composition of the atmosphere, while the local pressure and temperature determine the distribution of the elements across all the possible chemical species (CH$_4$, CO, H$_2$O, Na, LiH etc) in chemical equilibrium. In turn the composition affects the temperature structure through the opacities and heating rates. Non-equilibrium processes (transport and photochemistry) also have an influence on the chemical composition and thereby the temperature \citep[][]{DruTB16}.

Given the uncertainty in the elemental composition of exoplanetary atmospheres several studies \citep[e.g.][]{KreBD14b,MadCM14,WakSK17} attempt to fit for the abundances of certain strongly absorbing chemical species, such as H$_2$O, from observations which, for example, might provide a constraint on the oxygen abundance. However, such measurements are likely degenerate with other factors such as the temperature of the atmosphere and non-equilibrium processes.

Alternatively, several theoretical studies \citep[][]{MosLV13,Moses2013,AguVS2014,VenAS2014,Men12,KatSF14,ChaML15} have explored the parameter space by performing a range of simulations that vary the elemental composition and investigated the effect on the atmosphere properties, such as the temperature, wind velocities, chemical composition and simulated transmission and emission spectra.

To date the focus has largely involved varying either the metallicity or the carbon to oxygen ratio (C/O). In the former case the abundances of all elements heavier than H and He are typically scaled uniformly, meaning that the ratios between the heavy elements (e.g. C/O, C/N, O/N, amongst others) remain constant, usually at solar values. Other studies have focused on C/O, where the abundance of C (or O) is varied for a fixed abundance of O (or C). This quantity has particular interest as it is thought that C/O varies radially in the protoplanetary disc due to the progressive condensation of H$_2$O, CO$_2$ and CO, depending on the thermodynamical conditions (snow lines), thereby depleting the gas-phase O and C \citep[e.g.][]{ObeMB11,Mad12}. The formation mechanism and location, and subsequent migration and accretion of gas and dust, may therefore determine the C/O of the planetary atmosphere. 

Since the formation history and subsequent evolution of the planet is likely to determine the elemental composition, it is important to understand how this is likely to affect the properties of the atmosphere. In this study we investigate how the metallicity affects the thermal structure, circulation and expected emission flux, for the specific case of the atmosphere of the Neptune-mass exoplanet GJ~1214b. While we choose to adopt parameters relevant to a particular planet, we expect the general trends that we find to be applicable to the wider population of warm, Neptune-mass exoplanets.

The planet GJ~1214b has been an object of intense study since its discovery \citep{ChaBI09}. It is one of the few discovered sub-Neptune planets that has the potential to be characterised using current instruments \citep{MilF10} due to its orbit around a relatively bright and nearby M-dwarf. Efforts have been made in several published works to measure the transmission spectrum of the atmosphere which resulted in flat and featureless spectra using both ground-based \citep{BeaMH10,BeaDK11,CroBH11,deMBd12,WilCS14,CacKH14} and space-based instruments \citep{DesBM11,BerCD12,FraDG13}. The flat transmission spectrum was interpreted as being potentially due to thick clouds at low pressures that provide a grey absorption or, alternatively, due to an atmosphere with a high mean molecular weight and therefore small scale height. In either case, a cloud-free solar composition atmosphere was effectively ruled out. Only \citet{CroAJ11} found evidence of a non-flat spectrum with a significantly deeper transit depth at $\sim2.15$ \textmu m than at $\sim1.25$ \textmu m.

\citet{KreBD14} obtained a transmission spectrum with the Wide Field Camera 3 instrument (Hubble Space Telescope) that was able to confidently rule out a cloud-free atmosphere, as the measurements are expected to be of high enough precision to detect absorption features in a high mean molecular weight (cloud-free) atmosphere. While this result suggests that grey absorption due to clouds is responsible for the flat transmission spectrum the bulk composition of the atmosphere remains unconstrained, with both hydrogen-dominated and high mean molecular weight atmospheres being possible.

Various modelling efforts have investigated the atmosphere of GJ~1214b. \citet{HowB12} performed a series of 1D simulations assuming an isothermal atmosphere and an arbitrary composition of gas-phase species and haze species (sulphuric acid, Tholins etc). They found a best-fit to the available data to be a hydrogen-rich atmosphere with small scattering particles, though atmospheres dominated by other heavier molecules (H$_2$O, N$_2$) could not be discounted. \citet{MilZF12} found the effects of photochemistry and vertical mixing on the transmission spectrum to be small since the important changes in the mole fractions of the chemical species occured above the region probed by transmission spectroscopy, using a 1D chemical kinetics model.

The atmosphere of GJ~1214b has also been investigated with a number of General Circulation Models (GCMs) that simulate the 3D fluid flow of the atmosphere. \citet{Men12} considered both a hydrogen-dominated atmosphere (with 1$\times$ and 30$\times$ solar metallicity), as well as an atmosphere entirely composed of H$_2$O, using the Intermediate General Circulation Model (IGCM) that adopts a double-grey radiative transfer scheme. \citet{KatSF14} also investigated hydrogen-dominated atmospheres (with 1$\times$, 30$\times$ and 50$\times$ solar metallicity), as well as atmospheres dominated by H$_2$O and/or CO$_2$, with the SPARC/MITgcm \citep{Showman2009} that includes a non-grey radiative transfer scheme. Finally \citet{ChaML15,ChaMM15} used the Generic LMDZ GCM to simulate hydrogen-dominated (1$\times$, 10$\times$ and 100$\times$ solar metallicity) and H$_2$O-dominated atmospheres also with a non-grey radiative transfer scheme. In each case, it was found that as the metallicity is increased the circulation transitions towards a state with a stronger equatorial jet and with an enhanced zonal temperature gradient (day-night temperature contrast).

\citet{ZhaS17} performed a series of simulations of a `GJ~1214b-like' atmosphere with the MITgcm, with a Newtonian cooling scheme to represent the thermal evolution, adopting a range of bulk compositions. In particular they assumed the atmosphere to be entirely made of H$_2$, He, CH$_4$, H$_2$O, CO, N$_2$, O$_2$ or CO$_2$. Using a fixed radiative equilibrium temperature profile the effect of varying the mean molecular weight, heat capacity and radiative timescale on the dynamics and thermal structure was investigated. \citet{ZhaS17} found that increasing the mean molecular weight generally leads to decreasing zonal wind velocities in the equatorial jet, with a more banded structure, and a larger day-night temperature contrast. Variations in the heat capacity were shown to be less important.

In this study we use the Met Office Unified Model (UM) to simulate the atmosphere of GJ~1214b assuming various metallicities. To achieve this we have recently coupled a Gibbs energy minimisation scheme to the UM that calculates the chemical composition in each grid cell for a general set of elemental abundances for the local pressure and temperature. This represents the first flexible equilibrium chemistry scheme consistently coupled to a GCM applied to exoplanets.

%% file: model_description.tex

\section{Model Description and Setup}
\label{mod_desc}

\subsection{The Unified Model (UM)}

The UM is a complex atmosphere model used for both short-term, small-scale numerical weather prediction and long-term, large-scale climate simulations, originally developed for applications to the Earth atmosphere. Recent works have extended the models application to hot, hydrogen-dominated atmospheres \citep{MayBA14,MayBA14b,AmuMB16,MayDB17} and terrestrial exoplanet atmospheres \citep{BouMD17}.  

A description of the model setup for hydrogen-dominated atmospheres can be found in \citet{MayBA14} and \citet{AmuMB16}. Here we review some of the major aspects of the current model and provide specific details of the simulations presented in this work.

\subsubsection{Dynamics}
\label{section:mod_desc_dynamics}

The most fundamental component of the UM is the dynamical core that solves for the fluid flow of the atmosphere. The latest version of the UM dynamical core, ENDGame, solves the non--hydrostatic deep--atmosphere equations of motion on a rotating sphere \citep{WooSW14}. In contrast, most other GCMs so far applied to exoplanet atmospheres solve the hydrostatic primitive equations \citep[e.g.][]{Showman2009,MenR09,ThrC10,HenMP11}, that assume several approximations to simplify the equations. The details of these approximations, and their validity, are more fully described in \citet{MayBA14,MayBA14b}.

\citet{MayBA14,MayBA14b} performed a series of idealised tests to validate the application of the ENDGame dynamical core to solve for the 3D flow over long integration times. This included several tests under hot Jupiter conditions using a Newtonian cooling scheme to represent the thermal evolution of the atmosphere \citep{MayBA14,MayDB17}. In the current work we solve the `full' equations of motion and do not invoke the various approximations that lead to the primitive equations. The UM uses a geometric height-based (altitude) grid in the vertical domain. We refer the reader to \citet{MayBA14,MayBA14b} for a more complete description, however we repeat the equations below to highlight the role played by the heat capacity and the mean molecular weight, which both depend on the chemical composition. 

The set of five equations solved by the dynamical core describe the conservation of momentum (one for each directional component) and the conservation of mass, in addition to the thermodynamic equation, which are closed by the equation of state:

\begin{alignat}{6}
	\frac{{\rm D}u}{{\rm D}t} = \frac{uv\tan \phi}{r} - \frac{uw}{r} + fv - f'w - \frac{c_P\theta}{r\cos \phi}\frac{\partial \Pi}{\partial \lambda} + {\rm D}(u) \label{equation:u_wind}   \\
	\frac{{\rm D}v}{{\rm D}t} = -\frac{u^2\tan \phi}{r} - \frac{vw}{r} - uf - \frac{c_P\theta}{r}\frac{\partial \Pi}{\partial \phi} + {\rm D}(v) \label{eq:2} \\
	\delta \frac{{\rm D}w}{{\rm D}t} = \frac{u^2 + v^2}{r} + uf' - g(r) - c_P\theta\frac{\partial \Pi}{\partial r} \label{eq:3} \\
	\frac{{\rm D}\rho}{{\rm D}t} = -\rho\left[\frac{1}{r\cos\phi}\frac{\partial u}{\partial \lambda} + \frac{1}{r\cos\phi}\frac{\partial(v\cos\phi)}{\partial \phi} + \frac{1}{r^2}\frac{\partial(r^2w)}{\partial r}\right] \label{eq:4}\\
	\frac{{\rm D}\theta}{{\rm D}t} = \frac{Q}{\Pi} + {\rm D}(\theta) \label{equation:thermodynamic} \\
	\Pi^{\frac{1-\kappa}{\kappa}} = \frac{R\rho\theta}{P_0}. \label{equation:eos}
\end{alignat}

In the above, $u$, $v$ and $w$ are the wind velocity components in the longitudinal ($\lambda$), latitudinal ($\phi$) and radial ($r$) directions, respectively. $c_P$ is the specific heat capacity, $R$ is the specific gas constant, $Q$ is the heating rate, $D$ is the diffusion operator and $\kappa$ is the ratio $c_P/R$. $\delta$ is a switch (1 or 0) that enables a quasi-hydrostatic version of the equations. $P_0$ is a reference pressure, $\rho$ is the density and $g(r)$ is the height dependent gravity, $g(r) = g_{\rm p}\left(R_{\rm p}/r\right)^2$, where $g_{\rm p}$ and $R_{\rm p}$ are the surface gravity and planetary radius, respectively. $f$ and $f'$ are the Coriolis parameters, $f = 2\Omega\sin\phi$ and $f' = 2\Omega\cos\phi$, where $\Omega$ is the planetary rotation rate. 

$\theta$ and $\Pi$ are the potential temperature and Exner pressure, defined as
\begin{equation}
	\theta = T\left(\frac{P_0}{P}\right)^{\frac{R}{c_P}}
\end{equation}
and 
\begin{equation}
	\Pi = \left(\frac{P}{P_0}\right)^{\frac{R}{c_P}} = \frac{T}{\theta}.
\end{equation}
The material derivitive is defined as
\begin{equation}
	\frac{{\rm D}}{{\rm D}t} = \frac{\partial}{\partial t} + \frac{u}{r\cos\phi}\frac{\partial}{\partial \lambda} + \frac{v}{r}\frac{\partial}{\partial \phi} +w\frac{\partial}{\partial r}.
\end{equation}

We note that $c_P$ and $R$ depend on the chemical composition and therefore on the assumed metallicity of the atmosphere. Both quantities that appear in the above equations are treated as globally constant (spatially-invariant) model parameters. We will return to discuss these parameters in a following section.

A vertical damping `sponge' layer is applied to reduce the vertically propagating waves that result from the rigid boundaries of the model \citep{MayBA14}. The magnitude of the damping coefficient $R_w$ is given by
\begin{equation}
	R_w = 
\begin{cases}
	C\sin^2\left(\frac{1}{2}\pi\left(\eta-\eta_s\right)\left(\frac{1.0}{1.0-\eta_s}\right)\right), & \eta \ge \eta_s \\
	0, & \eta < \eta_s
\end{cases}
\end{equation}
where $C$ is a coefficient, $\eta$ is a non-dimensional height and $\eta_s$ is the start height for the damping. In previous works \citep{MayBA14,AmuMB16} $\eta$ was set to be a horizontally uniform function of height $z$,
\begin{equation}
\eta = \frac{z}{z_{\rm top}},
\end{equation}
where $z_{\rm top}$ is the height of the top of the atmosphere. Here we adopt a formulation of $\eta$ that is also a function of latitude,
\begin{equation}
	\eta(\phi) = \frac{z}{z_{\rm top}}\cos(\phi) + (1-\cos(\phi)),
\end{equation}
that effectively increases the vertical extent of the sponge in the polar regions. 

\begin{figure}
\centering
\begin{tabular}{c}
	\includegraphics[width=0.5\textwidth]{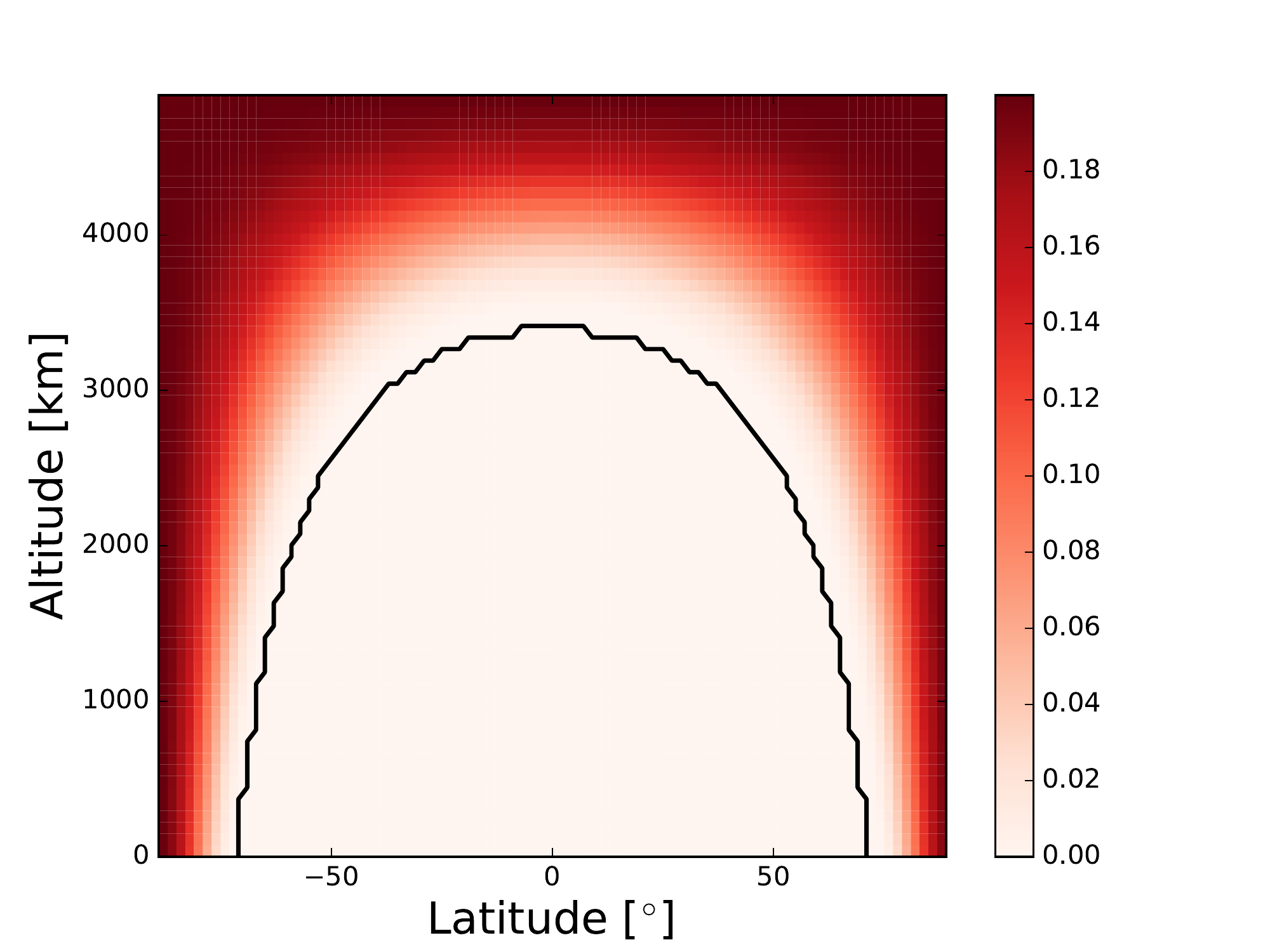}
\end{tabular}
\caption{Value of the vertical damping coefficient $R_w$ as a function of height and latitude for the R1 simulation. The shape of the damping is similar for other simulations but shifted in altitude depending on the height of the top of the atmosphere $z_{\rm top}$. The black contour indicates $R_w=0$.} 
\label{figure:sponge}
\end{figure}

To demonstrate the shape of the vertical damping profile we show the value of $R_w$ as a function of altitude and latitude in \cref{figure:sponge} for one of our simulations. While the height $z$ at which the vertical damping begins changes for each simulation, due to differing height grids (see \cref{section:mod_desc_params}), the damping starts at approximately the same pressure level. We note that varying the vertical damping has a small effect on the atmospheric flow \citep{MayBA14,MayBA14b}.

\subsubsection{Radiative transfer}

The UM includes a sophisticated radiative transfer scheme, the open-source Suite of Community Radiative Transfer codes based on Edwards and Slingo (SOCRATES) scheme\footnote{https://code.metoffice.gov.uk/trac/socrates} \citep{EdwS96,Edw96}, to calculate the radiative heating rates. SOCRATES, which takes the two-stream approximation and treats opacities using the correlated-$k$ method, has been adapted to hot Jupiter-like conditions \citep{AmuBT14,AmuTM17}. The UM including full radiative transfer calculations has previously been applied to the atmosphere of the hot Jupiter HD~209458b \citep{AmuMB16}.

In the correlated-$k$ method, the line-by-line opacities are approximated by a smaller number of $k$-coefficients to improve the computational efficiency. We have used the same $k$-coefficients as in \citet{AmuBT14,AmuMB16} and we refer the reader to that work for further details on the calculation of the $k$-coefficients and sources of the line-lists. The $k$-coefficients of the individual gases (H$_2$O, CH$_4$ etc) are combined on-the-fly in the model using the method of equivalent extinction \citep{Edw96,AmuTM17} which requires the mole fractions of the opacity species.

The $k$-coefficients used here were originally tested under hot Jupiter-like conditions and irradiation due to a HD~209458b star \citep{AmuBT14}. To test the accuracy for the temperature and composition conditions explored in this work we compared the heating rates derived from a calculation assuming the correlated-$k$ method with a calculation using the full line-by-line method. We found errors in the heating rate to be $\lesssim$10\%, similar to the errors found for hot Jupiter-like conditions \citep{AmuBT14,AmuTM17}.

During the main model integration the radiative transfer scheme splits the spectrum into 32 radiative bands. However, following the main model integration we restart the model and integrate for several planetary orbits using a higher spectral resolution (500 bands) in order to calculate the spectral emission from the top of the atmosphere, as detailed in \citet{BouMD17}.

The divergence of the radiative flux across each grid cell is converted into a heating rate (in K s$^{-1}$) using the heat capacity of the gas. In contrast to $c_P$ that appears in the set of equations solved by the dynamical core (\cref{section:mod_desc_dynamics}) the heat capacity here can be calculated locally in each grid cell. We therefore carefully distinguish between the heat capacity that appears in \cref{equation:u_wind,eq:2,eq:3,eq:4,equation:thermodynamic,equation:eos} that we call $c_{P,\rm{dyn}}$ with the heat capacity used to compute the heating rate that we call $c_{P,\rm{rad}}$. Of course, these quantities should ideally be consistent and identical.

\subsubsection{Chemistry}

In previous UM simulations of hydrogen-dominated atmospheres \citep{AmuMB16} we adopted the analytical solution to chemical equilibrium of \citet{Burrows1999}. This scheme provides a highly computationally efficient means of calculating the mole fractions of CO, CH$_4$, H$_2$O, NH$_3$ and N$_2$ under the assumption of a hydrogen-dominated mixture with roughly solar abundances of the elements and high temperatures. In addition, the abundances of the alkali species were estimated by assuming that the monatomic species (Na, K etc) were present in their solar elemental abundances above some critical temperature, and had zero abundances below \citep{AmuMB16} due to conversion into alkali chlorides (NaCl, KCl etc) at cooler temperatures. 

In this work we improve on this method by consistently coupling a Gibbs energy minimisation scheme to the UM to enable the calculation of the equilibrium mole fraction of a large number of chemical species in each model grid cell under a wide range of thermodynamical conditions and for a general set of elemental abundances. This scheme is more flexible than our previous method, as it can handle the chemistry of a much wider range of species, and is accurate for a significantly larger region of parameter space.

The Gibbs energy minimisation scheme has previously been described in \citet{DruTB16} and was originally developed within the 1D/2D atmosphere code ATMO \citep{Tremblin2015,Tremblin2016,TreCM17,DruTB16}. It is possible to include equilibrium condensation in the calculation, however in the present work we assume a gas-phase only composition. The mole fractions are calculated independently in each model grid cell. We compute the abundances of 41 gas-phase species in total: H$_2$, CO, H$_2$O, O$_2$ CH$_4$, CO$_2$, N$_2$, He, Ar, NH$_2$, NH$_3$, Na, NaH, NaOH, NaCl, K, KH, KOH, KCl, Cl, HCl, ClO, Cl$_2$, Ti, TiO, V, VO, F, HF, Li, LiCl, LiH, LiF, Cs, CsCl, CsH, CsF, Rb, RbCl, RbH and RbF.

Our decision to neglect equilibrium condensation is likely to have some effect on the final calculated mole fractions of several species. In particular, the formation of silicates and other oxygen containing condensates at high pressures and temperatures is expected to reduce the oxygen abundance at lower pressures/temperatures \citep[e.g.][]{Burrows1999}. Some previous studies take this effect into account simply by reducing the oxygen abundance by $\sim20\%$ \citep[e.g.][]{Moses2011,Venot2012}. However, the elemental abundance of exoplanet atmospheres is highly unconstrained and therefore here we simply adopt the solar elemental abundances and focus on the overall trend in the atmospheric properties as the metallicity varies.

In addition to potential effects on the oxygen abundance, the temperatures expected in the atmosphere of GJ~1214b lie around the condensation temperatures of various alkali species (e.g. Na$_2$S, KCl). Therefore, the neglect of condensation may lead to overpredictions of the gas-phase abundance of some alkali species. However, these species primarily absorb in the optical wavelengths ($\lambda<10^{-6}$ m) and since GJ~1214 is a relatively cool star the peak in the irradiation spectrum lies towards longer wavelengths and we do not expect absorption due to the alkali species to have a major influence on the temperature profile. Hot Jupiters, on the other hand, typically orbit hotter stars and absorption at short wavelengths due to the alkali species has significant influence on the temperature structure.

We assume the solar elemental abundances of \citet{Caffau2011}. To vary the metallicity we simply multiply the fractional abundances of each element, except H and He, by the relevant factor and renormalise so that the sum of the fractional abundances is unity.

We calculate the specific heat capacity of the mixture following \citet{KatSF14} as
\begin{equation}
	c_P(T) = \sum_i c_{P,i}(T)f_i,
\end{equation}
where $c_{P,i}(T)$ and $f_i$ are the specific heat capacity and mole fraction, respectively, of the species $i$ and the sum is over the total number of species in the mixture. Similarly, we compute the mean molecular weight $\mu$ as 
\begin{equation}
	\mu = \sum_i m_if_i,
\end{equation}
where $m_i$ is the molar mass of the species $i$. $R$ is related to $\mu$ as $R=\bar{R}/\mu$ where $\bar{R}$ is the molar gas constant.

\subsection{Simulations of GJ~1214b}

To investigate the effect of metallicity on the properties of the atmosphere we perform a series of simulations with 1$\times$, 10$\times$ and 100$\times$ solar composition. 

The metallicity fundamentally determines the mole fractions of the individual chemical species via the elemental abundances. The effect that this has on the atmosphere can be split into several mechanisms:
\begin{enumerate}
	\item{\textit{Dynamical}: $c_{p,{\rm dyn}}$ and $R$ that appear in the equations solved by the dynamical core (\cref{section:mod_desc_dynamics})}
	\item{\textit{Radiative heat capacity}: the heat capacity $c_{p,{\rm rad}}$ determines the temperature response of the atmosphere for a given heating rate}
	\item{\textit{Opacities}: the opacities control the magnitude and location of radiative absorption, emission and scattering and hence on the heating. }
\end{enumerate}

Previous studies \citep{Men12,KatSF14,ChaML15} investigated the effect of varying all three (1-3) simultaneously. \citet{ZhaS17}, on the other hand, considered the dynamical effect as well as varying the radiative timescale (effectively a parameterisation of the heating rates) (1-2) in isolation from the opacity effect. Here we design a set of simulations to investigate the importance of each of the above mechanisms consistently within the same model.

The simulations R1, R10 and R100 include the effect of all three mechanisms (1-3) for 1$\times$, 10$\times$ and 100$\times$ solar metallicity, respectively. We then investigate the dynamical effect (1) in isolation with simulations D10 and D100, for 10$\times$ and 100$\times$ solar metallicity, by only varying $c_{p,{\rm dyn}}$ and $R$. Finally, we determine the combined dynamical and radiative heat capacity effects (1-2) in the simulations C10 and C100, for 10$\times$ and 100$\times$ solar metallicity respectively. This set of simulations are summarised in \cref{table:sims}.

The model parameters $c_{P,{\rm dyn}}$ and $R$ are globally and temporally constant values in the current implementation of the UM. We estimate these values a priori for each simulation. For the R1, R10 and R100 simulations we compute the mole fractions of the set of chemical species in each grid cell for the set of elemental abundances consistent with 1$\times$, 10$\times$ and 100$\times$ solar, respectively. These mole fractions are then used to compute both the opacities and $c_{P,{\rm rad}}$ in each grid cell. On the other hand, for the D10 and D100 simulations we compute the mole fractions assuming solar elemental abundances, and compute the opacities and $c_{P,{\rm rad}}$ from these. Finally, in the C10 and C100 simulations we also compute the opacities using mole fractions that are consistent with 1$\times$ solar metallicity, but set $c_{P,{\rm rad}} = c_{P,{\rm dyn}}$.

We do not consider higher metallicities as our current model assumes the atmosphere to be hydrogen-dominated. For 100$\times$ solar composition hydrogen and helium comprise $\sim90\%$ of the gas mixture ($f_{\rm H_2} + f_{\rm He} \sim 90\%$). For significantly larger metallicities the atmosphere is no longer hydrogen-dominated as molecules such as H$_2$O and CO$_2$ become more abundant than H$_2$ and He \citep{MosLV13}. In this case line-broadening due to additional species is required \citep[e.g][]{HedM16}. This presents a challenge in terms of the availability of the data for the relevant pressures and temperatures \citep{ForRD16} as well as an increase in computational cost due to the addition of more opacity species.

\begin{table}
\caption{A summary of the simulations}
\centering
\label{table:sims}
\setlength\extrarowheight{2pt}
\begin{tabular}{l c c c c}
\hline\hline
& $c_{p,{\rm dyn}}$ & $R$ & $c_{p,{\rm rad}}$ & Opacities \\
\hline
R1 & 1$\times$ & 1$\times$ & 1$\times$ & 1$\times$\\ 
R10 & 10$\times$ & 10$\times$ & 10$\times$ & 10$\times$ \\ 
R100 & 100$\times$ & 100$\times$ & 100$\times$ &100$\times$\\ 
D10 & 10$\times$ & 10$\times$ & 1$\times$ & 1$\times$ \\ 
D100 & 100$\times$ & 100$\times$ & 1$\times$ & 1$\times$ \\ 
C10 & 10$\times$ & 10$\times$ & 10$\times$ & 1$\times$ \\
C100 & 100$\times$ & 100$\times$ & 100$\times$ & 1$\times$ \\
\hline
\end{tabular}
\end{table}

\subsection{Model setup and parameters}
\label{section:mod_desc_params}

\begin{figure}
\centering
\begin{tabular}{c}
\includegraphics[width=0.45\textwidth]{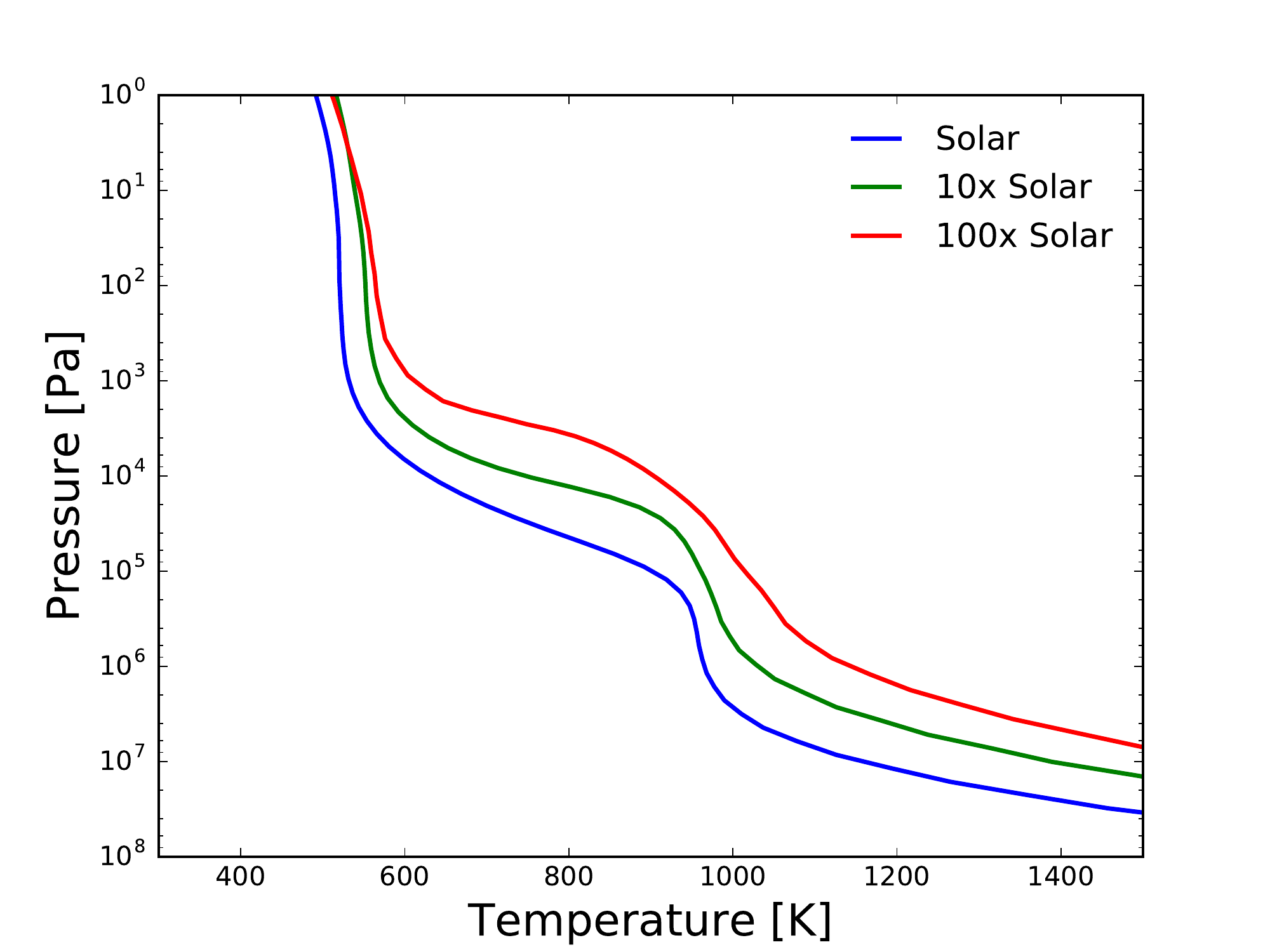} \\
\includegraphics[width=0.45\textwidth]{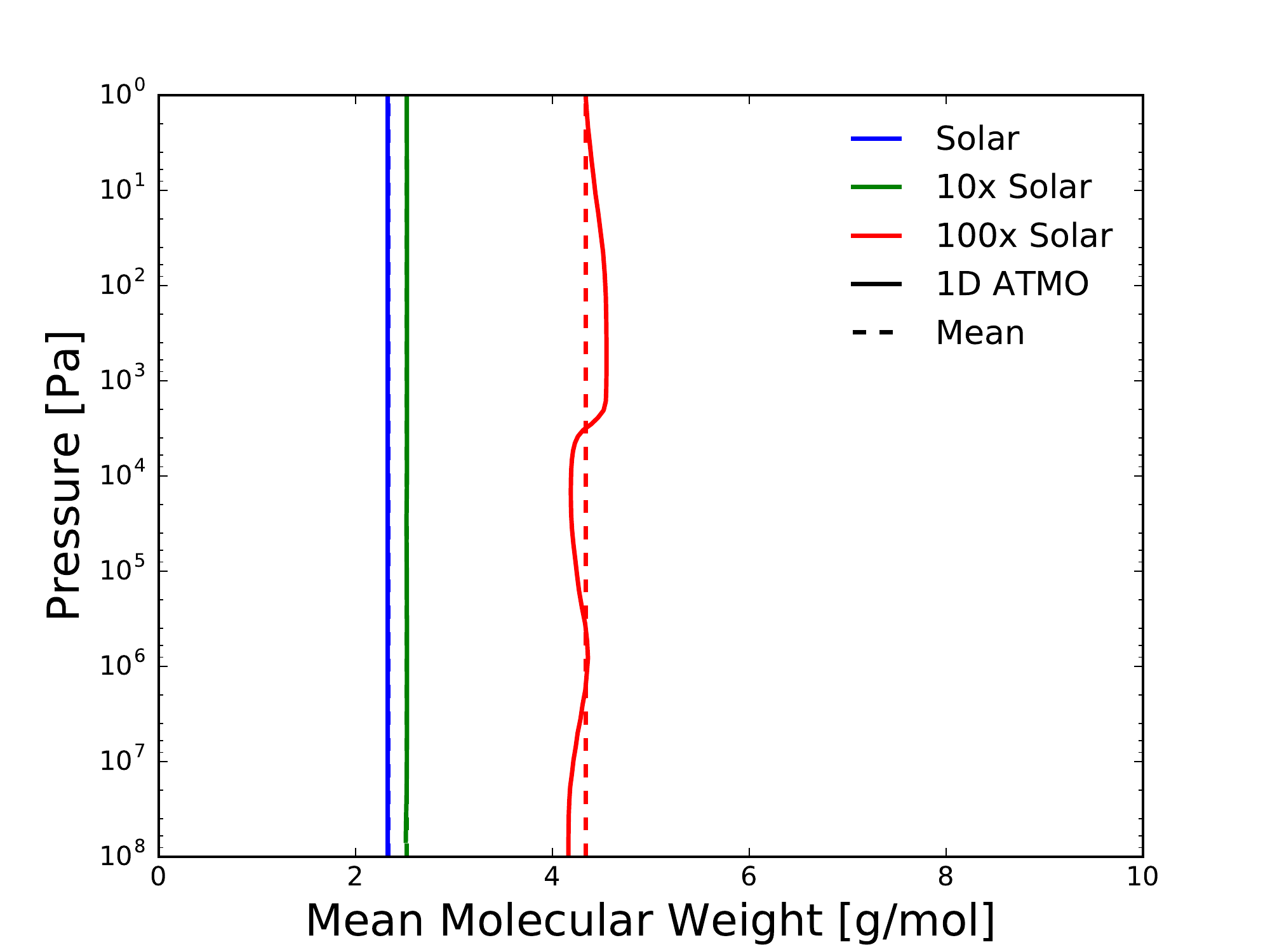} \\
\includegraphics[width=0.45\textwidth]{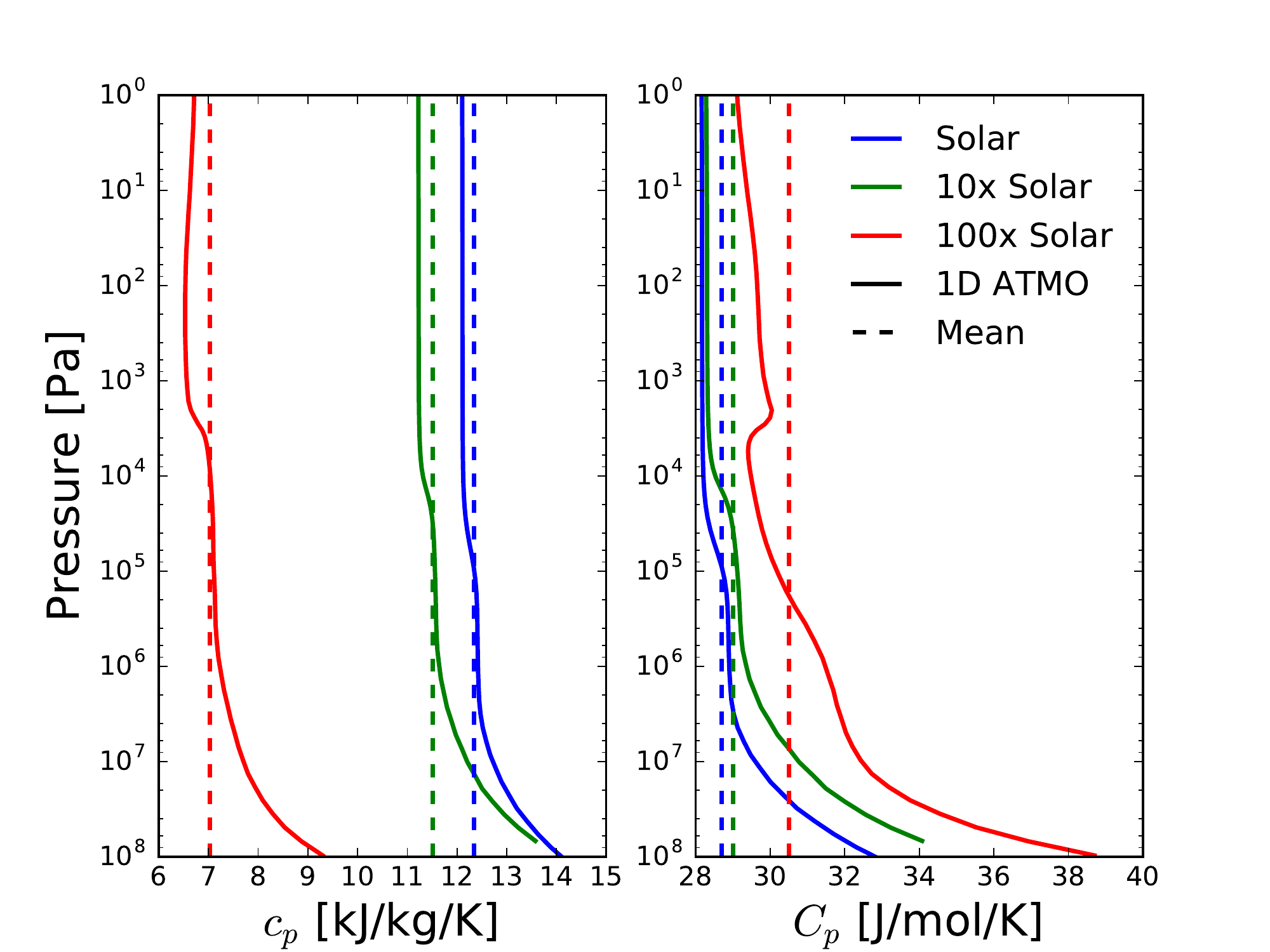}
\end{tabular}
\caption{Figures showing the pressure-temperature profile (top), mean molecular weight (middle) and heat capacity (bottom) calculated for GJ~1214b from the 1D ATMO model, assuming 1$\times$ (blue), 10$\times$ (green) and 100$\times$ (red) solar metallicity. Dashed lines indicate the arithmetic mean calculated from these profiles that are used in the 3D UM simulations.}
\label{figure:pt_cp_mmw}
\end{figure}

\begin{figure}
\centering
\begin{tabular}{c}
\includegraphics[width=0.45\textwidth]{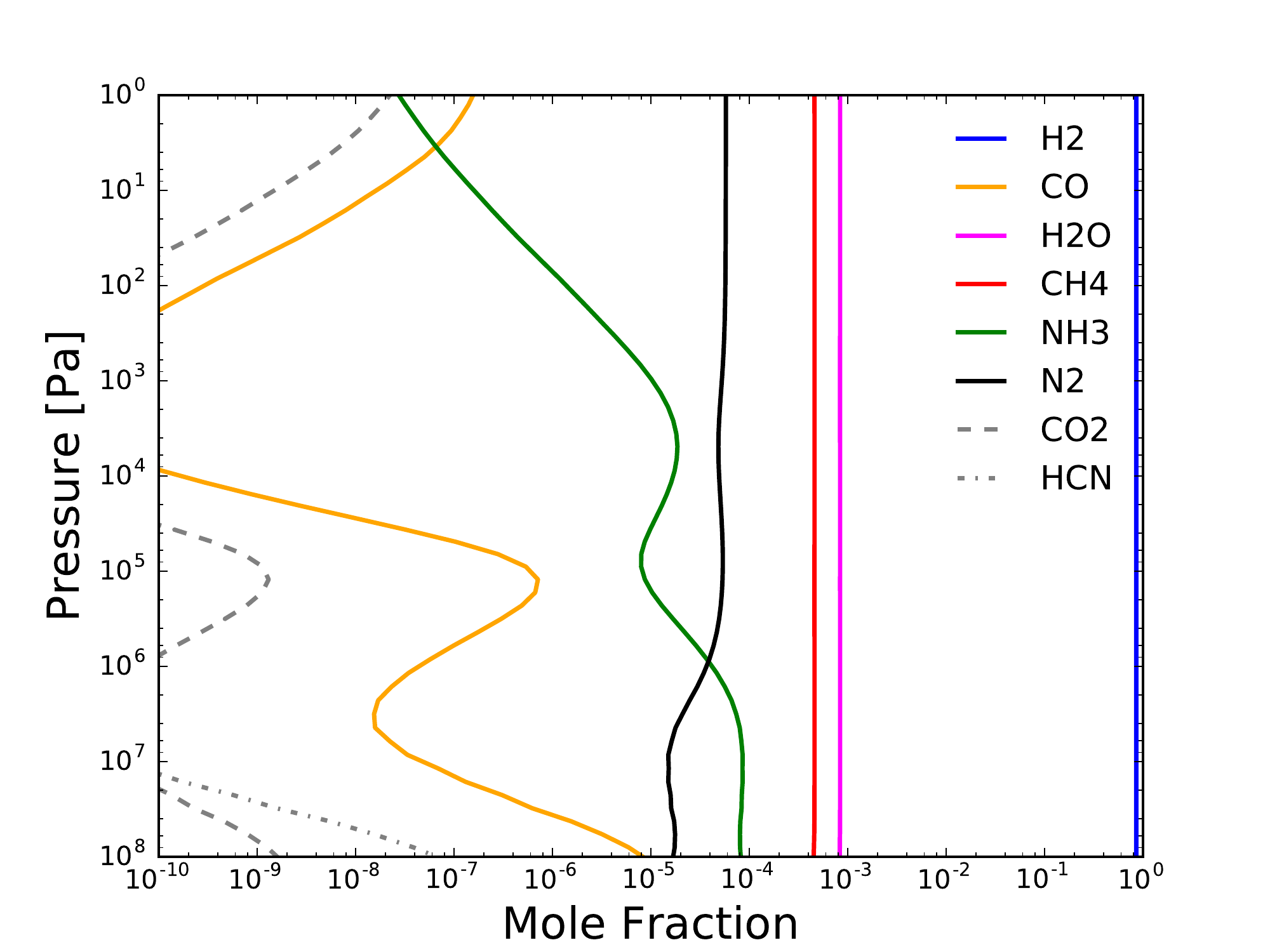} \\
\includegraphics[width=0.45\textwidth]{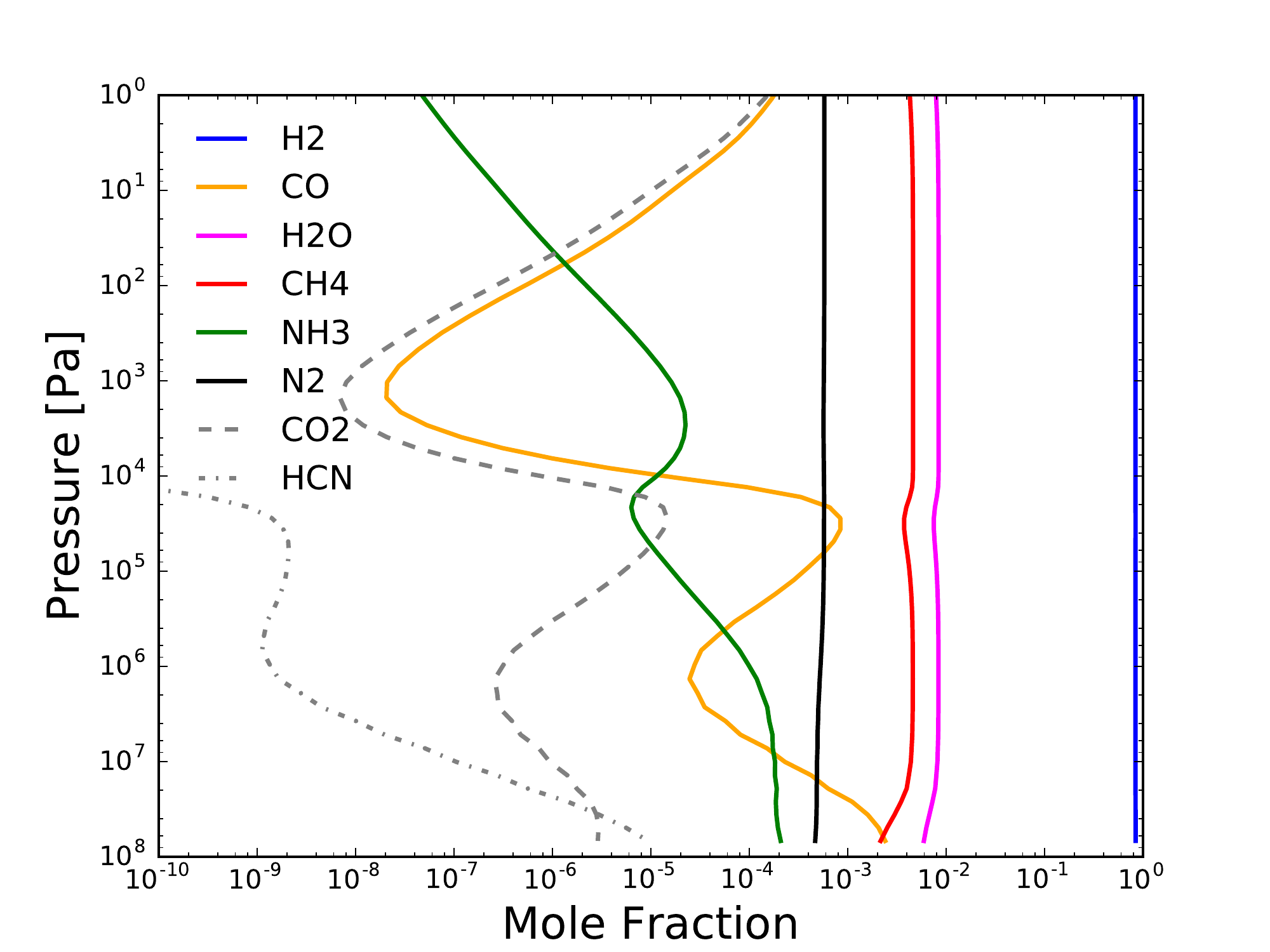} \\
\includegraphics[width=0.45\textwidth]{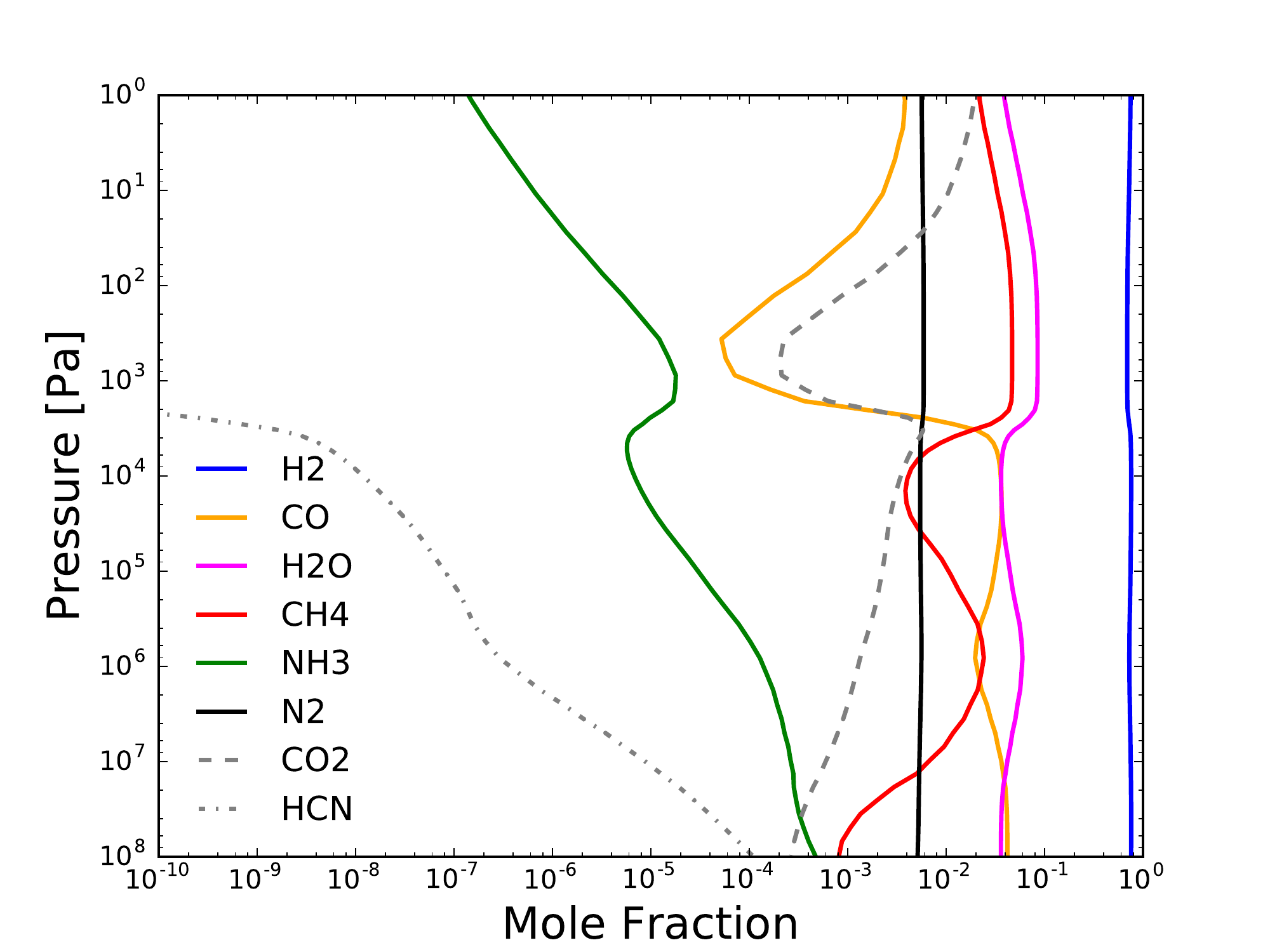}
\end{tabular}
\caption{Figures showing the chemical equilibrium abundance profiles of the major species for GJ~1214b from the 1D ATMO calculation, assuming 1$\times$ (top), 10$\times$ (middle) and 100$\times$ (bottom) solar metallicity.}
\label{figure:chem}
\end{figure}

We adopt the planetary, stellar and orbital parameters for the GJ~1214 system from \citet{Carter2011} and these are summarised in \cref{table:model_params}. For the stellar irradiation spectrum we assume a Phoenix BT-Settl model \citep{Allard2012} with effective temperature $T_{\rm eff} = 3200$ K, surface gravity $\log g = 5.0$ and solar metallicity.

We integrate each simulation for 800 Earth days ($\sim7\times10^7$ s) until the maximum wind velocities have ceased to evolve. Due to the very long radiative and dynamical timescales in the deep atmosphere this cannot be described as a true steady state of the system, and the deep atmosphere ($P\gtrsim10^6$ Pa) is almost entirely determined by the initial condition \citep{MayBA14,MayDB17}. \citet{TreCM17} recently showed using the 2D steady-state model ATMO that the deep atmosphere is warmer than predicted by 1D radiative-convective models, possibly due to the advection of potential temperature. It is not feasible to run GCMs for long enough integration times for the deep atmosphere to converge, though previous simulations have shown that the deep atmosphere does appear to evolve towards higher temperatures \citep{AmuMB16}.

In previous UM simulations \citep{AmuMB16} the chemical abundances were computed at each radiative time step. Since the coupled Gibbs energy minimisation scheme is significantly more computationally expensive than the previous methods it is desirable to perform the chemistry calculations less frequently. We tested a range of chemical time steps, from 150 s (equivalent to one radiative time step) to 3000 s (equivalent to twenty radiative time steps) and found negligible difference between the results; the zonal-mean zonal wind and global maximum wind velocities showed differences of $\lesssim0.1\%$. We therefore adopt the longest chemical time step that we tested which provides a significant increase in efficiency while retaining a high level of accuracy, in terms of the resulting wind velocities.

We use slightly different values for the vertical damping compared with \citet{MayBA14,MayBA14b} and \citet{AmuMB16}. We found that the R100 simulation failed in the early stages of the calculation, while the model is far from a steady-state, due to a transient numerical instability with rapidly oscillating features in the vertical wind field. The exact cause of this remains unclear but the features coincide with a sharp gradient in the H$_2$O, CH$_4$ and CO mole fractions (\cref{figure:chem} illustrates such a feature) and may therefore be due to sharp vertical gradients in the heating rates. We found that adjusting the magnitude ($C=0.20$), vertical extent ($\eta_s$ = 0.70) and shape (polar-dipped, rather than horizontally-uniform) of the sponge layer was sufficient to avoid this numerical instability. We do not expect this to significantly effect the resulting wind velocities \citep{MayBA14,MayBA14b}. Eventually the oscillations reduce, and do not reappear, as the model is integrated to longer times. For consistency we adopt the same sponge parameters for all simulations presented here. We note that such an instability is unlikely to occur in the case of non-equilibrium chemistry where the abundances are expected to be well-mixed \citep{CooS06,AguPV14}, removing sharp gradients in the abundances.

The mean molecular weight of the atmosphere increases with metallicity thereby causing the pressure to decrease with altitude more rapidly. As the UM is a geometric height-based model we adjust the height of the upper boundary $z_{\rm top}$ to capture similar pressure ranges in each simulation. The values of $z_{\rm top}$ used for each simulation are shown in \cref{table:cp_mu}. In each case the resulting minimum pressure at the upper boundary is $\sim$1 Pa and the vertical damping begins at approximately the same pressure level in each simulation.

\begin{table}
\caption{Model parameters for GJ~1214b}
\setlength\extrarowheight{2pt}
\label{table:model_params}
\begin{tabular}{l r}
\hline\hline
Parameter & Value \\
\hline
Radius, $R_{\rm p}$ & 1.45$\times10^7$ m \\
Mass, $M_{\rm p}$ & 3.80$\times10^{26}$ kg \\
Semi major axis, $a$ & 1.23$\times10^{-2}$ AU \\
Surface gravity, $g_{\rm surf}$ & 12.2 m s$^{-2}$ \\
Intrinsic temperature, $T_{\rm int}$ & 100 K \\
Lower boundary pressure, $P_{\rm bottom}$  & 2$\times10^7$ Pa \\
Rotation rate, $\Omega$ & 4.60$\times10^{-5}$ s$^{-1}$ \\
Vertical damping coefficient, $R_w$ & 0.20 \\
Vertical damping extent, $\eta_s$ & 0.70 \\
Horizontal resolution & 144$\times$90 \\
Vertical resolution & 66 \\
Dynamical time step & 30 s \\
Radiative time step & 150 s \\
Chemical time step & 3000 s \\
\hline
\end{tabular}
\end{table}

\subsection{Initial Conditions}
\label{init}

Each simulation is initialised with zero winds and a horizontally uniform thermal profile, as done for most previous GCM simulations of hot exoplanets \citep[e.g.][]{AmuMB16,Showman2009,KatSF14}. We initialise the UM with a 1D ATMO thermal profile, computed using the same planetary and stellar parameters (\cref{table:model_params}); we assume a zenith angle of $\cos{\theta}=0.5$ and an efficient redistribution of heat by reducing the incoming irradiation by a factor 0.5.

We calculate 1D ATMO thermal profiles assuming 1$\times$, 10$\times$ and 100$\times$ solar metallicity which are used to initialise the R1, R10 and R100 UM simulations, respectively, and are shown in \cref{figure:pt_cp_mmw}. As the metallicity is increased the dayside-average thermal profile shows an overall increase in temperature, a trend also seen in previous 1D models of hot exoplanet atmospheres \citep[e.g.][]{AguVS2014}, due to increases in the overall atmospheric opacity. The D10, D100, C10 and C100 simulations are all initialised with the 1$\times$ solar thermal profile.

The chemical equilibrium abundances of several important species are shown in \cref{figure:chem} for each of these initial thermal profiles. Generally the composition is dominated by H$_2$ and He (the latter not shown) with H$_2$O, CH$_4$ and N$_2$ being the most abundant trace species, and CO and NH$_3$ having smaller abundances. As the metallicity increases the abundances of these species increases at the expense of H$_2$ and He. For the 100$\times$ solar case CO becomes more abundant than CH$_4$ for some pressures. In \cref{figure:chem} we show the abundances of CO$_2$ and HCN, to illustrate their increasing abundance with metallicity. However, we do not currently include these species in the opacity calculations in our 3D model.

We estimate the global values of $c_{P,{\rm dyn}}$ and $R$ from these 1D profiles. We first calculate $\mu$ and $c_P$ for each level of the 1D profile and then perform an arithmetic mean (not weighted) along the profile to estimate a mean global value. The 1D profiles of $\mu$ and $c_P$ as well as the arithmetic mean for each metallicity case are shown in \cref{figure:pt_cp_mmw}. The arithmetic mean values used in the UM simulations are also shown in \cref{table:cp_mu}.

As the metallicity is increased $\mu$ increases due to the relatively larger abundances of species heavier than H$_2$ and He. As the metallicity increases from 1$\times$ to 100$\times$ solar $\mu$ increases from 2.33 g mol$^{-1}$ to 4.34 g mol$^{-1}$. The specific heat capacity decreases from $\sim1.2\times10^4$ ${\rm J}~{\rm kg}^{-1}~{\rm K}^{-1}$ to $\sim7\times10^3$ ${\rm J}~{\rm kg}^{-1}~{\rm K}^{-1}$. 

\begin{table*}
\caption{Simulation specific parameters} 
\centering 
\setlength\extrarowheight{2pt}
\label{table:cp_mu}
\begin{tabular}{l c c c c c} 
\hline\hline 
& R1 & R10 & R100 & D10 and C10 & D100 and C100 \\
\hline 
Specific heat capacity, $c_{p,{\rm dyn}}$  [${\rm J}~{\rm kg}^{-1}~{\rm K}^{-1}$] & 12343 & 11518 & 7025 & 11518 & 7025   \\ 
Specific gas constant, $R$ [${\rm J}~{\rm kg}^{-1}~{\rm K}^{-1}$] & 3573.5 & 3299.1 & 1917.3 & 3299.1 & 1917.3 \\
Height of upper boundary, $z_{\rm top}$ [$\times10^6$ m] & 4.9 & 5.5 & 2.4 & 4.5 & 2.3 \\
\hline
\end{tabular}
\end{table*}

%% file: results.tex


\section{Results}
\label{sec:results}

In this section we present the results from our GCM simulations of GJ~1214b, focussing on the thermal structure, the circulation and the emission from the top of the atmosphere.

\subsection{The thermal structure}
\label{section:results_temp}

\begin{figure*}
\centering
\begin{subfigure}[b]{0.37\textwidth}
\includegraphics[width=\textwidth]{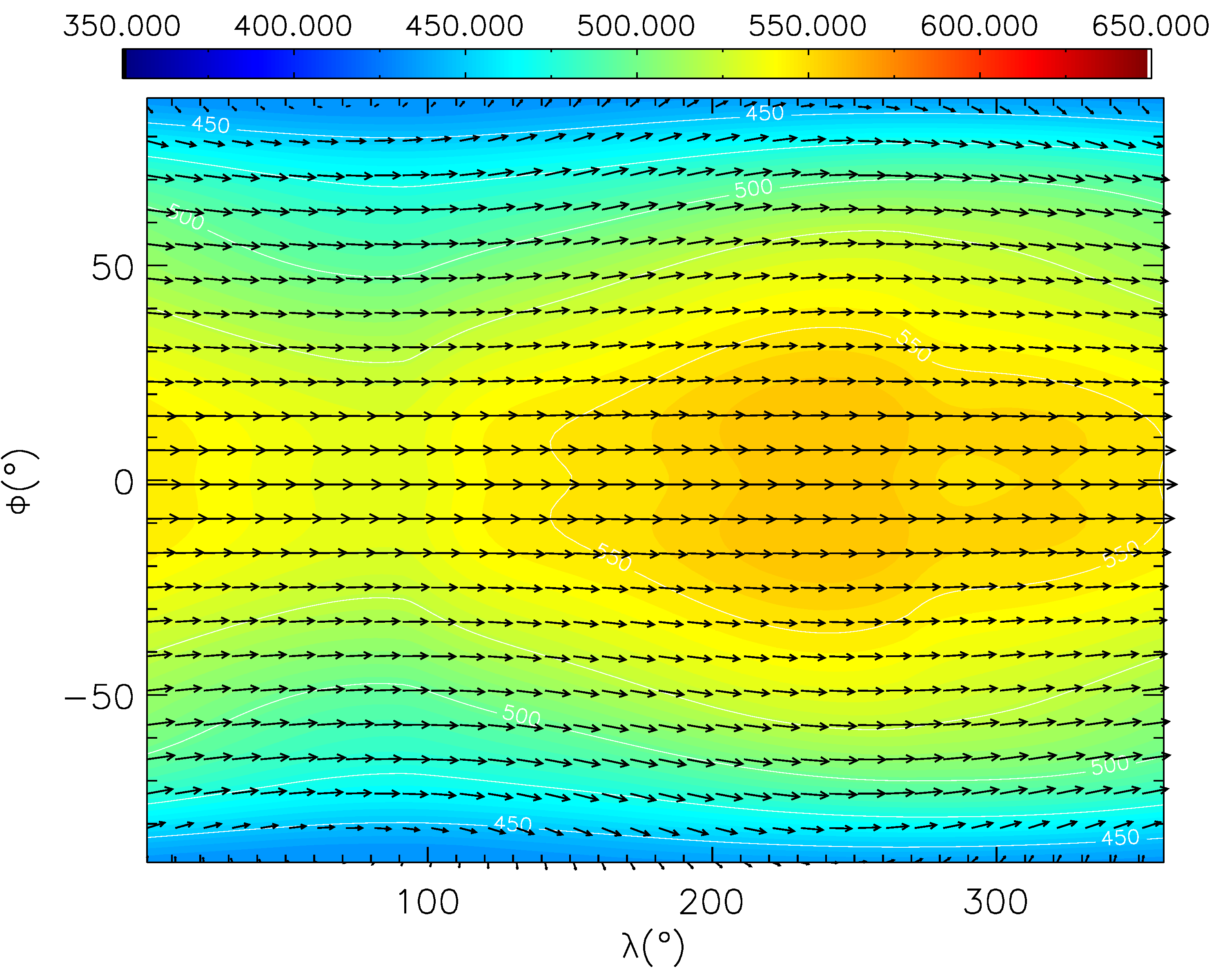} 
\caption{R1}
\end{subfigure}

\begin{subfigure}[b]{0.37\textwidth}
\includegraphics[width=\textwidth]{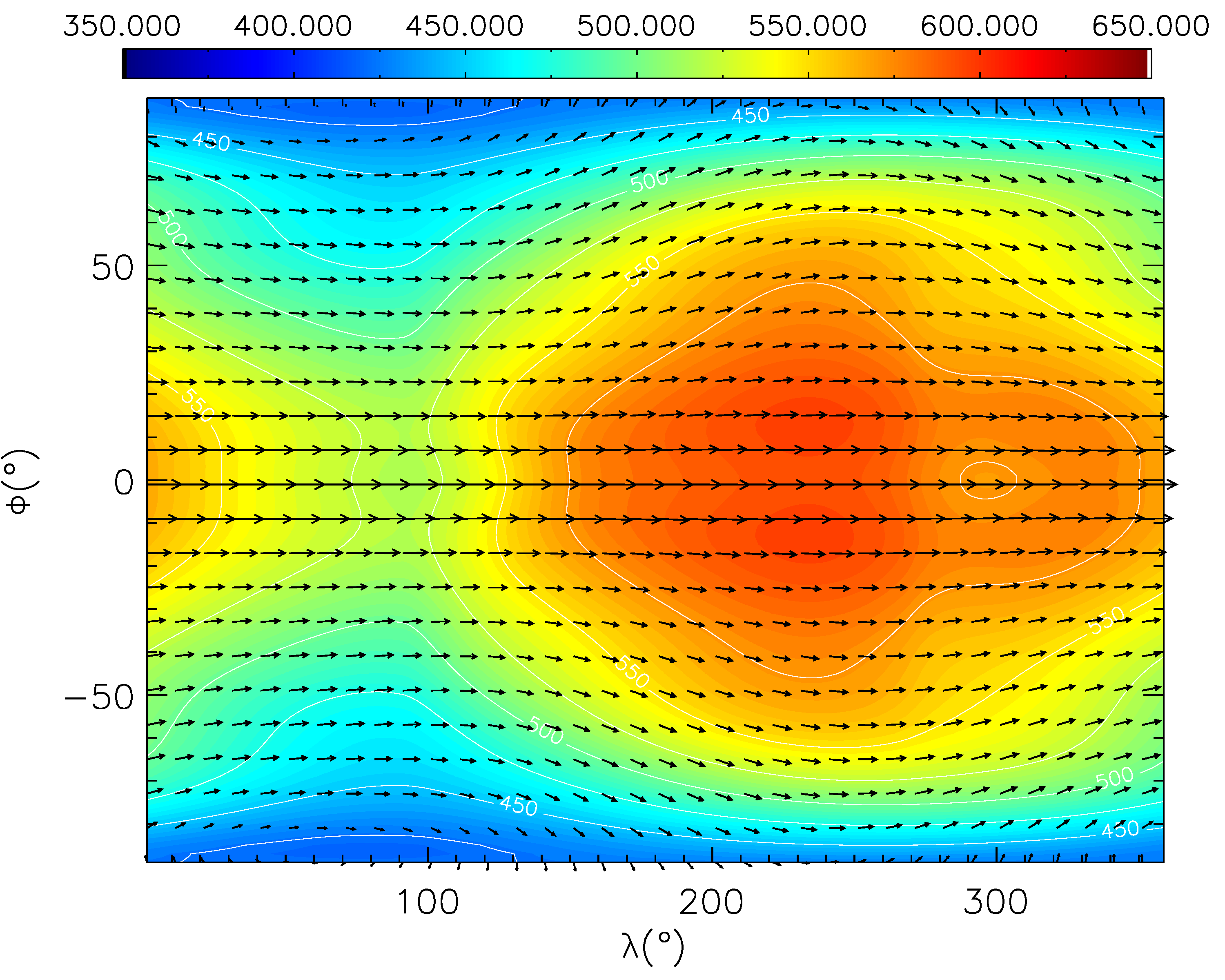} 
\caption{R10}
\end{subfigure}
~
\begin{subfigure}[b]{0.37\textwidth}
\includegraphics[width=\textwidth]{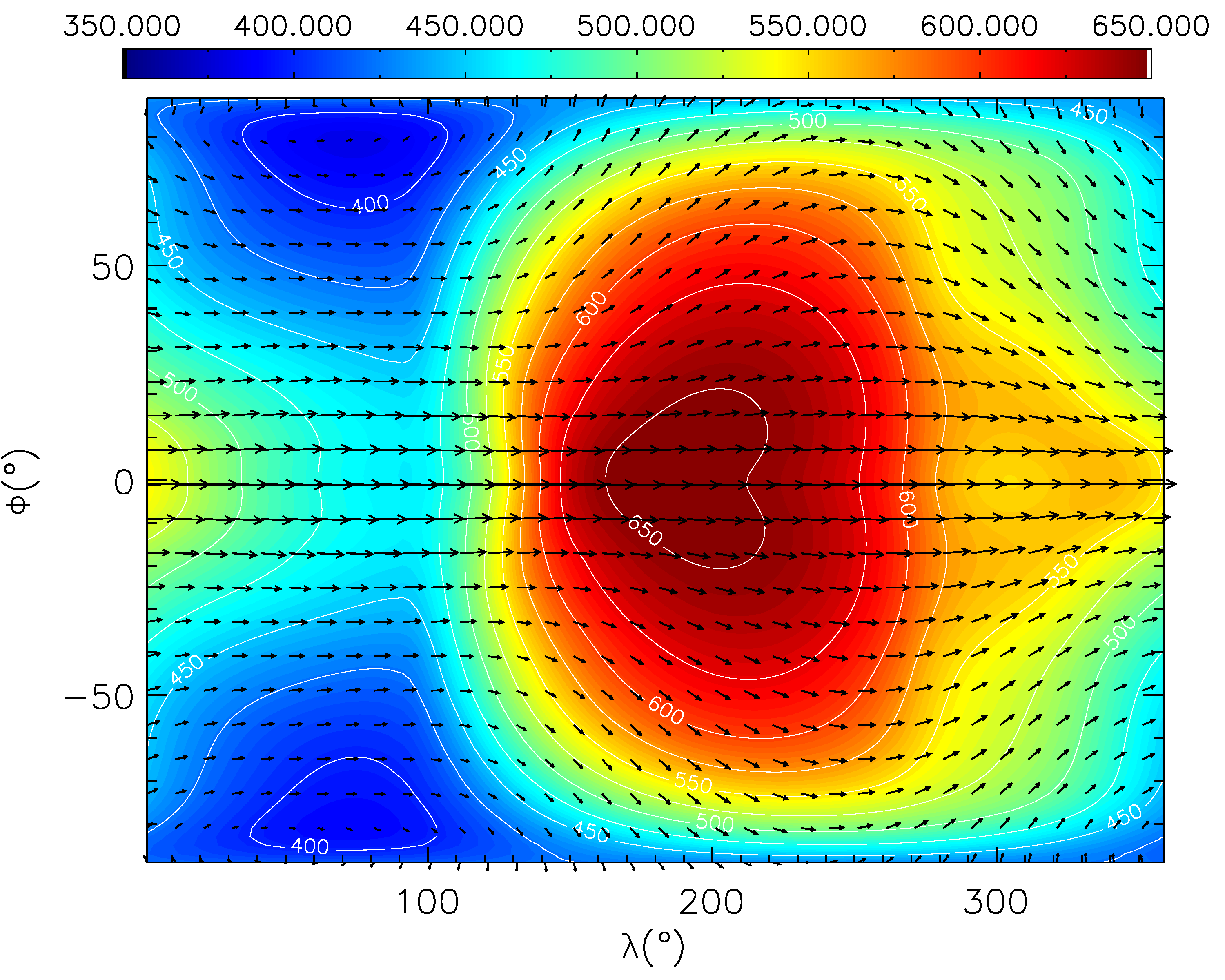} 
\caption{R100}
\end{subfigure}

\begin{subfigure}[b]{0.37\textwidth}
\includegraphics[width=\textwidth]{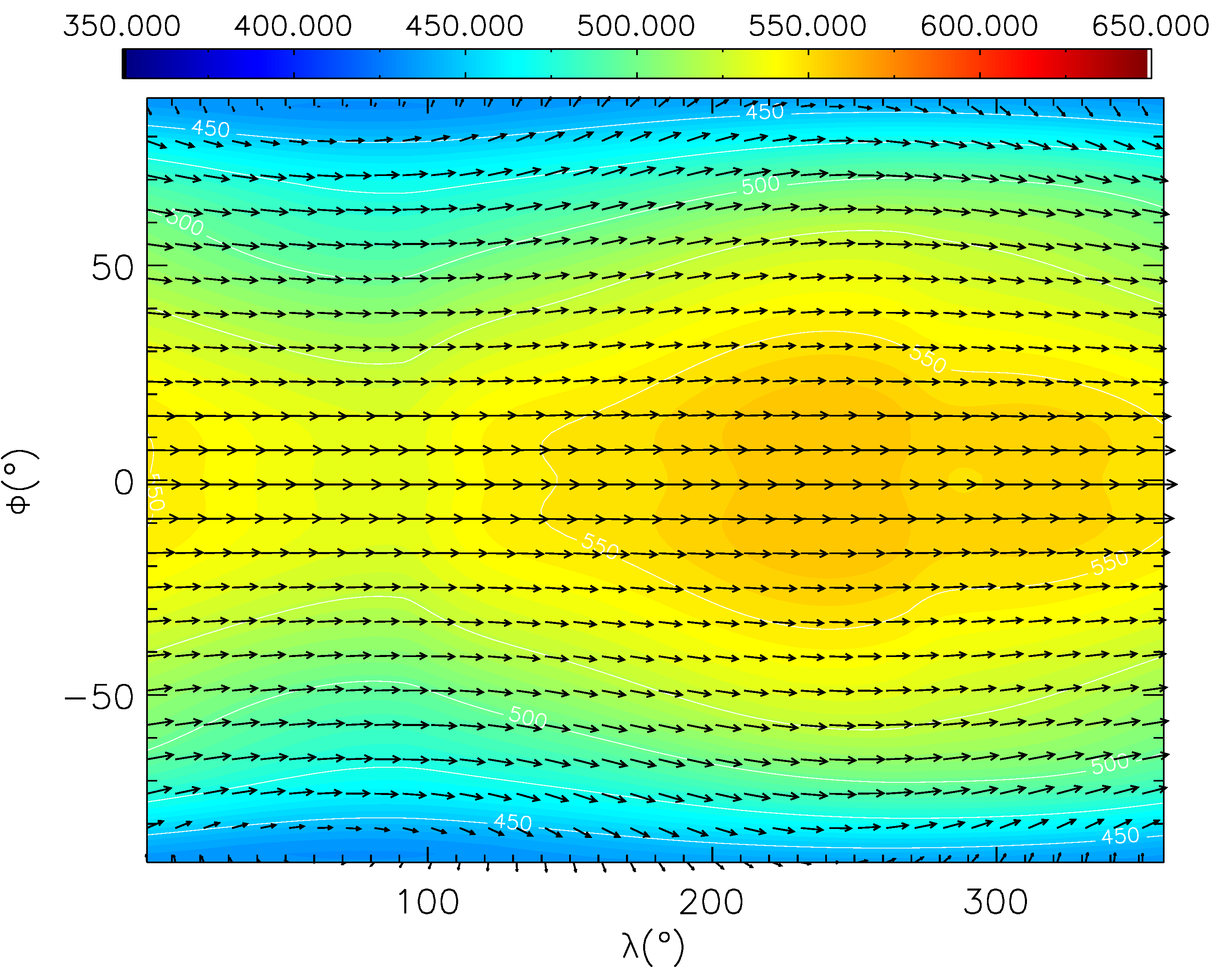} 
\caption{D10}
\end{subfigure}
~
\begin{subfigure}[b]{0.37\textwidth}
\includegraphics[width=\textwidth]{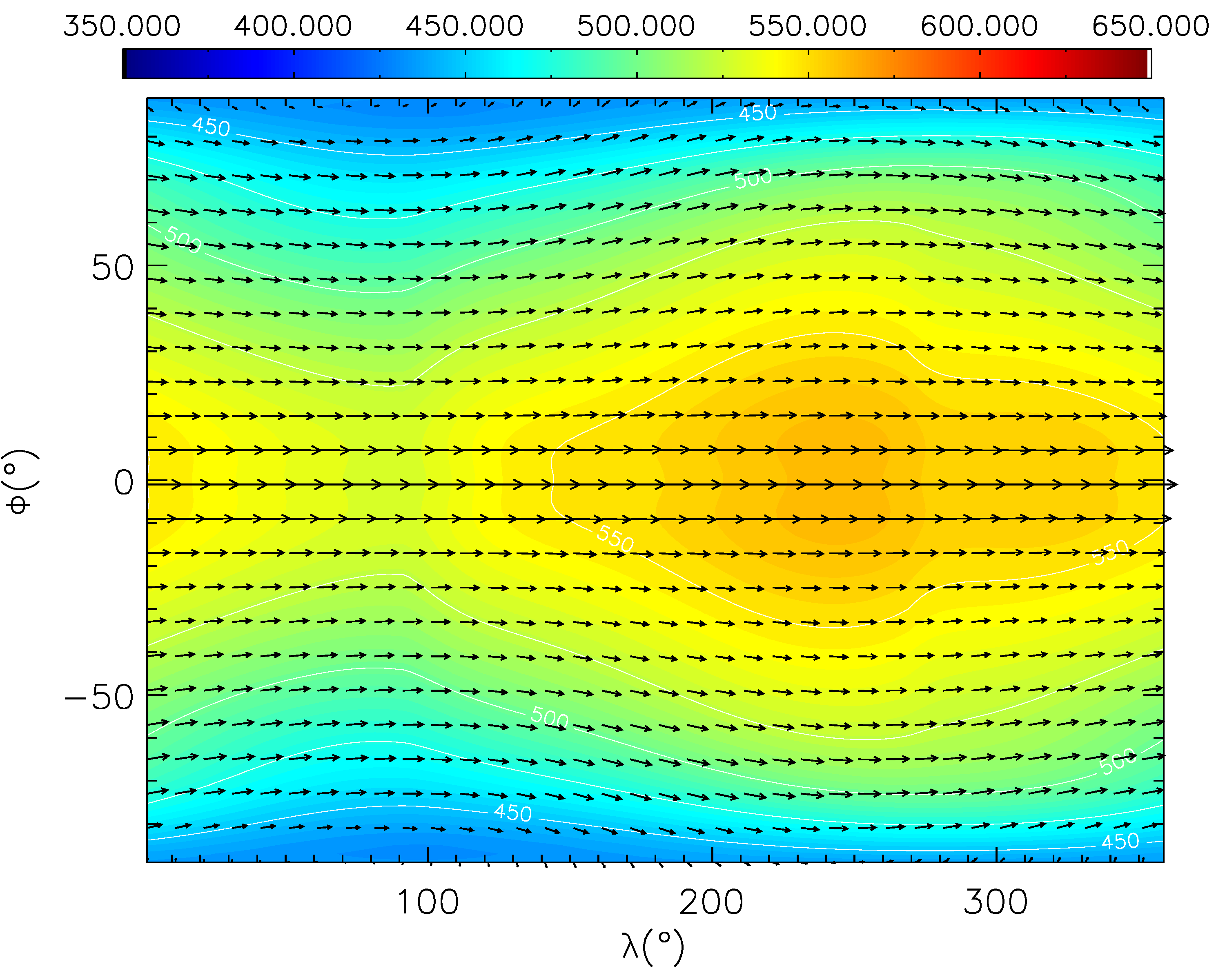} 
\caption{D100}
\end{subfigure}

\begin{subfigure}[b]{0.37\textwidth}
\includegraphics[width=\textwidth]{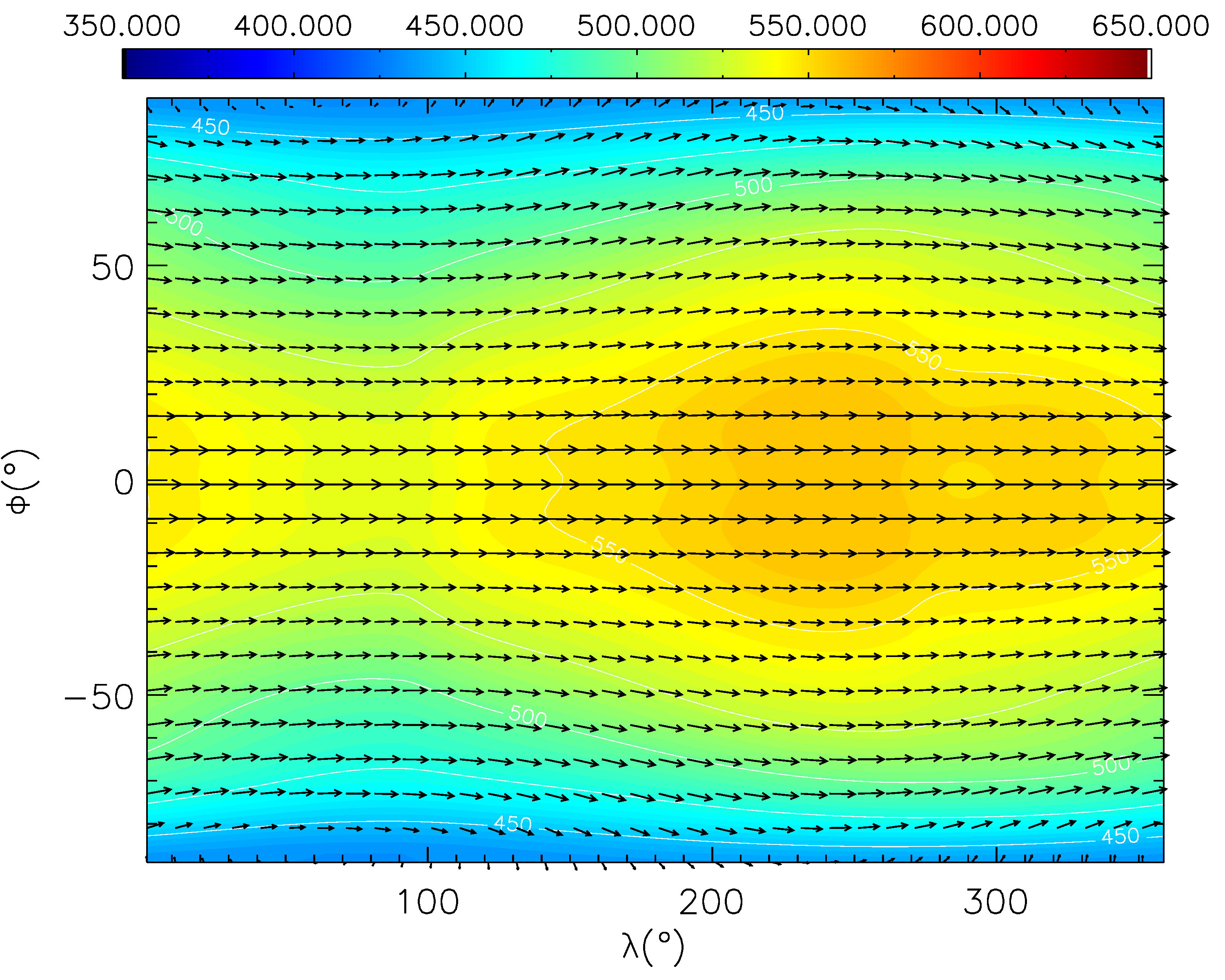} 
\caption{C10}
\end{subfigure}
~
\begin{subfigure}[b]{0.37\textwidth}
\includegraphics[width=\textwidth]{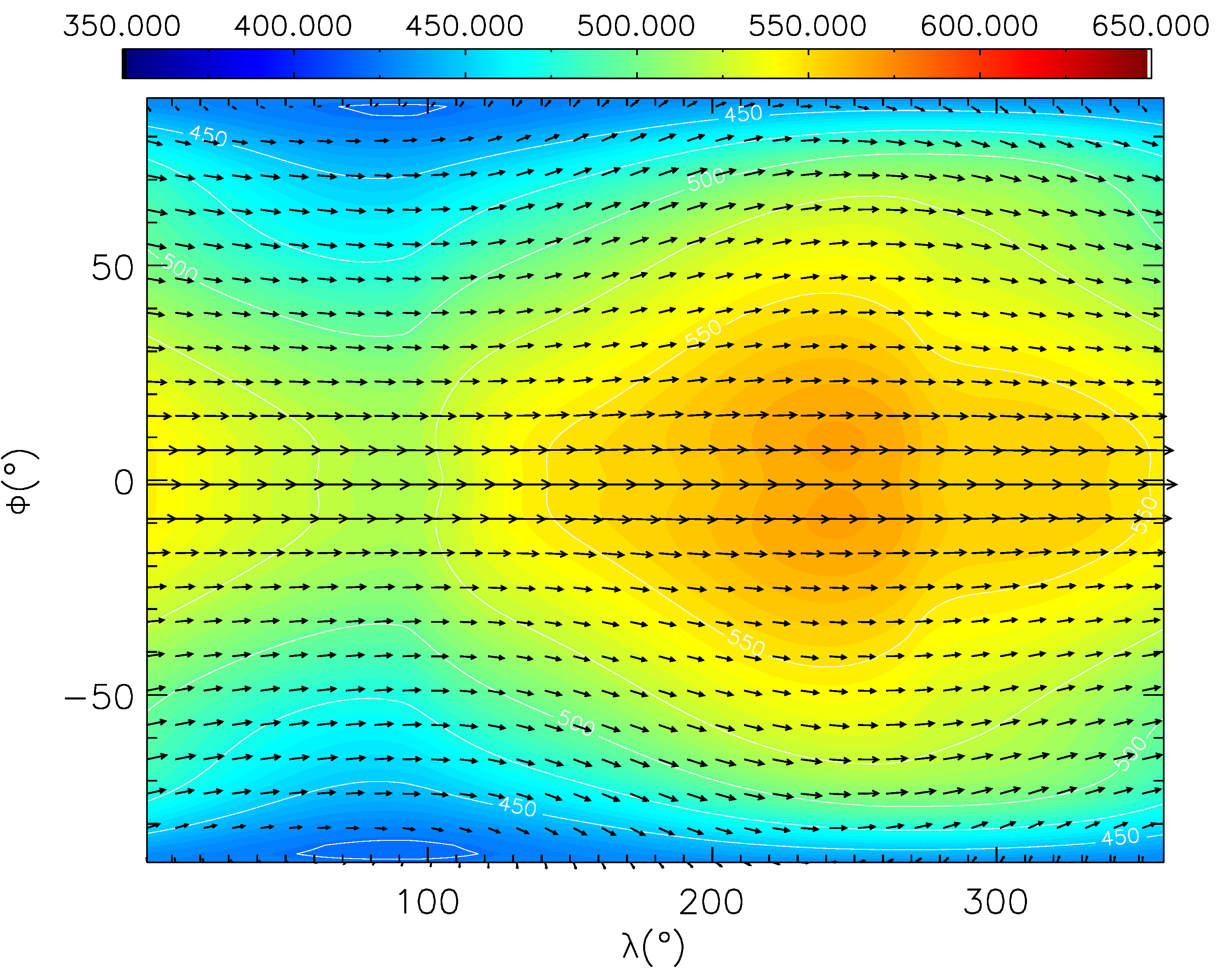} 
\caption{C100}
\end{subfigure}

\caption{Temperature (colour scale) and horizontal wind velocities (vector arrows) on a surface of constant pressure at $P=100$ Pa for each simulation after 800 days.}
\label{figure:temperature_100Pa}
\end{figure*}

\begin{figure*}
\centering
\begin{subfigure}[b]{0.37\textwidth}
\includegraphics[width=\textwidth]{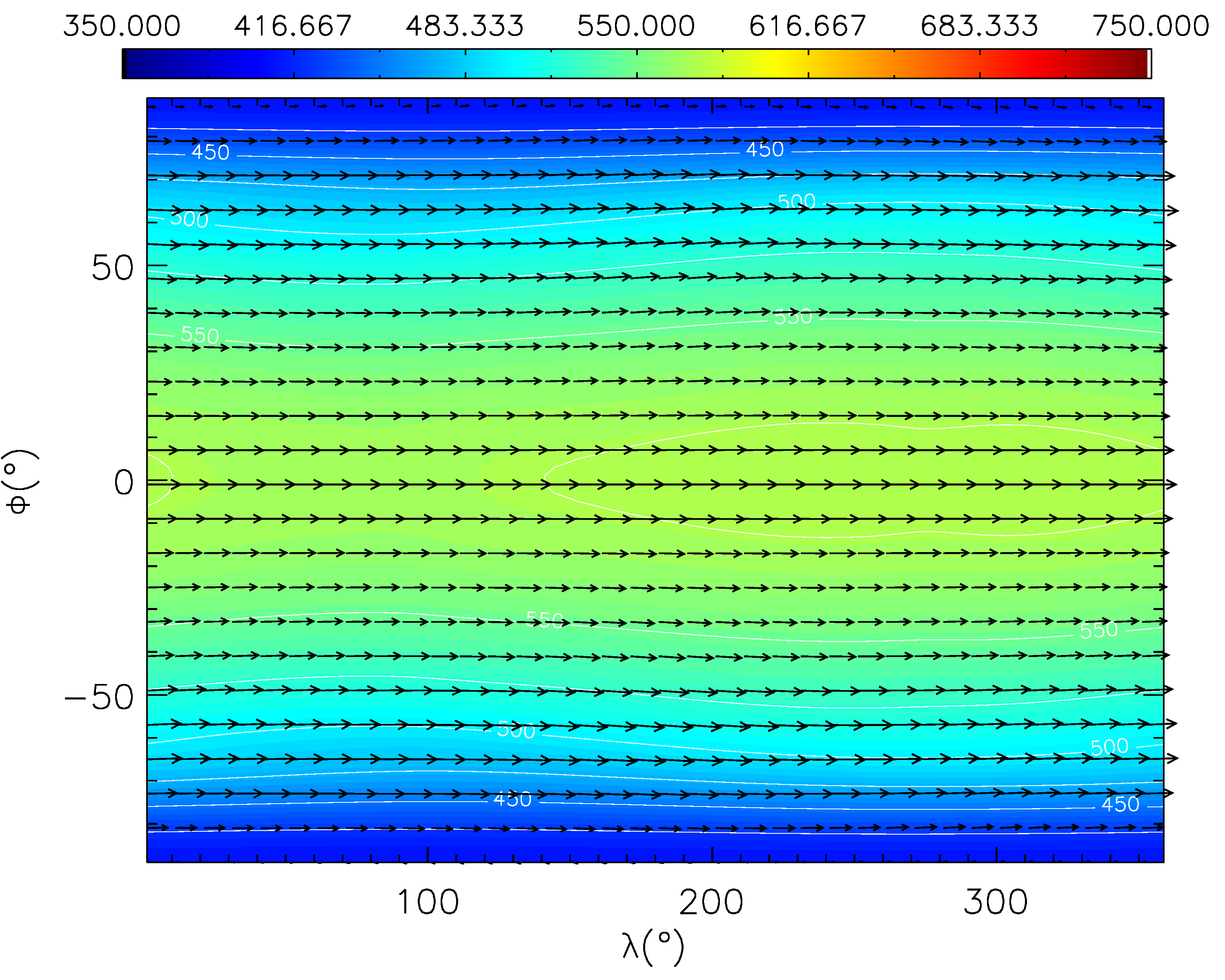} 
\caption{R1}
\end{subfigure}

\begin{subfigure}[b]{0.37\textwidth}
\includegraphics[width=\textwidth]{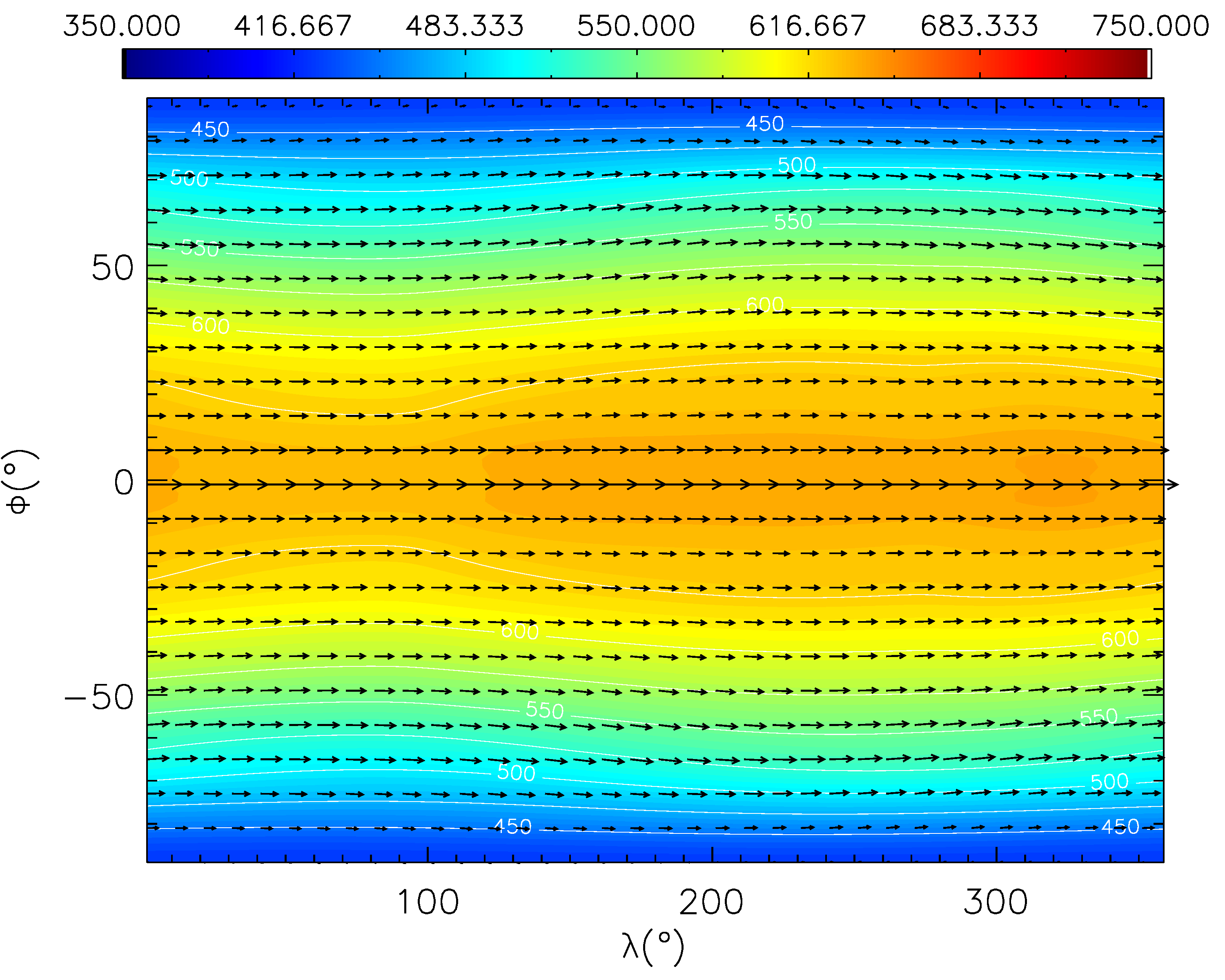} 
\caption{R10}
\end{subfigure}
~
\begin{subfigure}[b]{0.37\textwidth}
\includegraphics[width=\textwidth]{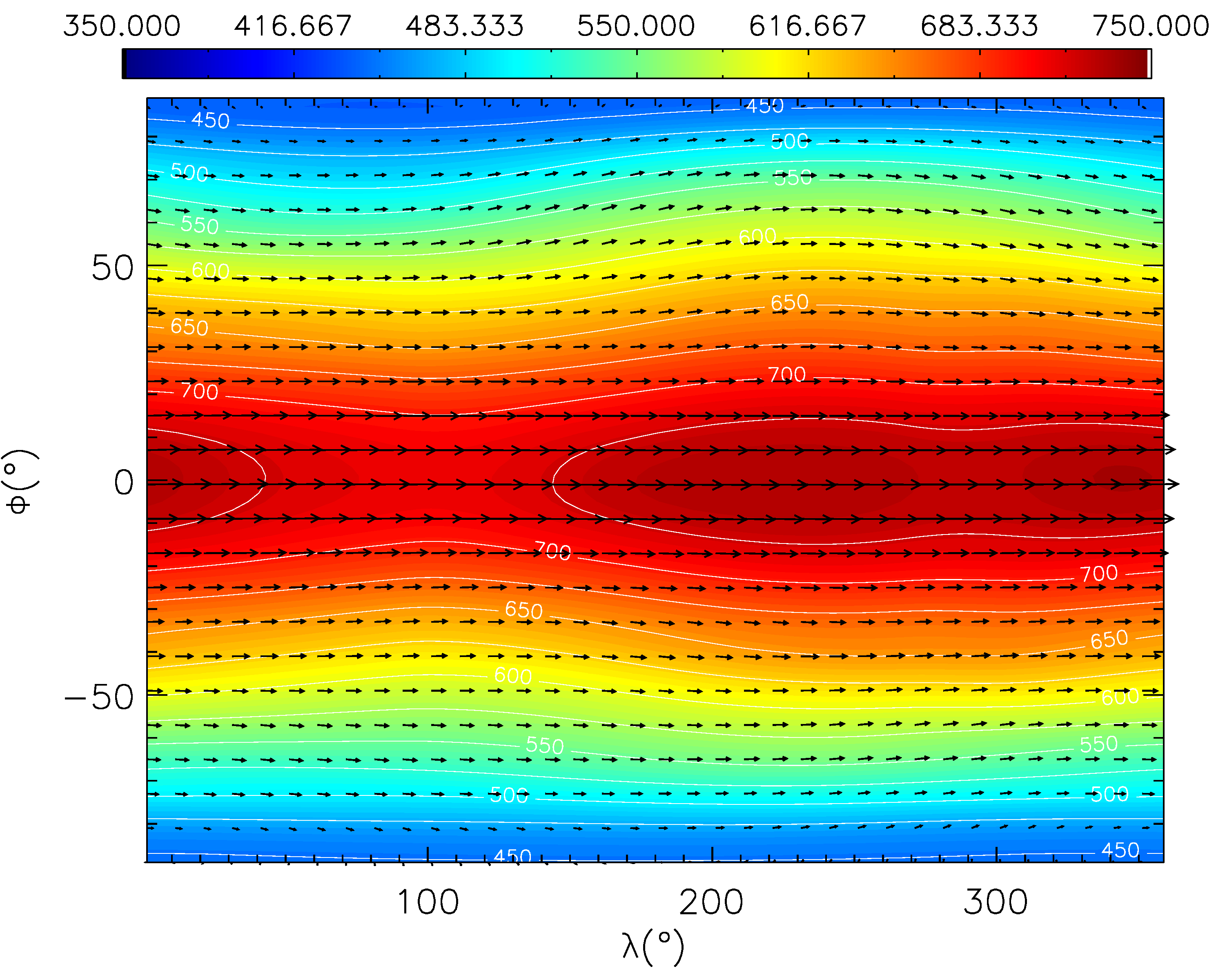} 
\caption{R100}
\end{subfigure}

\begin{subfigure}[b]{0.37\textwidth}
\includegraphics[width=\textwidth]{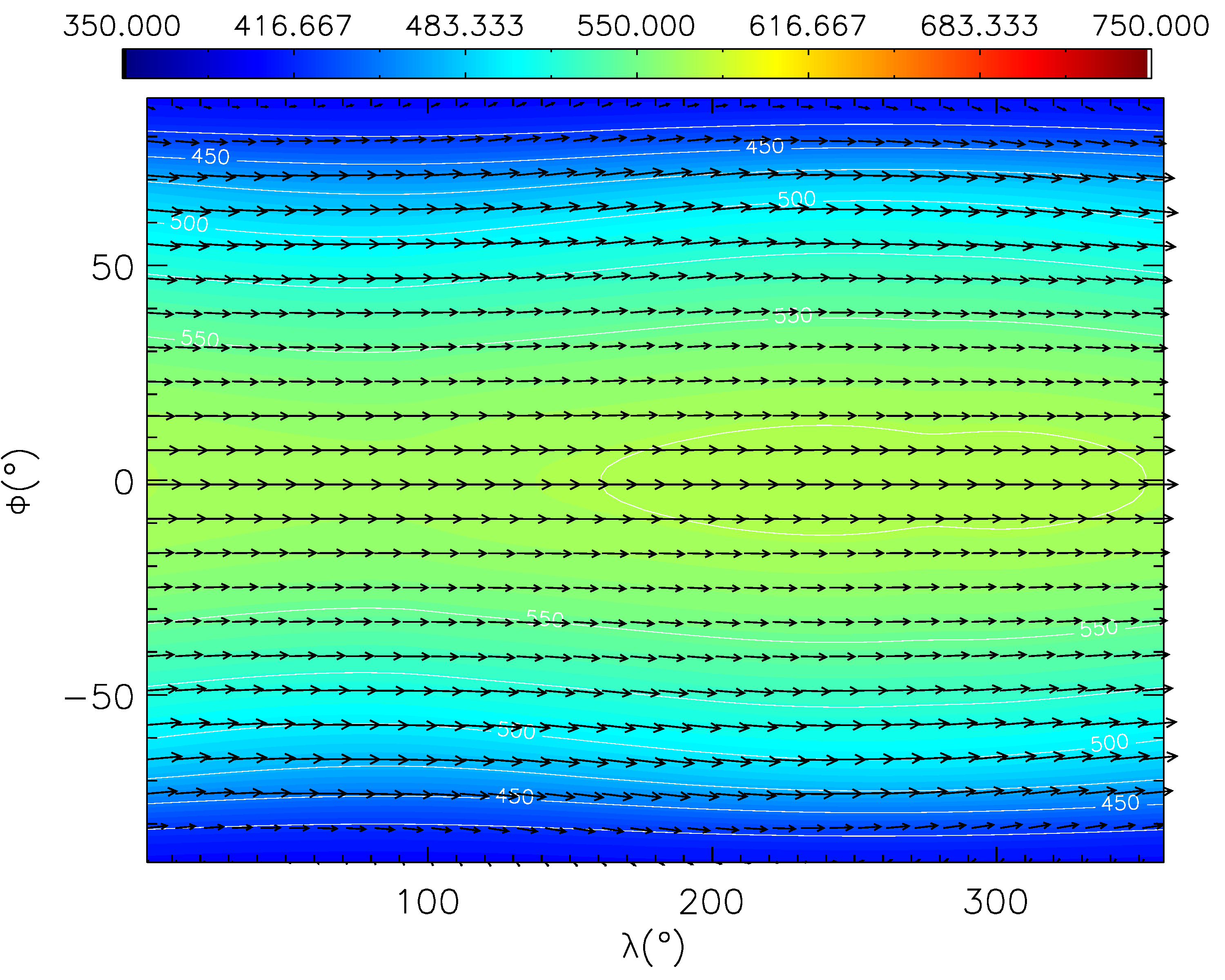} 
\caption{D10}
\end{subfigure}
~
\begin{subfigure}[b]{0.37\textwidth}
\includegraphics[width=\textwidth]{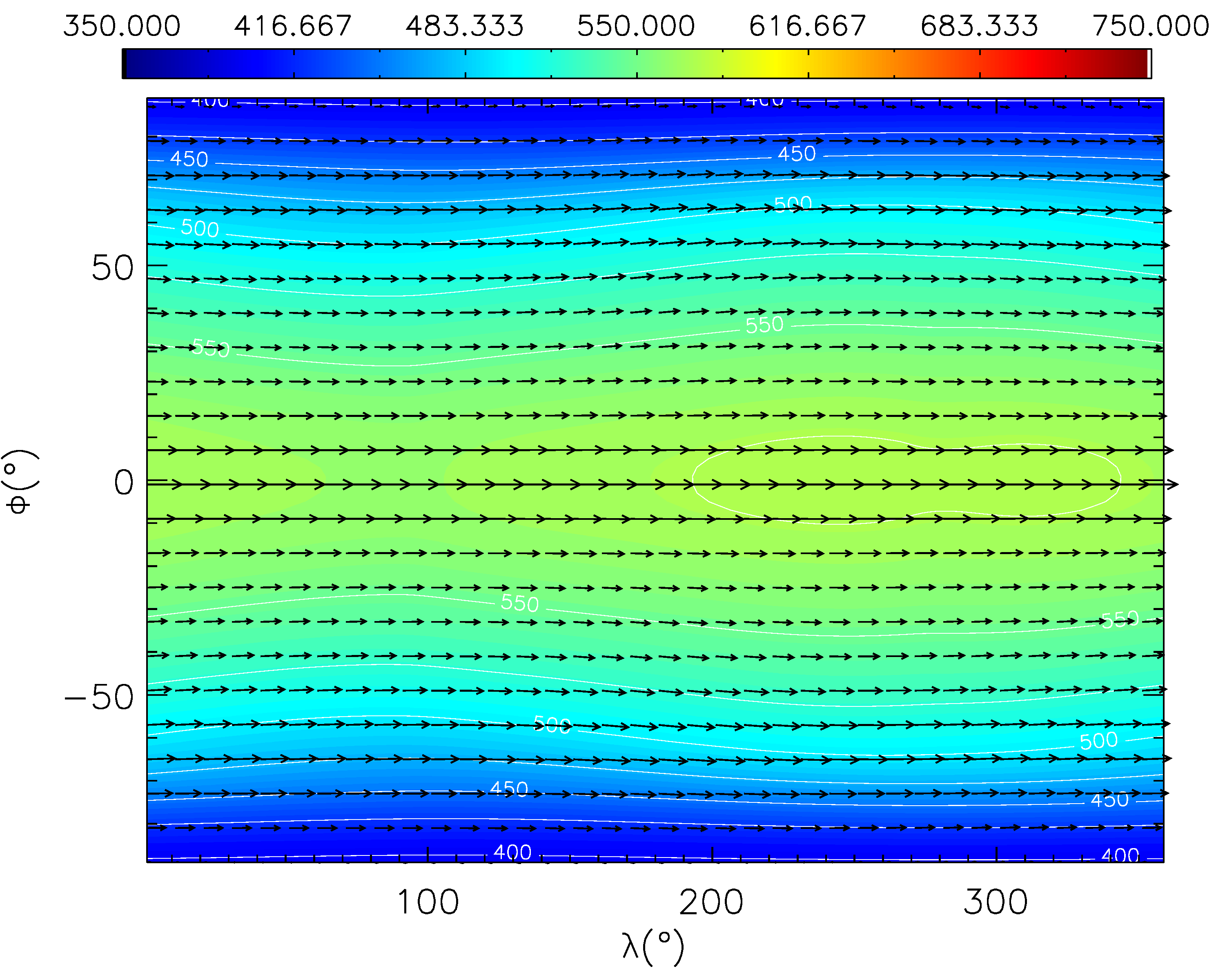} 
\caption{D100}
\end{subfigure}

\begin{subfigure}[b]{0.37\textwidth}
\includegraphics[width=\textwidth]{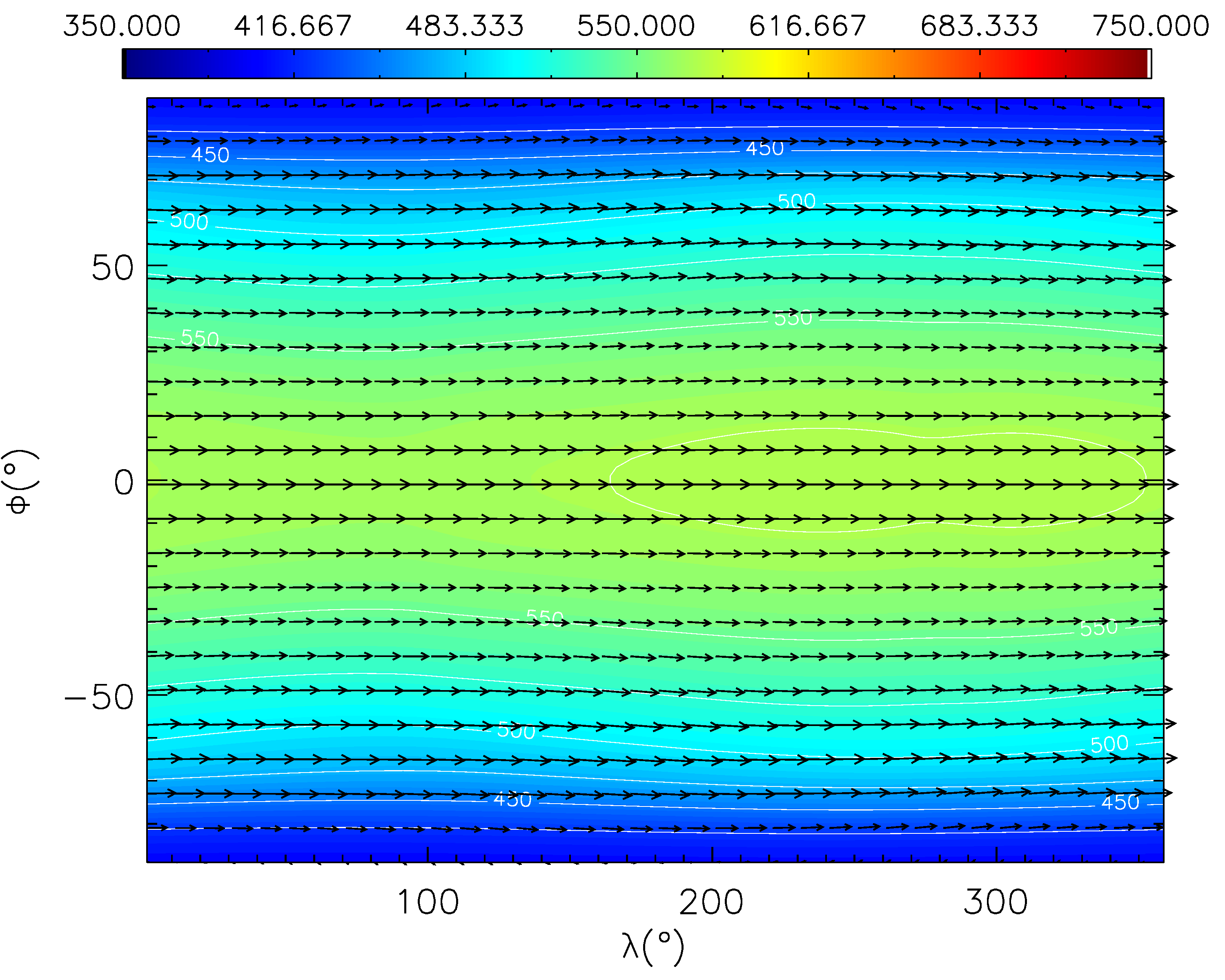} 
\caption{C10}
\end{subfigure}
~
\begin{subfigure}[b]{0.37\textwidth}
\includegraphics[width=\textwidth]{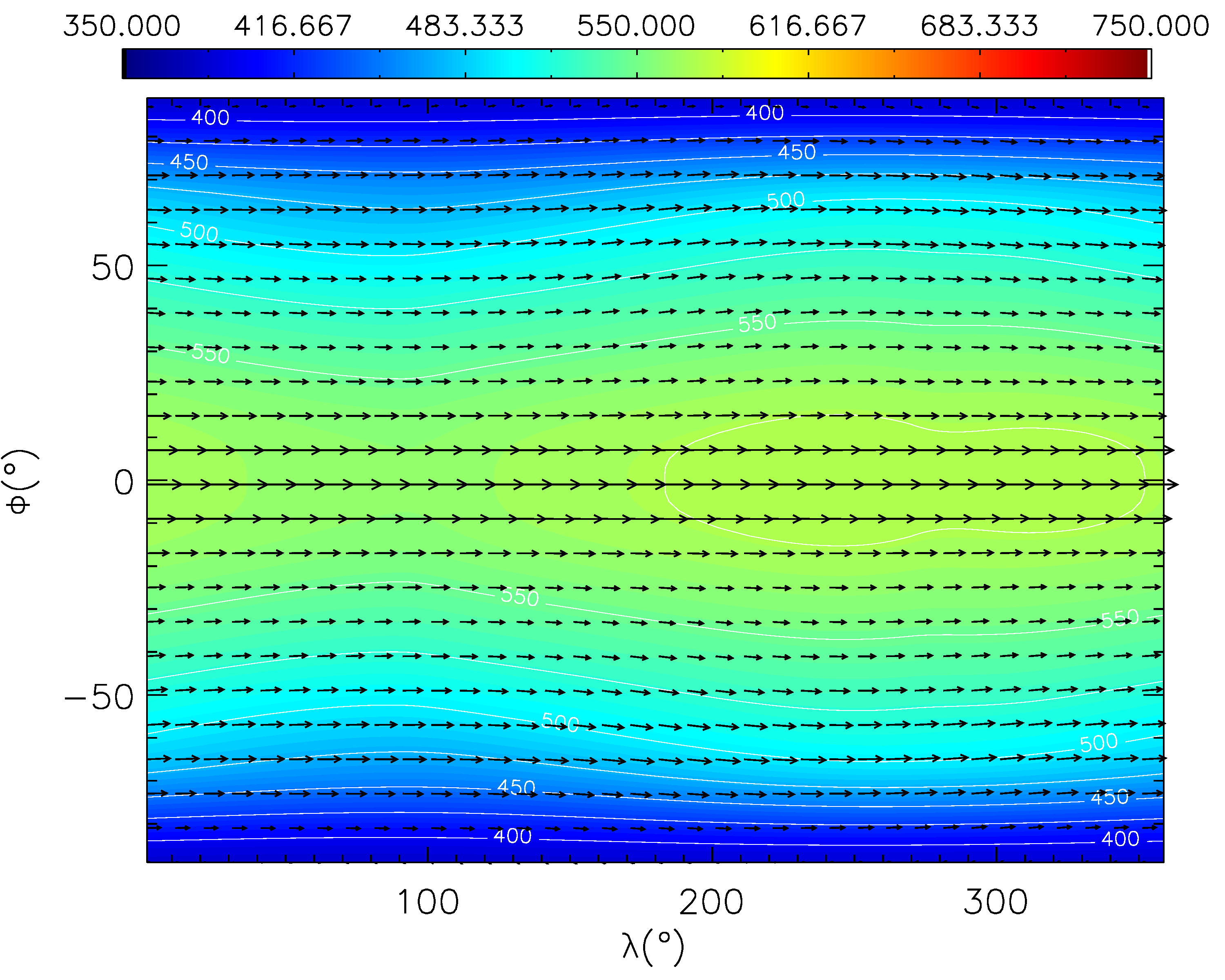} 
\caption{C100}
\end{subfigure}

\caption{As \cref{figure:temperature_100Pa} but at $P=3000$ Pa.}
\label{figure:temperature_3000Pa}
\end{figure*}

\cref{figure:temperature_100Pa} shows the temperature and horizontal wind velocity vectors on a surface of constant pressure at 100 Pa for each simulation. For R1 the dayside hemisphere is generally warmer than the nightside hemisphere. The highest temperatures occur at a longitude of $\sim230^{\circ}$, which is shifted eastwards from the sub-stellar point ($180^{\circ}$). The meridional temperature gradient is more significant than the zonal temperature gradient for R1 at this pressure level. The high latitude regions are more than 100 K cooler than the equatorial regions, while the maximum zonal temperature difference is $\sim50$ K. The wind velocity vectors show a rather uniform eastward flow for most of the latitude range.

The R10 and R100 simulations show the trend of a progressively warmer dayside and cooler nightside as the metallicity is increased. This results in a significantly larger zonal temperature gradient for R100 compared with R1. The location of the hot spot has also shifted westwards ($\sim190^{\circ}$), closer to the sub-stellar point. A change is also apparent in the horizontal wind vectors with a more significant meridional (poleward) component for latitudes $>50^{\circ}$ in R100, compared with the largely eastward flow in R1.

The D10 and D100 simulations allow us to isolate the dynamical effect from the radiative heat capacity and opacity effects. There is a negligible variation in temperature and horizontal wind vectors at 100 Pa between the simulations R1, D10 and D100. This test indicates a negligible impact of the dynamical effect on the thermal structure when the metallicity is varied between 1$\times$ and 100$\times$ solar. 

We investigate the combined dynamical and radiative heat capacity effects in the C10 and C100 simulations. The effect of increasing the metallicity has a small impact on the thermal structure with a slight increase in the zonal temperature gradient. This indicates that the radiative heat capacity effect is more important than the dynamical effect, though less important than the opacity effect. 

\cref{figure:temperature_3000Pa} shows temperature and horizontal wind velocity vectors at a deeper pressure level of 3000 Pa. At this pressure we find all simulations have no significant zonal temperature gradient with the largest temperature gradient occuring between the poles and the equator. This meridional temperature gradient increases significantly as the equatorial region becomes hotter as the metallicity is increased in the R10 and R100 simulations. The D10, D100, C10 and C100 show negligible difference with the R1 simulation at this pressure level. This demonstrates the much greater importance of the opacity effect over both the dynamical and radiative heat capacity effects.

\begin{figure}
\centering
\begin{tabular}{c}
	\includegraphics[width=0.45\textwidth]{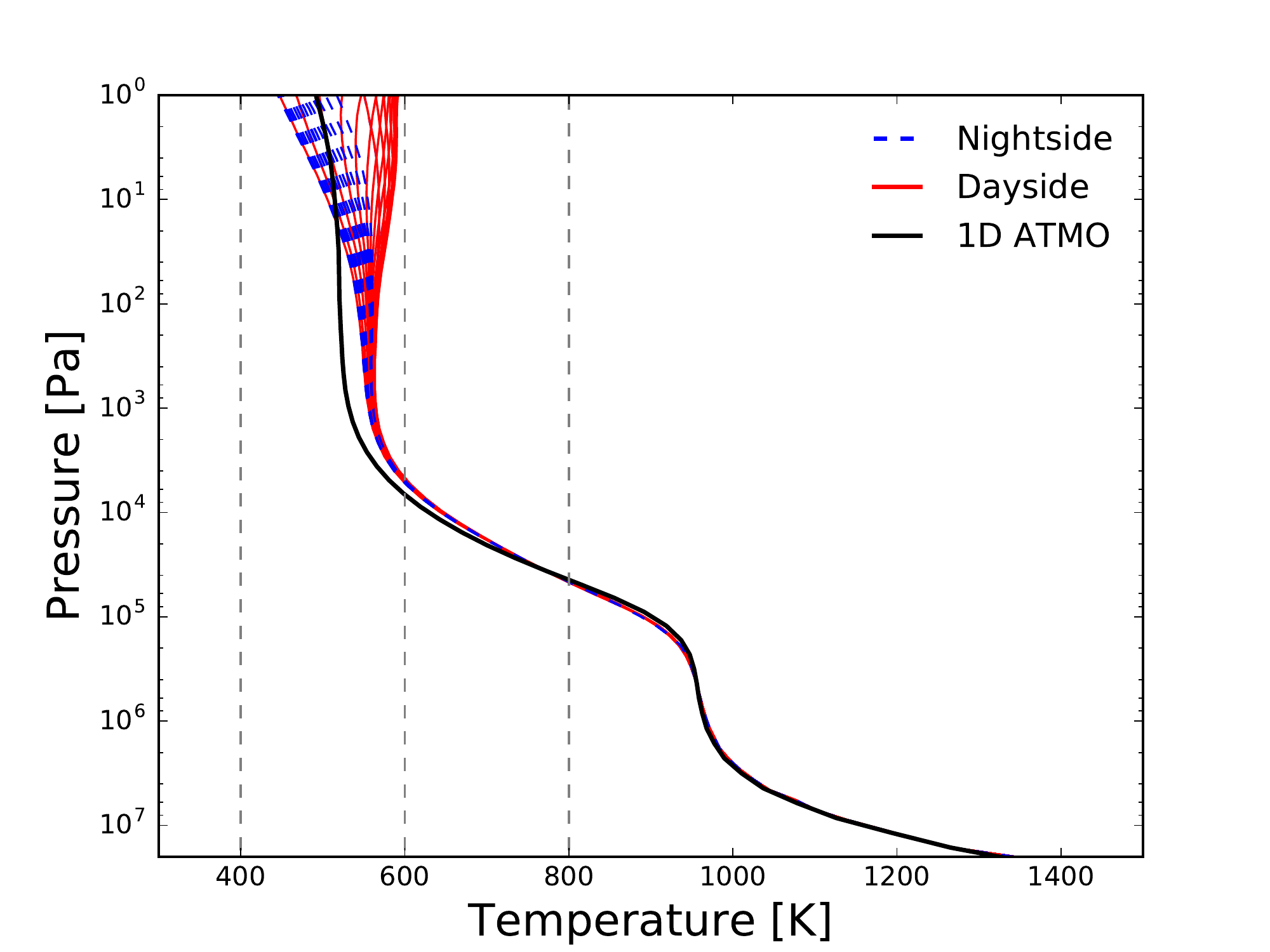} \\
	\includegraphics[width=0.45\textwidth]{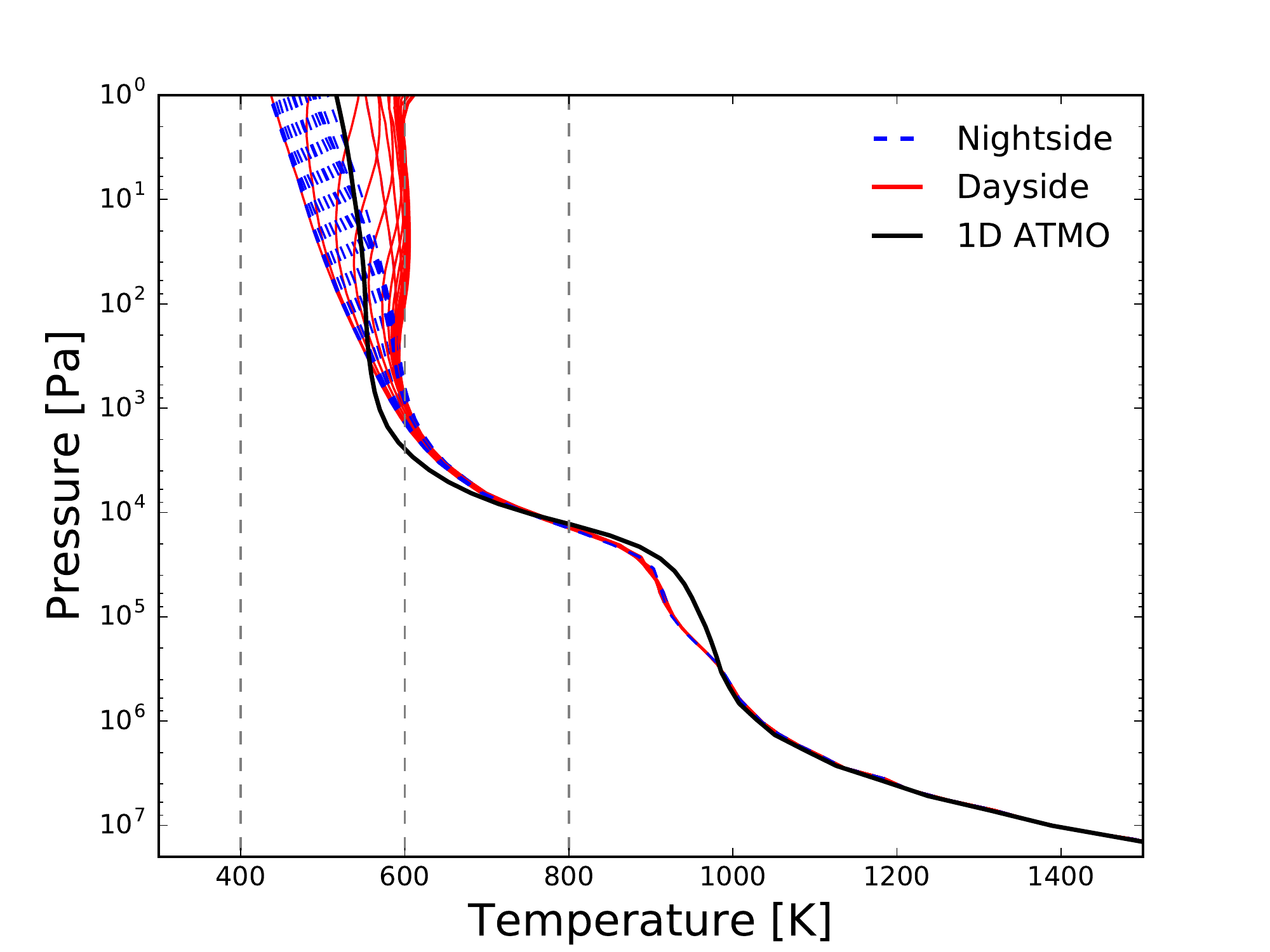} \\
	\includegraphics[width=0.45\textwidth]{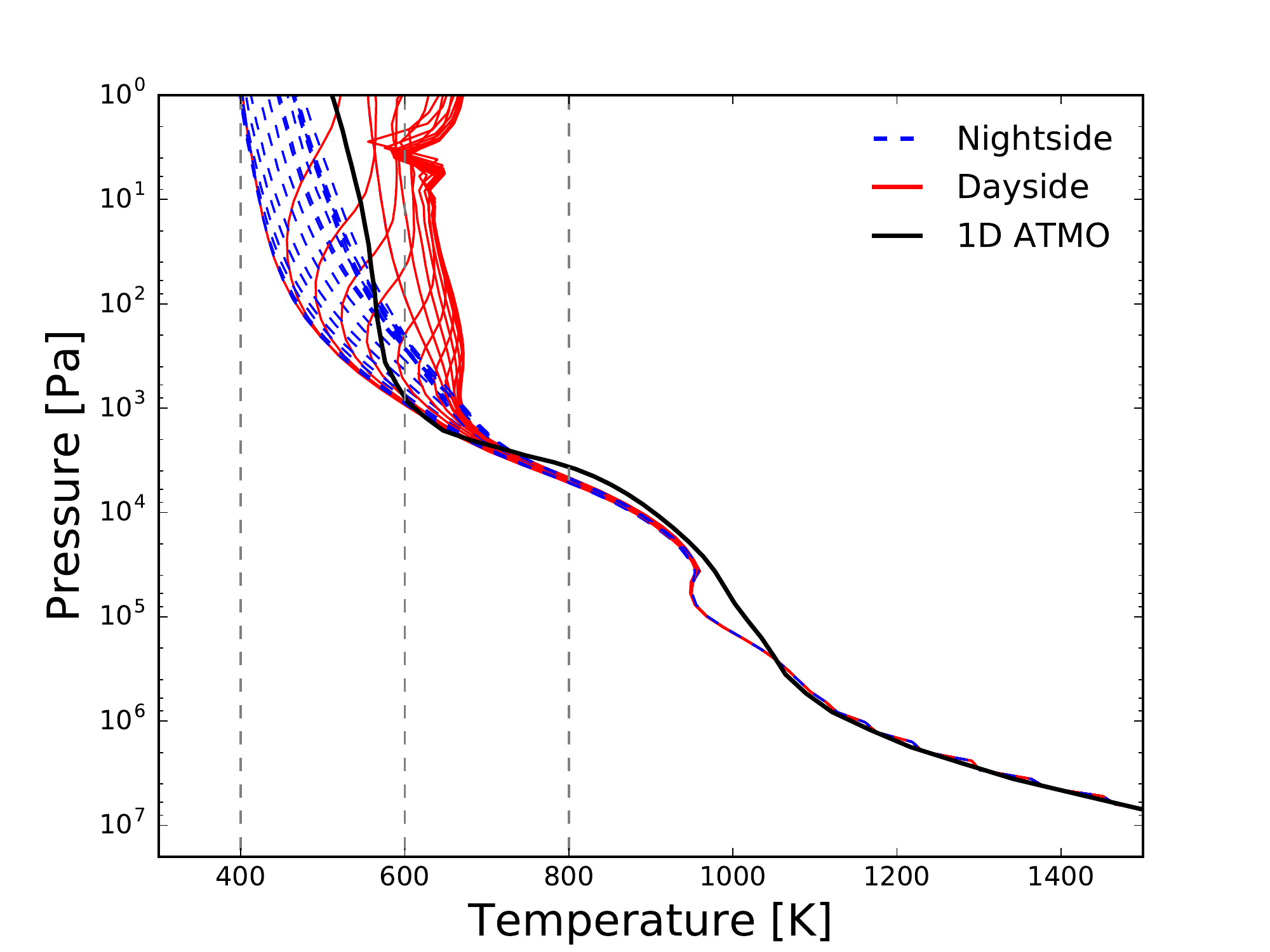} 
\end{tabular}
\caption{Pressure-temperature profiles extracted from the 3D grid at the equator ($\phi=0^{\circ}$) for various longitudes, for the R1 (top), R10 (middle), R100 (bottom) simulations. Profiles taken from the dayside are shown in solid red lines while profiles from the nightside are shown in dashed blue lines. The black solid line is the 1D ATMO pressure-temperature profile used to initialise the 3D simulation.}
\label{figure:pt_longitudes}
\end{figure}

\cref{figure:pt_longitudes} shows a series of thermal profiles extracted from the 3D grid around the equator ($\phi = 0^{\circ}$) for a series of longitude points for the simulations R1, R10 and R100. The 1D ATMO thermal profiles used to initialise each 3D simulation are also shown. Generally we find that the temperature is uniform around the equator for $P>10^3$ Pa.  For smaller pressures the zonal temperature gradient tends to increase with decreasing pressure. As the metallicity is increased (when included all three of the dynamical, heat capacity and opacity effects) the temperature of the dayside typically becomes warmer while the nightside becomes cooler, between $1<P<10^3$ Pa.

The thermal profiles follow the 1D ATMO thermal profile reasonably closely for most of the pressure range. At low pressures the profiles sampled from the dayside are typically warmer than the 1D ATMO profile while the profiles sampled from the nightside are typically cooler. For high pressures $P\gtrsim10^6$ Pa the UM thermal profiles follow the initial 1D ATMO profile almost exactly. As previously noted, at these pressures the dynamical and radiative timescales are very long and the results of the 3D simulation here are essentially set by the initial condition.

\begin{figure}
\centering
\begin{tabular}{c}
	\includegraphics[width=0.45\textwidth]{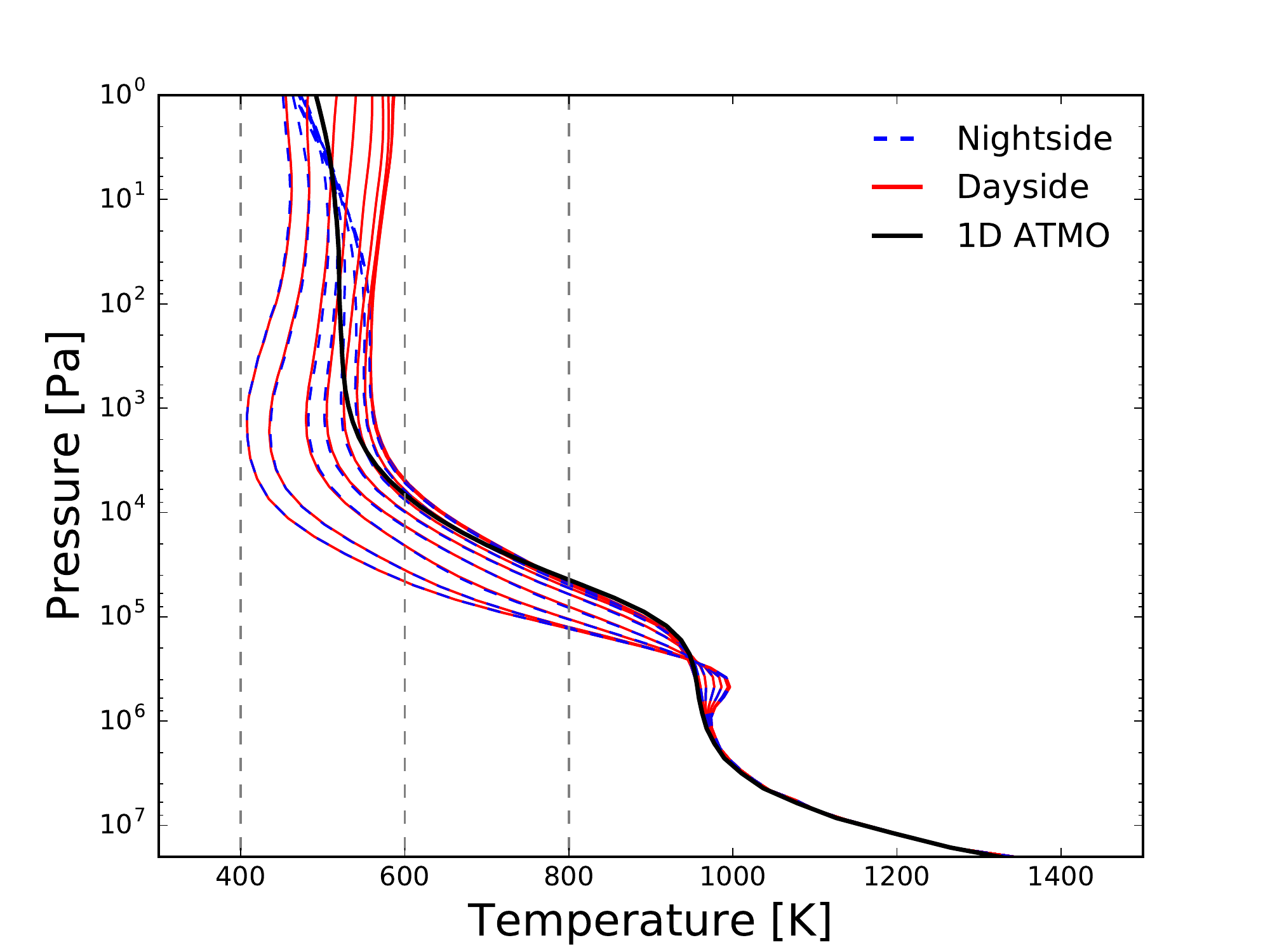} \\
	\includegraphics[width=0.45\textwidth]{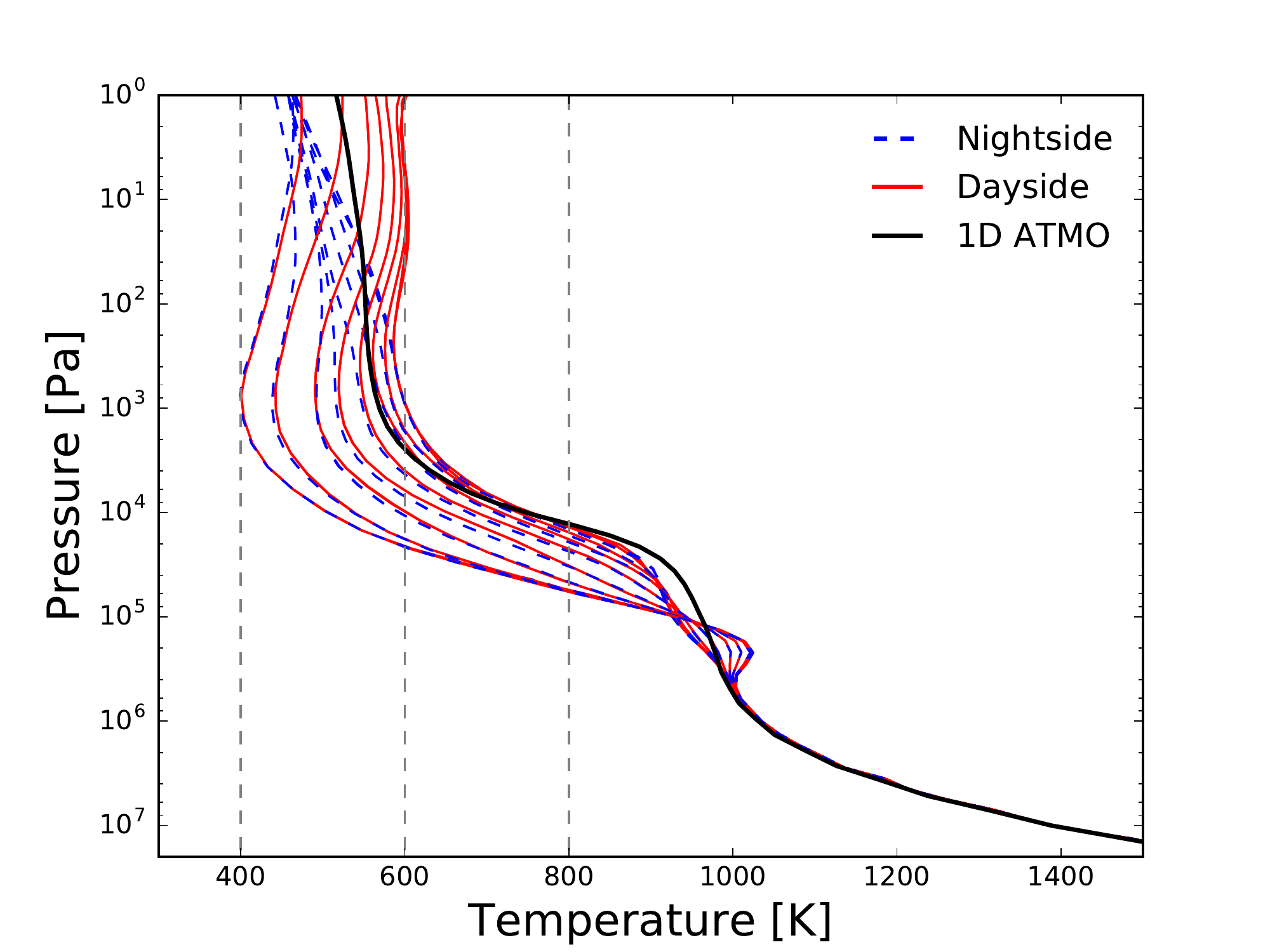} \\
	\includegraphics[width=0.45\textwidth]{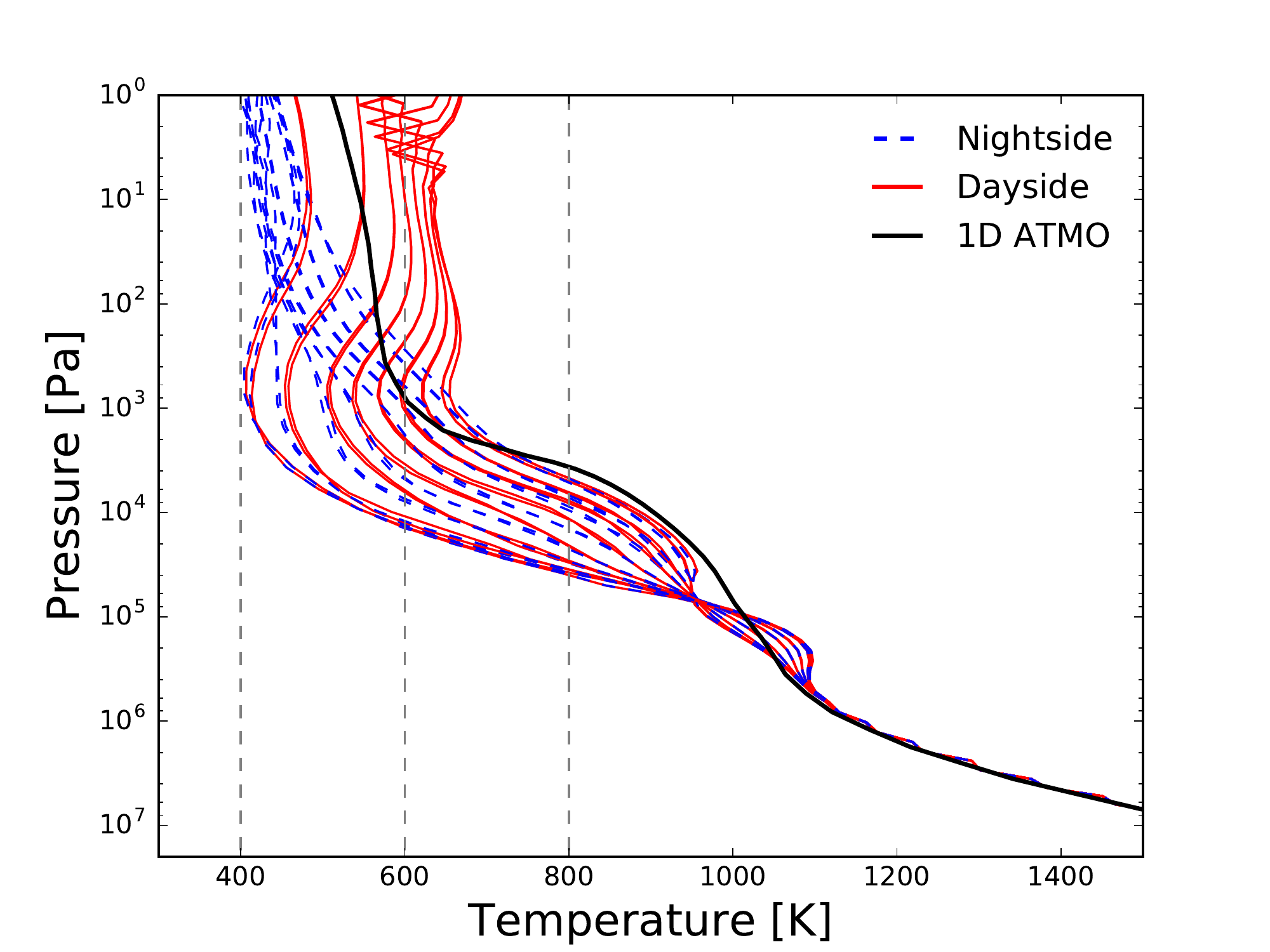} 
\end{tabular}
\caption{As \cref{figure:pt_longitudes} but showing profiles over a series of latitudes for longitudes of $\lambda=180^{\circ}$ (dayside, solid red) and $\lambda = 0^{\circ}$ (nightside, dashed blue).}
\label{figure:pt_latitudes}
\end{figure}

\cref{figure:pt_latitudes} shows a series of thermal profiles extracted from a series of latitude points on the dayside ($\lambda = 180^{\circ}$) and on the nightside ($\lambda=0^{\circ}$). Comparing with \cref{figure:pt_longitudes}, it is apparent that the meridional temperature gradient is generally more significant than the zonal temperature gradient for GJ~1214b, for each of the R1, R10 and R100 simulations. In each case the variation in temperature with latitude remains significant down to a pressure of $P\sim10^5$ Pa. As the metallicity increases, the meridional temperature gradient increases across a broad range of pressures.

\begin{figure*}
\centering
	\begin{tabular}{c c}
	\includegraphics[width=0.45\textwidth]{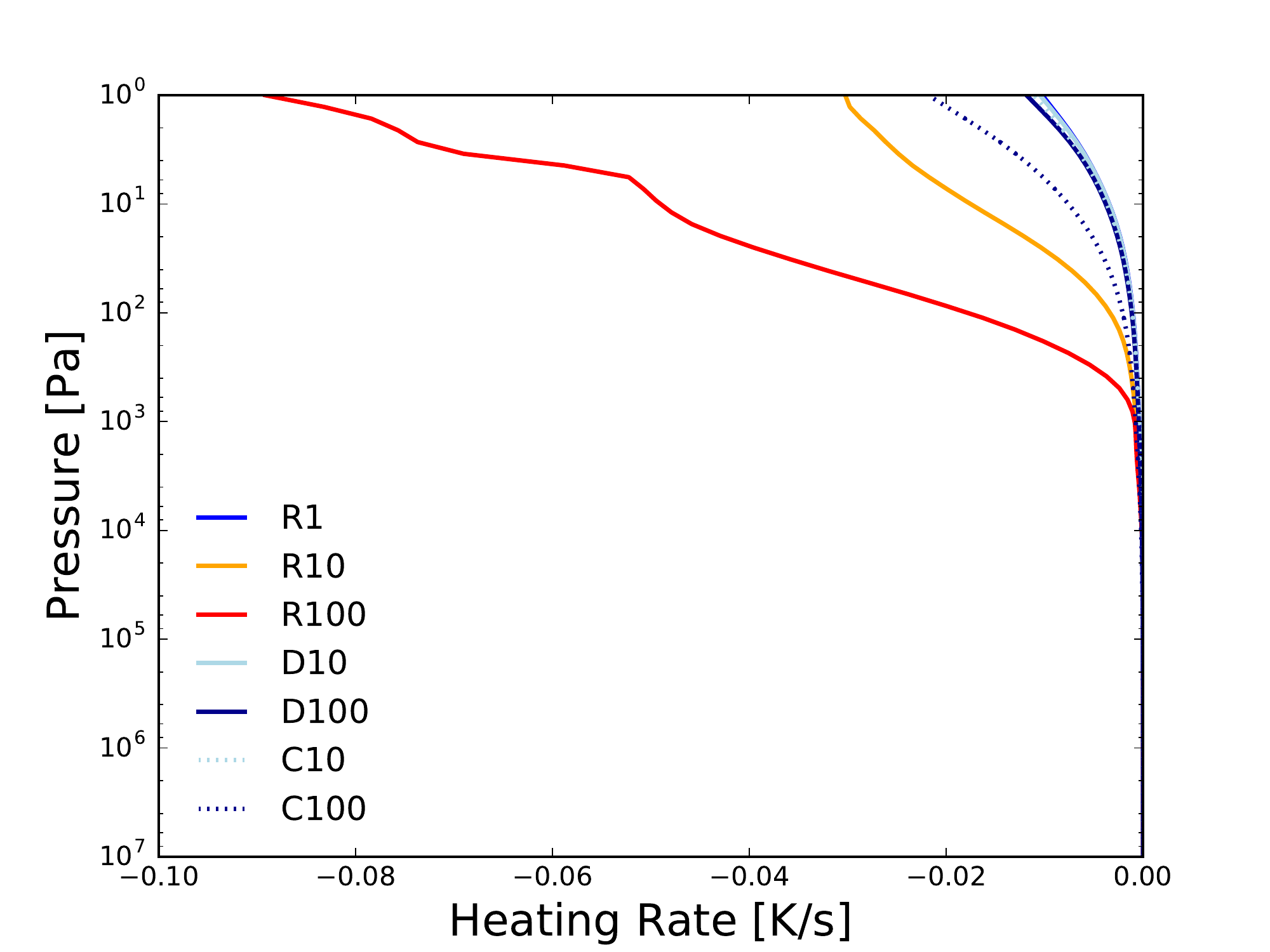} & 
	\includegraphics[width=0.45\textwidth]{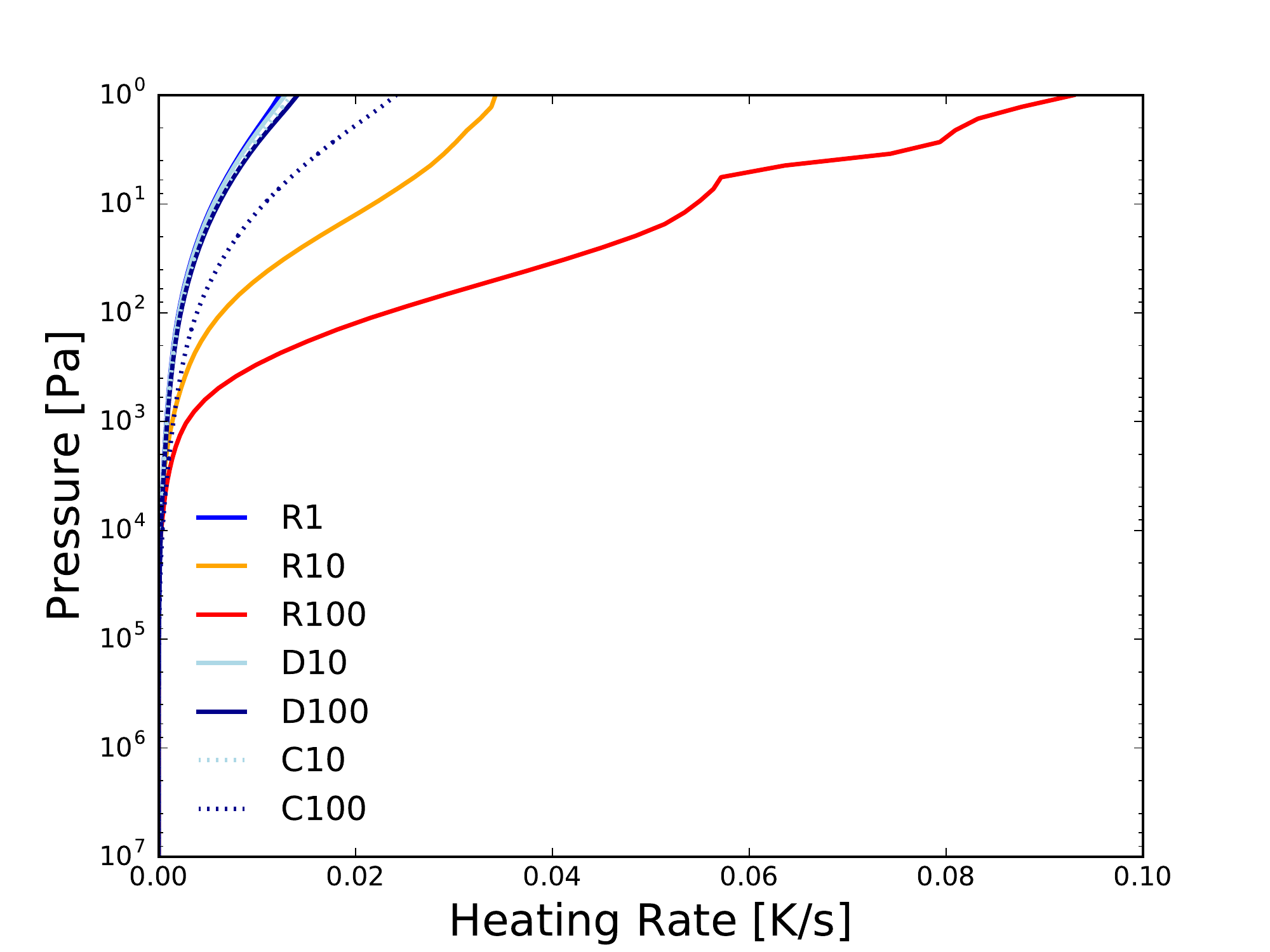} 
	\end{tabular}
	\caption{Dayside average longwave (left) and shortwave (right) heating rates.}	
	\label{figure:heating_rates}
\end{figure*}

\begin{figure}
\centering
	\includegraphics[width=0.45\textwidth]{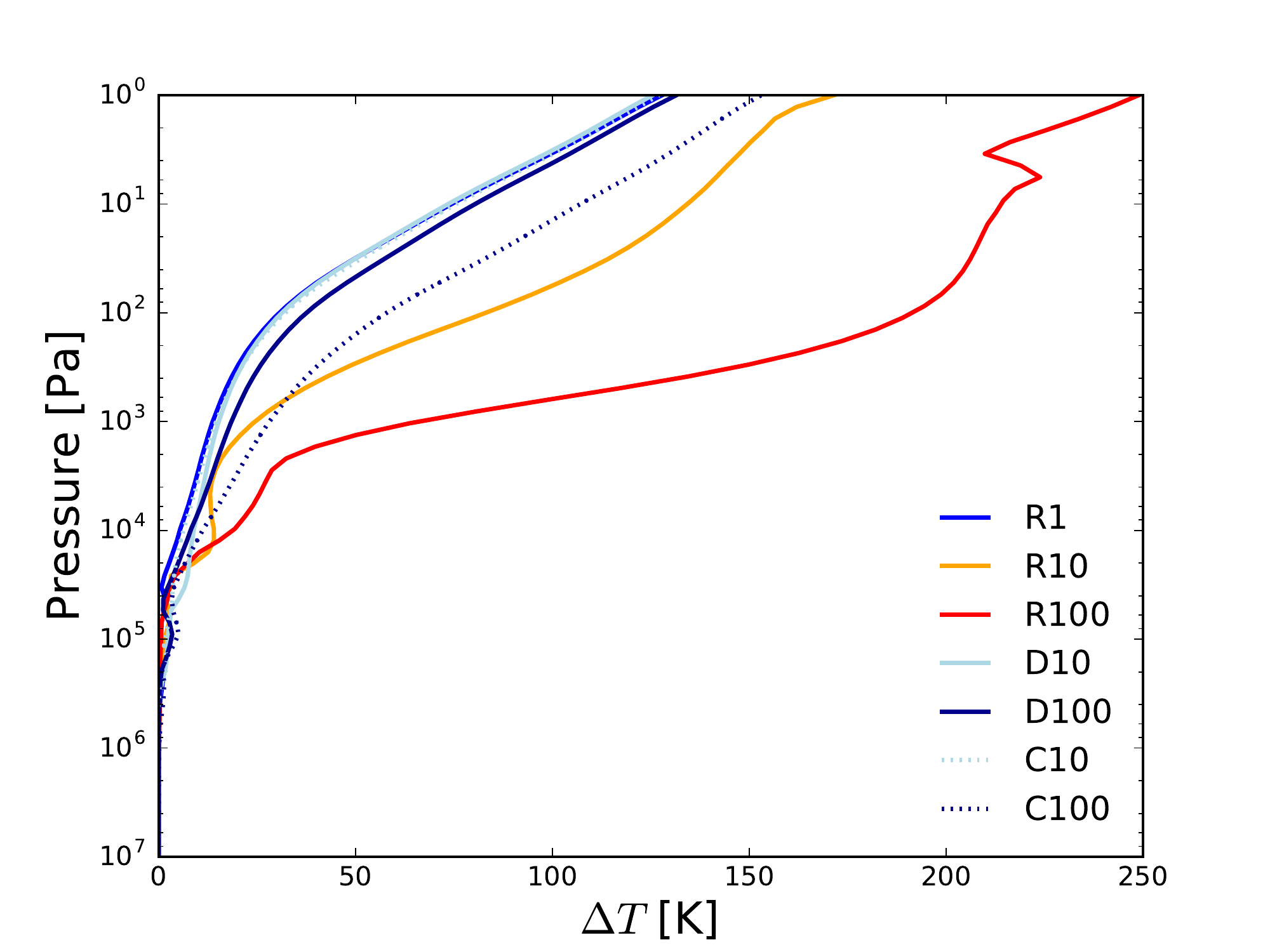}
	\caption{Difference between the maximum and minimum latitudinally-averaged temperature as a function of pressure.}
	\label{figure:delta_t}
\end{figure}

\cref{figure:delta_t} shows the difference between the maximum and minimum latitudinally-averaged temperature for each simulation. The largest temperature difference occurs for the R100 simulation with $\Delta T\sim200$ K at 100 Pa, followed by the R10 simulation with $\Delta T\sim100$ K. The temperature difference for the D10 and D100 simulations does not change significantly compared with R1. The maximum zonal temperature is increased in C100 with $\Delta T\sim 100$ K at 100 Pa, compared to $\Delta T\sim25$ K for R1.

The dayside average heating rates are shown in \cref{figure:heating_rates}. The heating rates increase significantly in both R10 and R100 compared to R1, highlighting the importance of the opacity effect. In contrast, the heating rate increase is smaller for C100 and there is negligible change in the heating rates for D10, D100 and C10. These heating rate profiles are consistent with the changes in the thermal structure.

In summary, we find that as the metallicity is increased the combined result of the dynamical, radiative heat capacity and opacity effects is to increase the zonal temperature gradient, as the dayside becomes warmer and the nightside becomes cooler. In addition, the hot spot moves closer to the sub-stellar point as the metallicity increases. However, the dominant temperature gradient for the atmosphere of GJ~1214b is generally in the meridional rather than zonal direction.

We find that the opacity effect is much more important than both the dynamical and radiative heat capacity effects, and leads to the largest changes in the thermal structure as the metallicity is increased. The dynamical effect was found to have a negligible impact on the temperature. The radiative heat capacity effect produced similar results to the combined effect, with an increasing zonal temperature gradient, but to a smaller degree. The opacity effect was therefore found to be the most important factor in determining the thermal structure, as the metallicity is increased.

\subsection{Zonal-mean zonal wind}
\label{section:zon_wind}

\begin{figure*}
\centering
\begin{subfigure}[b]{0.37\textwidth}
\includegraphics[width=\textwidth]{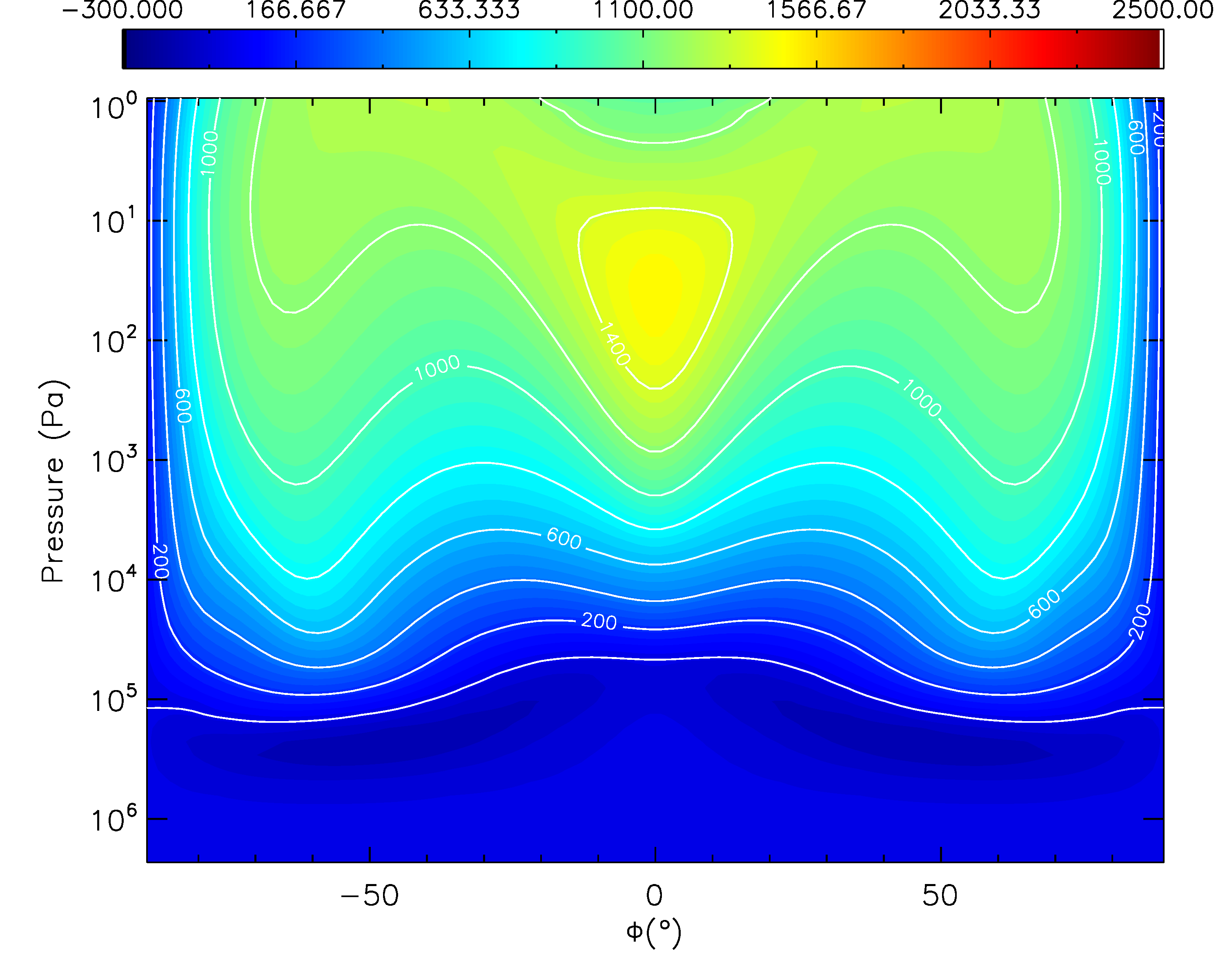} 
\caption{R1}
\end{subfigure}

\begin{subfigure}[b]{0.37\textwidth}
\includegraphics[width=\textwidth]{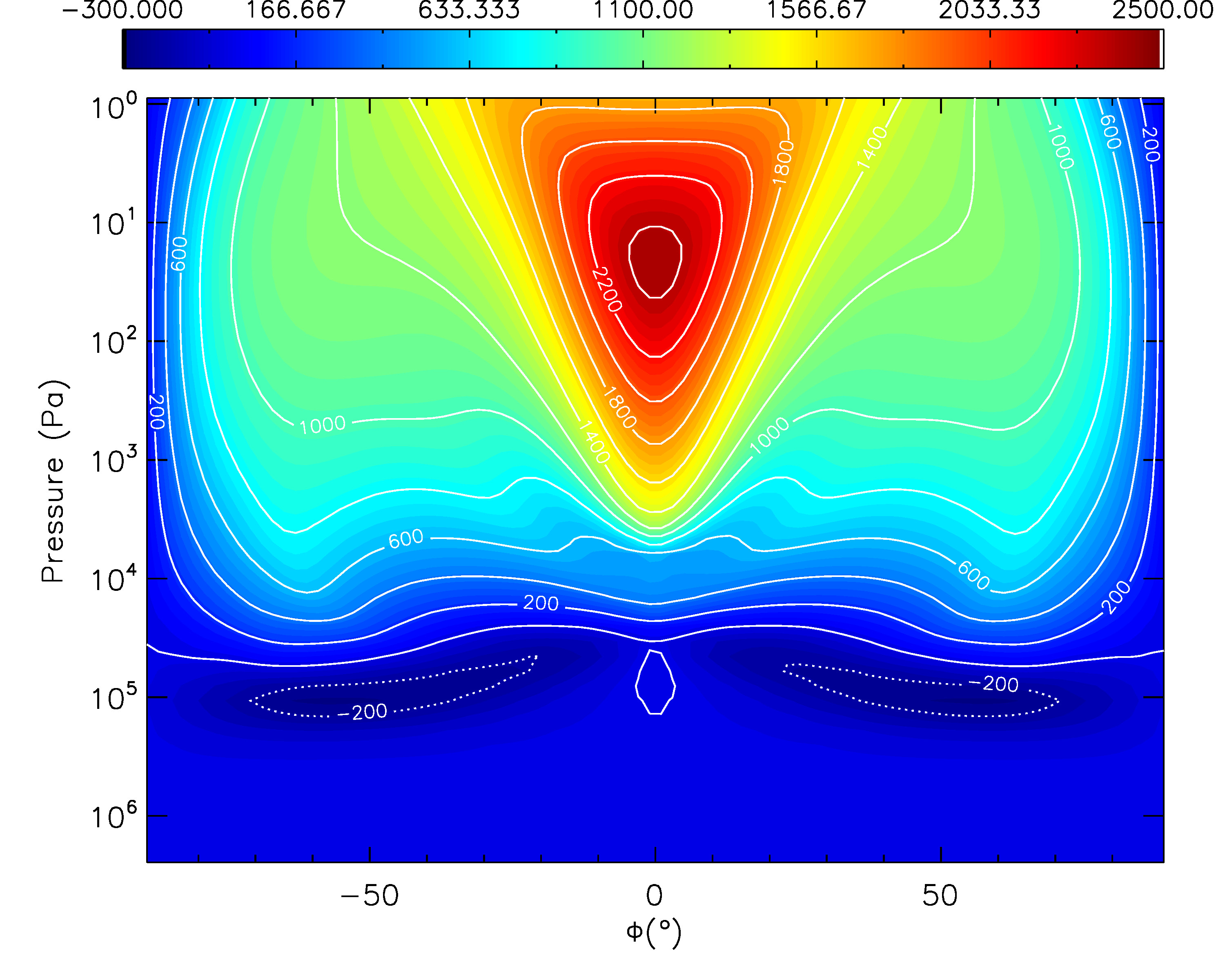} 
\caption{R10}
\end{subfigure}
~
\begin{subfigure}[b]{0.37\textwidth}
\includegraphics[width=\textwidth]{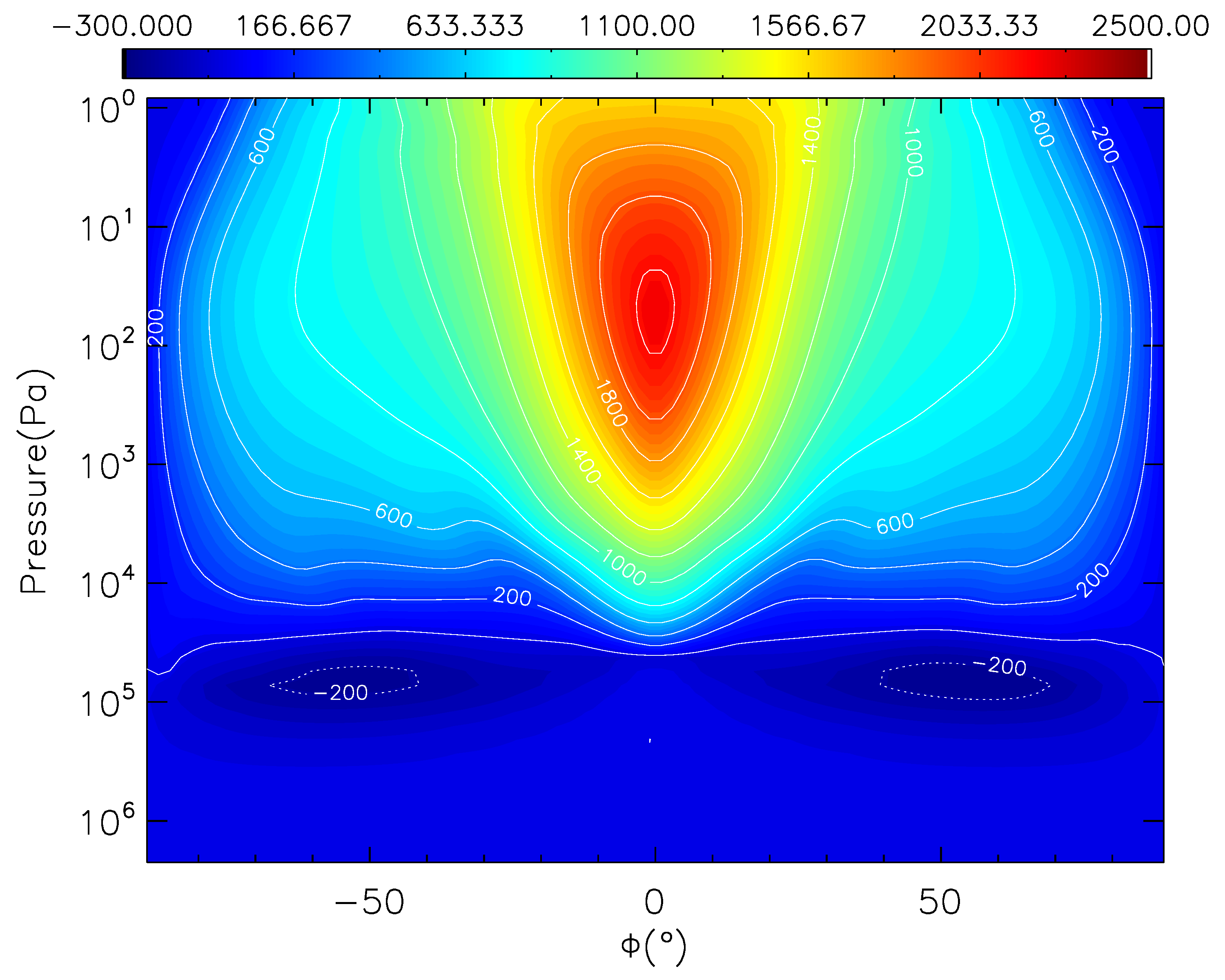} 
\caption{R100}
\end{subfigure}

\begin{subfigure}[b]{0.37\textwidth}
\includegraphics[width=\textwidth]{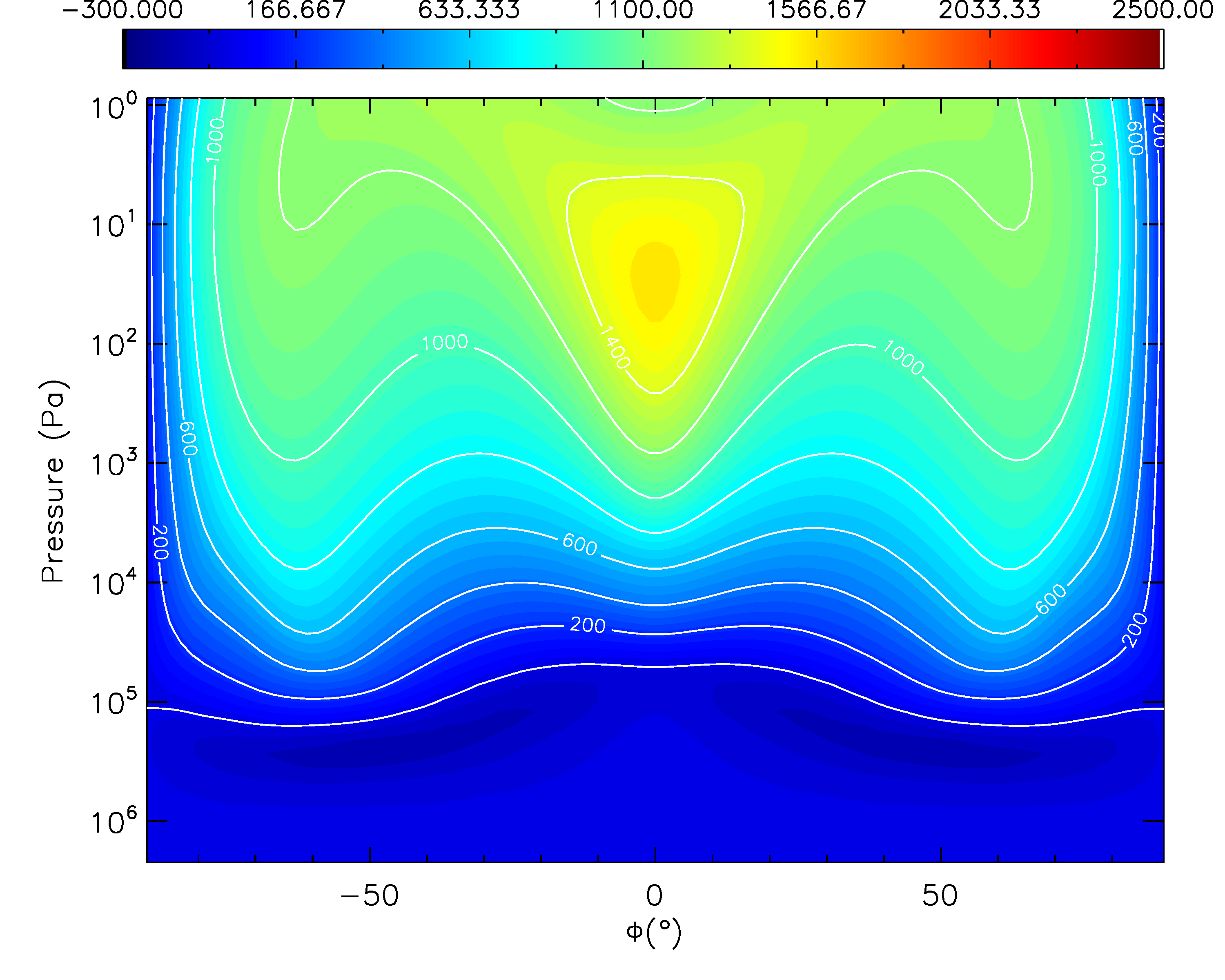} 
\caption{D10}
\end{subfigure}
~
\begin{subfigure}[b]{0.37\textwidth}
\includegraphics[width=\textwidth]{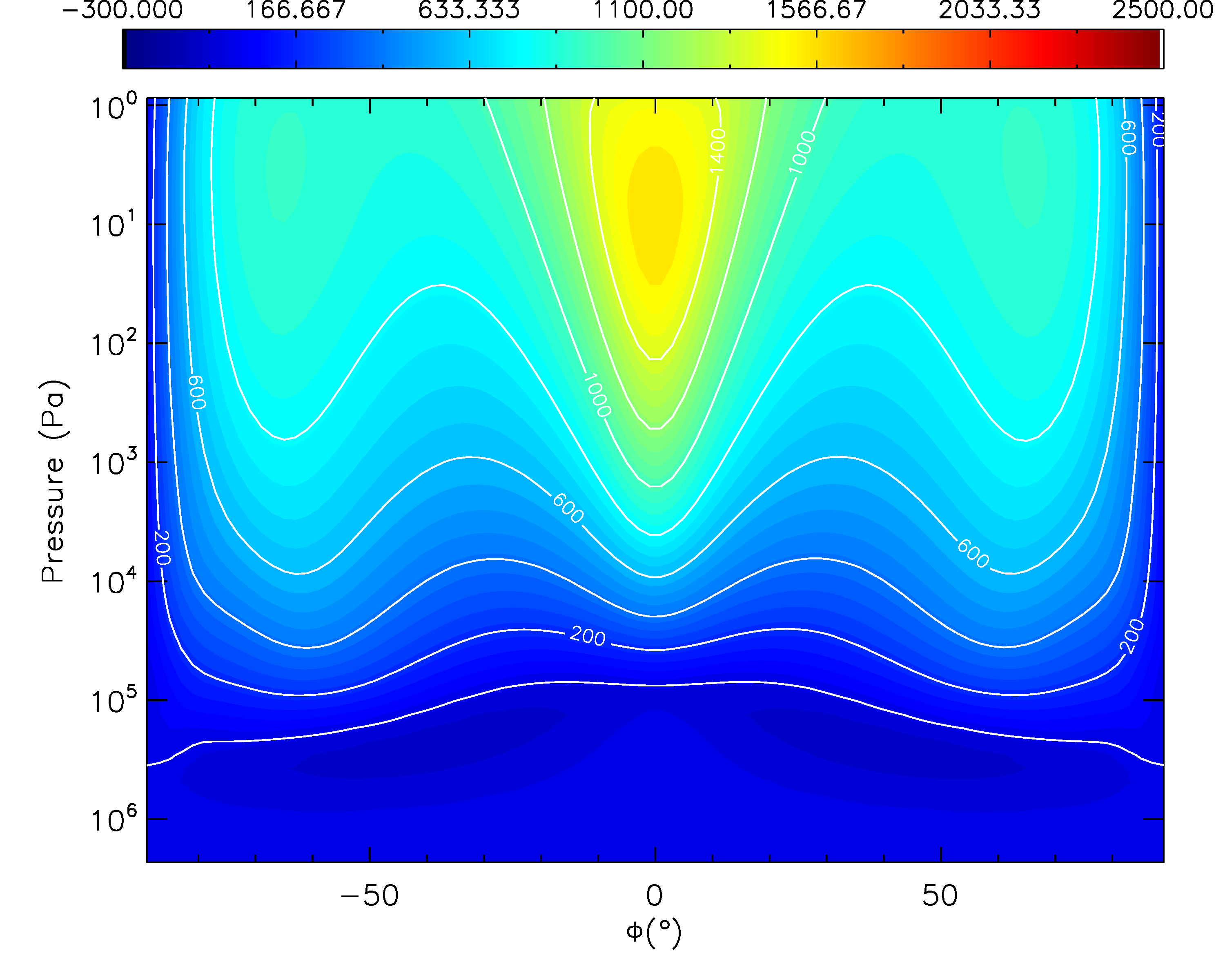} 
\caption{D100}
\end{subfigure}

\begin{subfigure}[b]{0.37\textwidth}
\includegraphics[width=\textwidth]{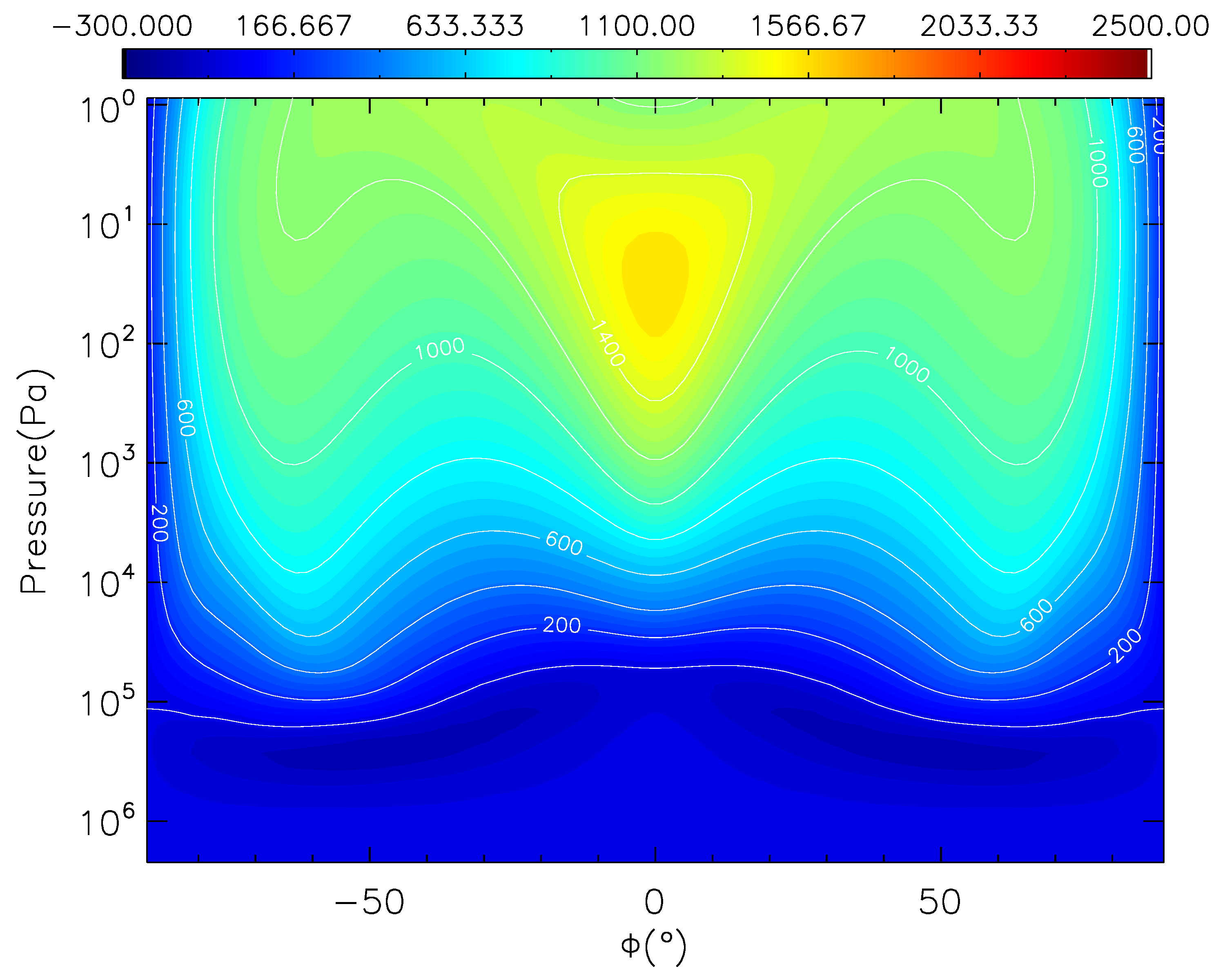} 
\caption{C10}
\end{subfigure}
~
\begin{subfigure}[b]{0.37\textwidth}
\includegraphics[width=\textwidth]{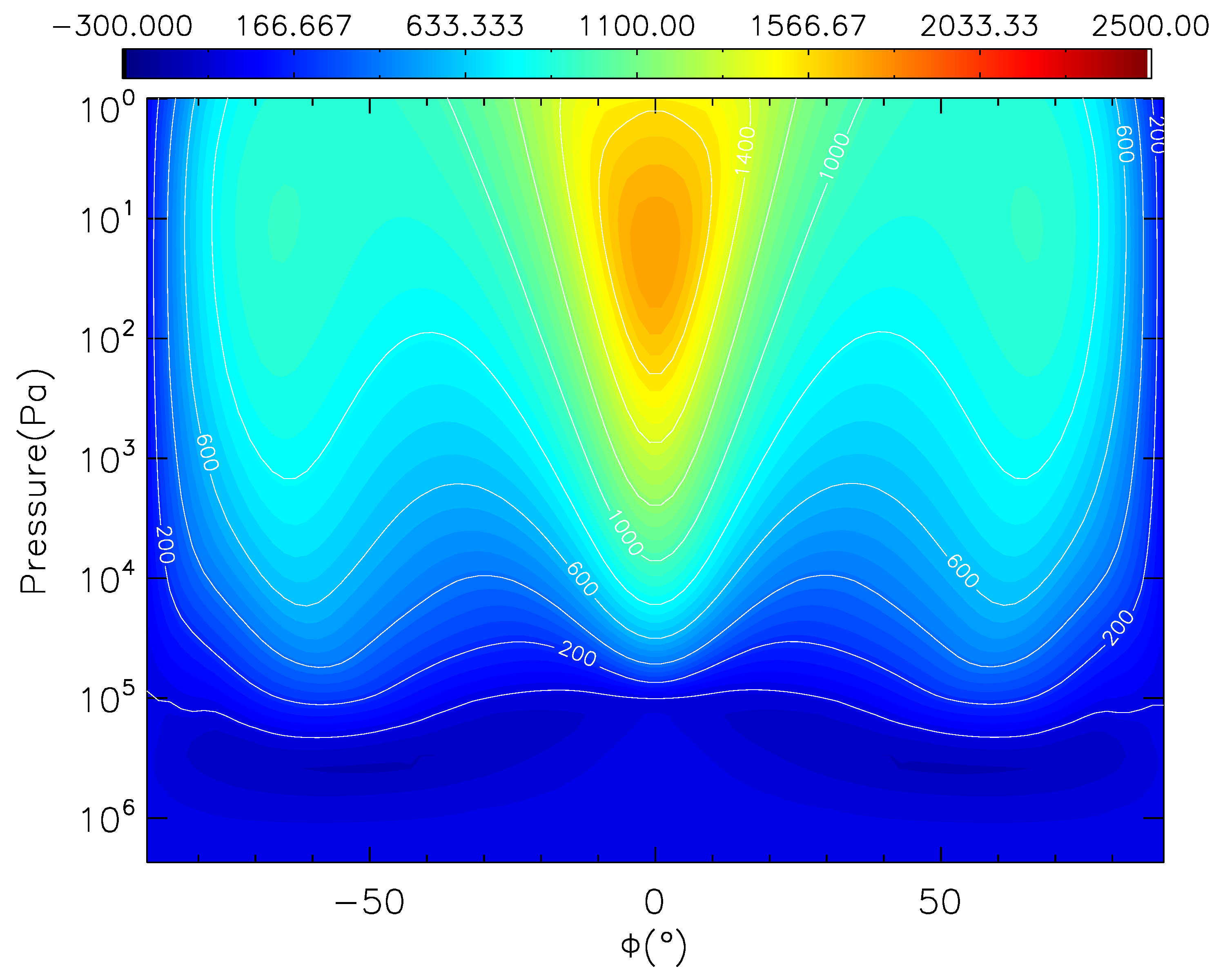} 
\caption{C100}
\end{subfigure}

\caption{Zonal-mean zonal wind for each simulation. The figures show a temporal average between 600 and 800 days.}
\label{figure:zonal_wind}
\end{figure*}

A very informative quantity often used to interpret the large scale flow is the zonal-mean zonal wind ($\bar{u}$) pattern, shown in \cref{figure:zonal_wind} as a function of pressure and latitude for each simulation. This is the zonal component of the wind averaged over all longitudes. We also perform a temporal average between 600 and 800 days; by 600 days the wind velocities for $P\lesssim10^6$ Pa have reached a steady-state. We note that the total axial angular momentum was conserved to within 99.9\% for all simulations.

The R1 simulation, with 1$\times$ solar metallicity, shows a large scale eastward (prograde) circulation across a wide latitude and pressure range with typical wind velocities of $\bar{u}\sim1000$ m s$^{-1}$. There are three maxima in $\bar{u}$, one each at high latitudes ($\phi \sim \pm 70^{\circ}$) and another at the equator. Within the equatorial jet is a maximum $\bar{u}>1400$ m s$^{-1}$. At high pressures is a region of westward flow with relatively small velocities of a few hundred m s$^{-1}$.

The velocities in the equatorial jet increase significantly in the R10 simulation, compared to R1, with $\bar{u}\sim2400$ m s$^{-1}$ in the core of jet (the jet core being the region where $\bar{u}$ is a maximum). The zonal wind velocities at other latitudes outside the equatorial jet are similar to R1. In the R100 simulation, the wind velocities generally decrease, both within the jet and for higher latitudes, compared to R10. However, the jet penetrates to deeper pressure levels compared with both R1 and R10. The maximum equatorial wind velocities are $\bar{u}\sim2200$ m s$^{-1}$ in the R100 simulation. Generally, as the metallicity is increased, the atmosphere transitions from one with a large-scale eastward flow at solar metallicities to one with a dominant equatorial jet at 10 and 100$\times$ solar metallicity. 

There is also some evolution in the zonal-mean zonal wind for the D10 and D100 simulation, considering only the dynamical effect. As the metallicity is increased $\bar{u}$ also increases within the equatorial jet, but decreases for higher latitudes. The latitudinal width of the equatorial jet appears to slightly decrease in the D100 simulation compared with the R1 simulation.

When the radiative heat capacity effect is included, in addition to the dynamical effect, the impact of increasing the metallicity is larger. In the C100 simulation, the equatorial wind velocities are larger than in the D100 simulation. The vertical extent of the equatorial jet is also greater with the base of the jet reaching $P\sim10^5$ Pa, compared to $P\sim10^4$ Pa in the D100 simulation.

\subsection{Emission phase curves}

In this section we present the emission phase curves for each simulation to assess their dependence on the metallicity of the atmosphere.

\begin{figure*}
\centering
\begin{tabular}{cc}
	\includegraphics[width=0.5\textwidth]{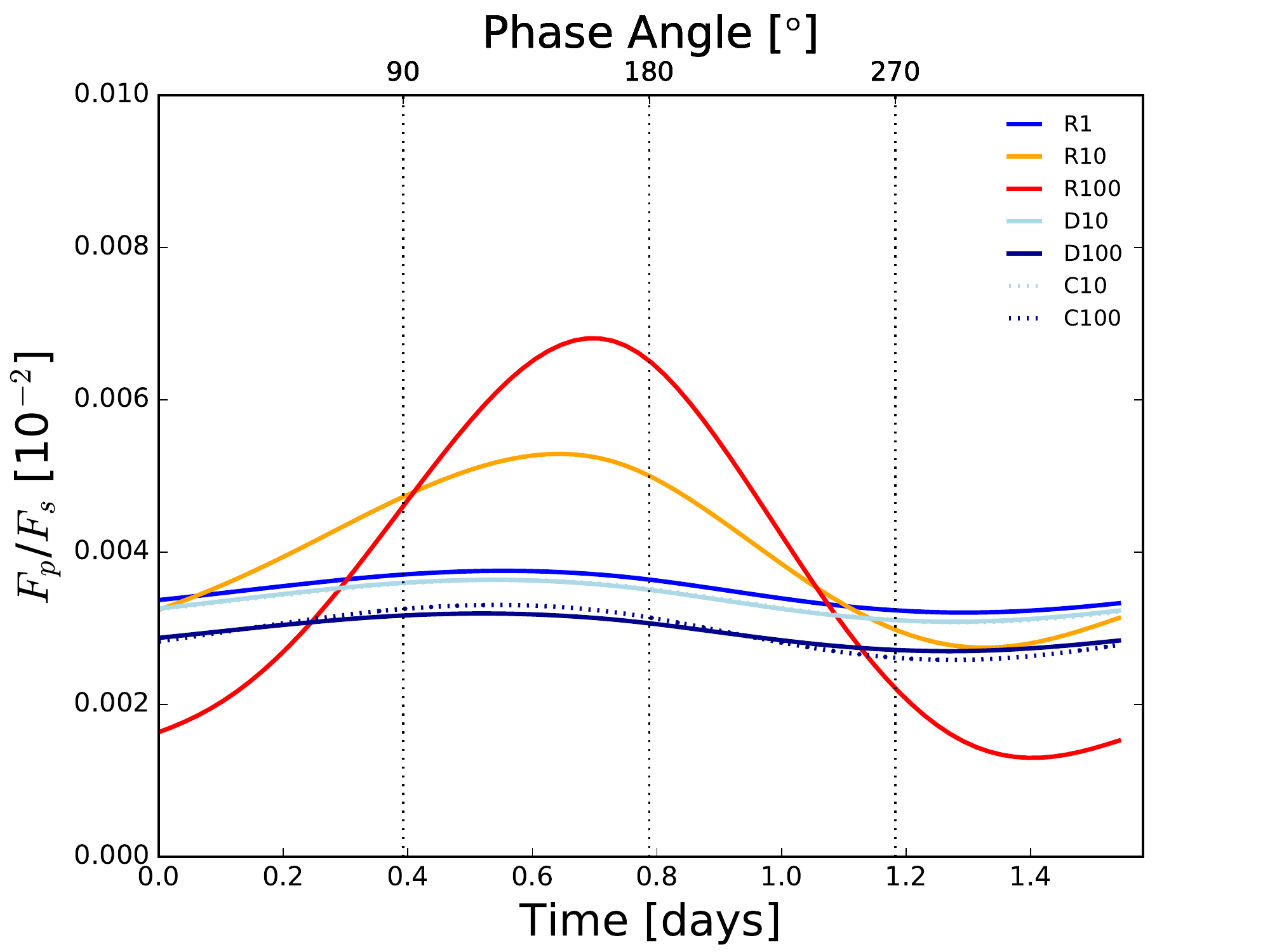} &
	\includegraphics[width=0.5\textwidth]{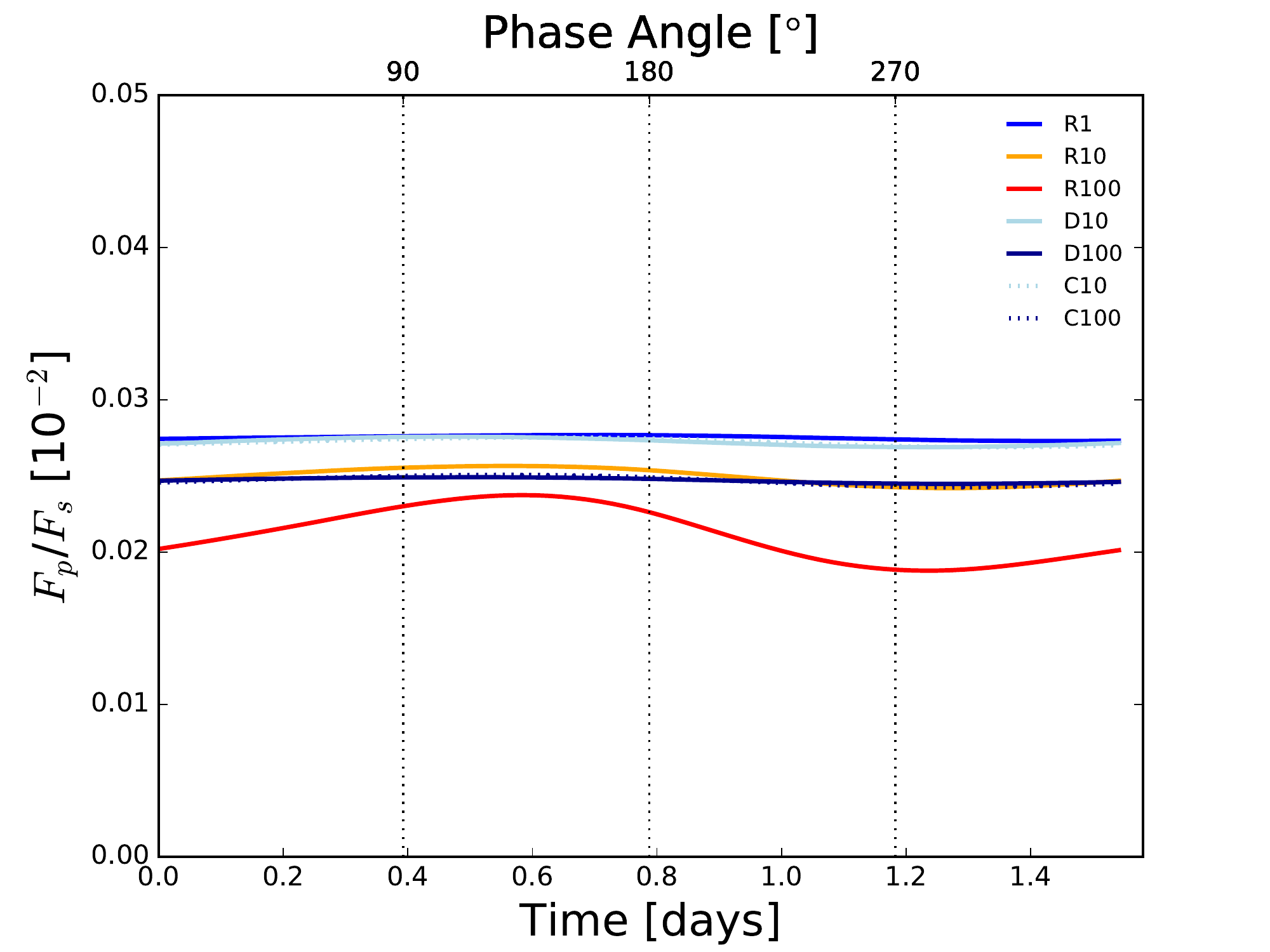} \\
\end{tabular}
\caption{Emission phase curves for each simulation for the {\it Spitzer}/IRAC 3.6 $\si{\micro\metre}$ ({\it left}) and 4.5 $\si{\micro\metre}$ ({\it right}) channels.}
\label{figure:phasecurve}
\end{figure*}

\cref{figure:phasecurve} shows the calculated emission phase curve of each simulation in the 3.6 and 4.5 $\si{\micro\metre}$ {\it Spitzer}/IRAC channels. We show the ratio of the planet-to-star flux $F_{\rm p}/F_{\rm s}$ as a function of time and phase angle. The 3.6 $\si{\micro\metre}$ channel covers a methane absorption feature while the 4.5 $\si{\micro\metre}$ channel contains absorption due to CO and CO$_2$ (however CO$_2$ is not included as an opacity source in the present model).

In the 3.6 $\si{\micro\metre}$ channel increasing the metallicity leads to a progressive increase in the peak amplitude of the phase curve when the opacity effect is included (i.e. from R1 to R100). Since $F_{\rm p}$ is strongly dependent on the temperature of the atmosphere in the region of the photosphere, this is a result of the increasing zonal temperature gradient with metallicity; and therefore a warmer dayside and cooler nightside. In addition, the peak in the emission phase curve shifts closer towards secondary eclipse (a phase angle of 180$^{\circ}$) which corresponds to the hot spot shifting closer to the sub-stellar point.

When including only the dynamical effect (D10 and D100) the peak amplitude and phase offset are not effected by metallicity, however at all phase angles there is a progressive decrease in $F_{\rm p}/F_{\rm s}$. The C10 and C100 phase curves are similar to D10 and D100, except C100 shows a slight increase in amplitude caused by a slight increase in the zonal temperature gradient due to the radiative heat capacity effect.

In the 4.5 $\si{\micro\metre}$ channel the phase curves for all simulations are generally flatter than at 3.6 $\si{\micro\metre}$. An increase in the peak amplitude is seen moving from the R1 to R100 simulations, as at 3.6 $\si{\micro\metre}$, but to a smaller degree. The phase offset does not appear to vary significantly as the metallicity is increased. $F_{\rm p}/F_{\rm s}$ decreases with increasing metallicity for the D10 and D100 (as well as C10 and C100) simulations, similar to the case at 3.6 $\si{\micro\meter}$.

%% file: discussion.tex

\section{Discussion}
\label{sec:disc}

\subsection{The effect of metallicity on the atmosphere}

In this work we have investigated the effect of metallicity on the circulation and thermal structure of the atmosphere by breaking the mechanism into three components:
\begin{enumerate}
 	\item{the dynamical effect}
	\item{the radiative heat capacity effect}
	\item{the opacity effect.}
\end{enumerate}

Our results indicate that the opacity effect is the dominant mechanism that leads to changes in both the dynamics and thermal structure as the metallicity is varied.

\subsubsection{The dynamical effect}

The dynamical effect is due to the appearance of the composition dependent variables $R=\bar{R}/\mu$ and $c_P$ in the set of equations solved by the dynamical core (\cref{section:mod_desc_dynamics}). From our simulations we find a negligible impact on the thermal structure due to the dynamical effect but a small impact on the zonal-mean zonal wind. As the metallicity is increased from 1$\times$ to 100$\times$ solar $\bar{u}$ slightly increases in the equatorial jet and decreases for other latitudes. The latitudinal width of the equatorial jet also appears to decrease.

The Rossby deformation radius ($L_R$) is an informative length scale of the atmosphere that characterises at what point rotational effects become as important as gravity and bouyancy effects. This length scale indicates the typical size of dynamical features in the atmosphere.  $L_R$ can be approximated at the equator \citep[e.g.][]{ZhaS17} as
\begin{equation}
\label{equation:rossby}
	L_R \sim \left(\frac{NH}{\beta}\right)^{1/2} \sim \left[\frac{\bar{R}T}{\beta^2\mu}\left(  \frac{\bar{R}}{C_P} - \frac{\partial \ln T}{\partial \ln P} \right)  \right]^{1/4},
\end{equation}
where $N$ is the Brunt-V{\"a}is{\"a}il{\"a} frequency, $H$ is the scale height and $\beta$ is the meridional gradient of the Coriolis parameter. Here $C_P$ is the molar heat capacity and $\bar{R} = 8.314$ J mol$^{-1}$ K$^{-1}$ is the molar gas constant, not to be confused with $c_P$ and $R$.

Using \cref{equation:rossby} we can see that as $\mu$ increases $L_R$ decreases and the size of the typical dynamical features should decrease. In addition, for a fixed profile of $P$ and $T$, $L_R$ decreases as the molar heat capacity $C_P$ increases. In our simulations, both $\mu$ and $C_P$ increase as the metallicity is increased, therefore reducing $L_R$. This is reflected in the zonal-mean zonal wind (\cref{figure:zonal_wind}) as the equatorial jet becomes narrower in latitudinal extent as the metallicity is increased.

\subsubsection{The radiative heat capacity effect}

The second mechanism is due to the composition dependent heat capacity (in our simulations $c_{P,{\rm rad}}$) that is used to calculate the heating rate of the atmosphere in K s$^{-1}$. Our results show that the radiative heat capacity effect has a small impact on the thermal structure, with the dayside becoming warmer and nightside becoming cooler as the metallicity is increased from 1$\times$ to 100$\times$ solar. In addition, the radiative heat capacity effect results in larger $\bar{u}$ in the equatorial jet.

As the metallicity is increased $c_{P,{\rm rad}}$ decreases resulting in a larger temperature response of the atmosphere to a given net energy change. A decrease in $c_{P,{\rm rad}}$ therefore naturally leads to a warmer dayside and cooler nightside if all other parameters are similar. In addition, since it is the day-night temperature contrast that is suggested to drive the equatorial jet \citep{ShoP11,MayDB17}, an increase in the zonal temperature gradient leads to a larger $\bar{u}$ in the jet.

\subsubsection{The opacity effect}

As the metallicity of the atmosphere varies the most fundamental consequence is to change the mole fractions of the individual chemical species. The opacity effect is due to the varying abundances of the absorbing species. Our simulations show the opacity effect to be the most important of the three effects that we considered, leading to the largest changes in the dynamics and thermal structure as the metallicity is increased. The day-night temperature contrast increases significantly as the metallicity is increased from 1$\times$ to 100$\times$ solar. In turn, the increased zonal temperature gradient leads to changes in the circulation, which in our simulations generally results in a deeper and faster equatorial jet with weaker zonal-mean zonal winds at higher latitudes.

As the metallicity is increased the relative abundances of strongly absorbing gas-phase species such as H$_2$O, CH$_4$ and CO increase, at the expense of H$_2$ and He. Overall, this increases the opacity of the atmosphere leading to shallower heating. Since the radiative timescale scales with the pressure ($\tau_{\rm rad}\propto P$) the heating is occuring in a region with a faster radiative timescale and heat is less efficiently transported from the dayside to the nightside. Ultimately this results in a warmer dayside, cooler nightside and a hot spot that is closer to the sub-stellar point.

\subsubsection{Mechanisms that affect the phase curve}

Changes in the dynamics and thermal structure due to the dynamical, radiative heat capacity and opacity effects have an impact on the expected emission of the atmosphere. In particular, when including all three of the effects, the peak amplitude and phase offset change significantly in the 3.6 $\si{\micro\metre}$ {\it Spitzer}/IRAC channel, with a smaller effect in 4.5 $\si{\micro\metre}$ channel. Since the emission is strongly dependent on the temperature of the atmosphere at the photosphere this indicates a change in the temperature structure in that region.

\begin{figure*}
\centering
\begin{tabular}{c c}
	\includegraphics[width=0.5\textwidth]{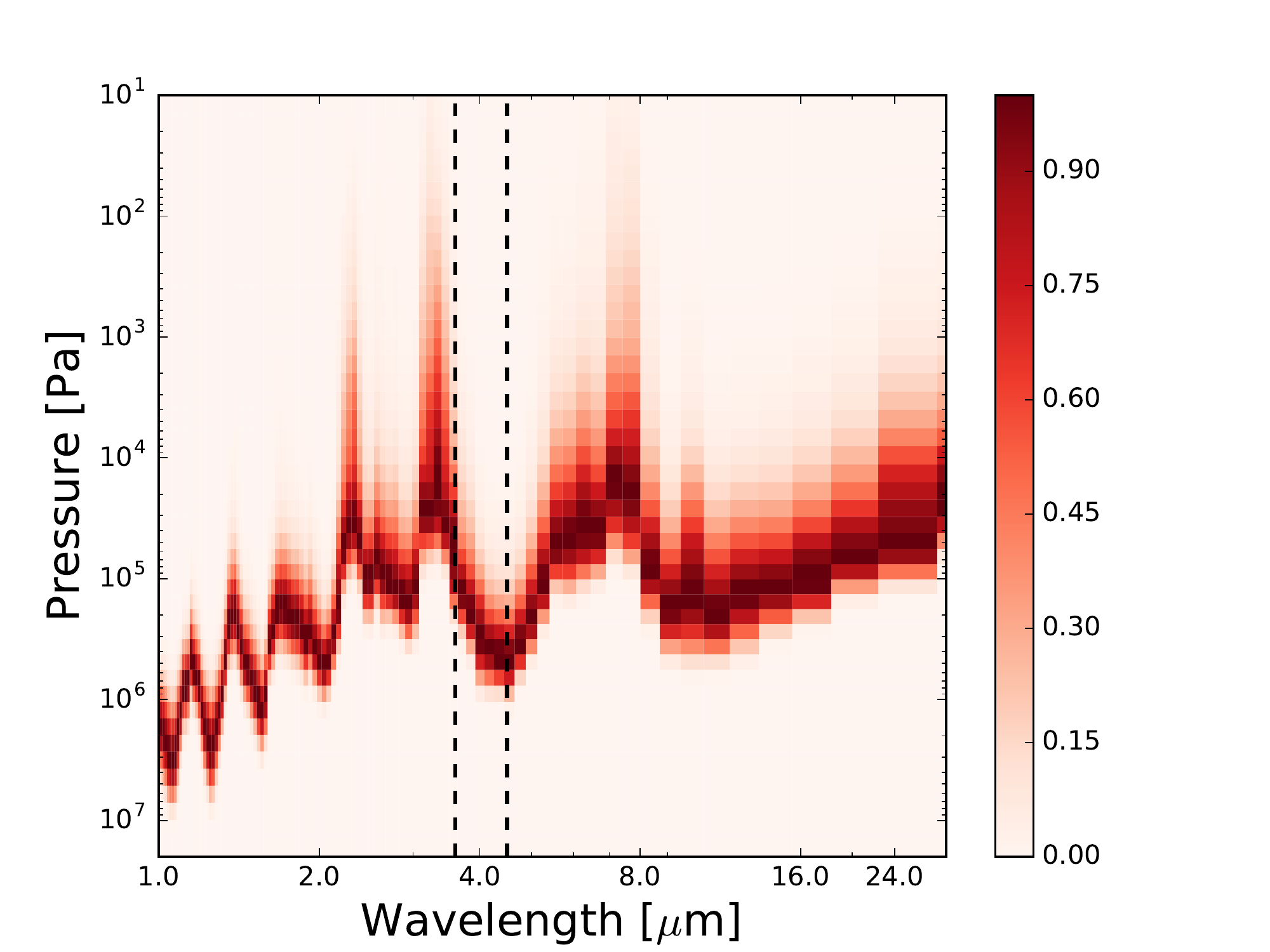} &
	\includegraphics[width=0.5\textwidth]{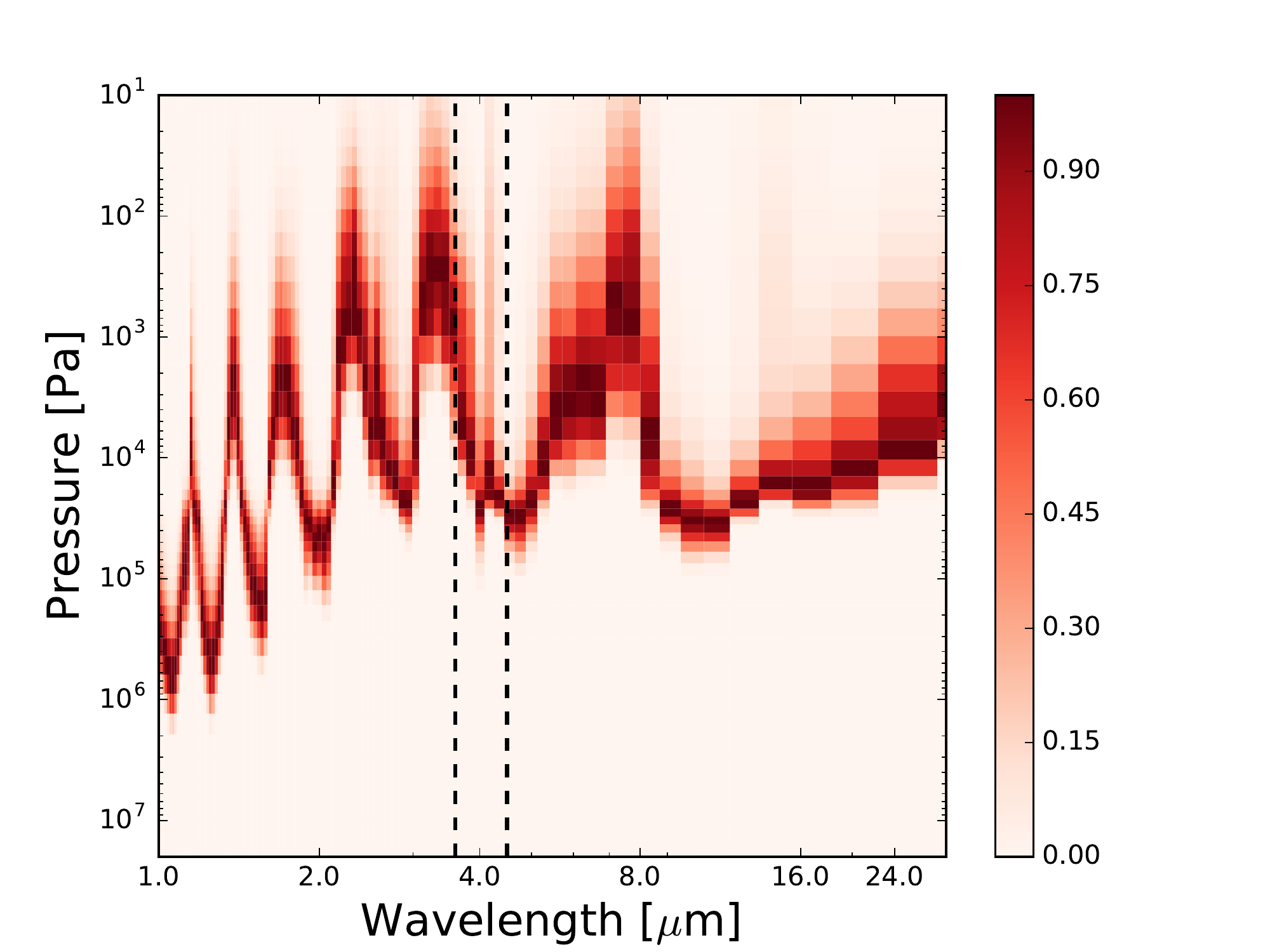} 
\end{tabular}
\caption{Spectral contribution functions as a function of pressure calculated from the 1D ATMO models assuming 1$\times$ (left) and 100$\times$ solar metallicity. Black dashed lines indicate 3.6 and 4.5 $\si{\micro\metre}$. The figures are plotted using the resolution (pressure and wavelength) of the model.}
\label{figure:cf}
\end{figure*}

In \cref{figure:cf} we show the contribution (or weighting) function \citep[e.g.][]{Griffith1998,KnuCA08,DruTB16} for the 1$\times$ and 100$\times$ solar 1D ATMO simulations. The maximum of the contribution function can be used to estimate the pressure level of the photosphere $P_{\rm phot}$, from which the top of atmosphere flux originates from. We use these contribution functions to approximate $P_{\rm phot}$ for each of the 3D simulations, which are shown in \cref{table:photosphere} along with the corresponding altitude $z_{\rm phot}$. We estimate $P_{\rm phot}$ for R1, R10 and R100 using the 1$\times$, 10$\times$ and 100$\times$ solar metallicity ATMO results, respectively, while we use the 1$\times$ solar metallicity ATMO result for D10, D100, C10 and C100.

In the 3.6 $\si{\micro\metre}$ channel $P_{\rm phot}\sim10^3$ Pa and $P_{\rm phot}\sim10^{2}$ Pa for the R1 and R100 simulations, respectively. Comparing these pressures with the thermal profiles of each simulation (\cref{figure:pt_longitudes}) it is clear that at these pressure levels there is a significant zonal temperature gradient, with a maximum day-night temperature contrasts of $\sim200$ K in the R100 simulation. On the other hand, the emission in the 4.5 $\si{\micro\metre}$ channel originates from deeper in the atmosphere with $P_{\rm phot}\sim10^4$ and $P_{\rm phot}\sim10^3$ for R1 and R100, respectively. For these pressure levels the zonal temperature gradient remains relatively small (\cref{figure:pt_longitudes}) and we would expect smaller peak amplitudes in the emission phase curves.

\begin{table*}
\centering
\setlength\extrarowheight{2pt}
\begin{tabular}{l l c c c c c c c}
\hline\hline
 & & R1 & R10 & R100 & D10 & D100 & C10 & C100 \\
\hline
$z_{\rm phot}$ (3.6 $\si{\micro\metre}$) & [$\times10^6$ $\si{\metre}$] & 2.66 & 3.50 & 2.29 & 2.42& 1.32 & 2.43 & 1.32 \\
$P_{\rm phot}$ (3.6 $\si{\micro\metre}$) & [$\times10^3$ Pa] & 4.21 & 0.25 & 0.06 & 4.21 & 4.21 & 4.21 & 4.21  \\
$z_{\rm phot}$ (4.5 $\si{\micro\metre}$)  & [$\times10^6$ $\si{\metre}$] & 2.04 & 2.60 & 1.81 & 1.87 & 1.03 & 1.87 & 1.03 \\
$P_{\rm phot}$ (4.5 $\si{\micro\metre}$) & [$\times10^3$ Pa] &  41.36 & 9.28 & 1.83 & 42.40 & 42.40 & 42.40 & 42.40 \\
\hline
\end{tabular}
\caption{Estimated values of $P_{\rm phot}$ and $z_{\rm phot}$ for each simulation in the 3.6 and 4.5 $\si{\micro\meter}$ {\it Spitzer}/IRAC channels}
\label{table:photosphere}
\end{table*}

\begin{figure}
\centering
\includegraphics[width=0.5\textwidth]{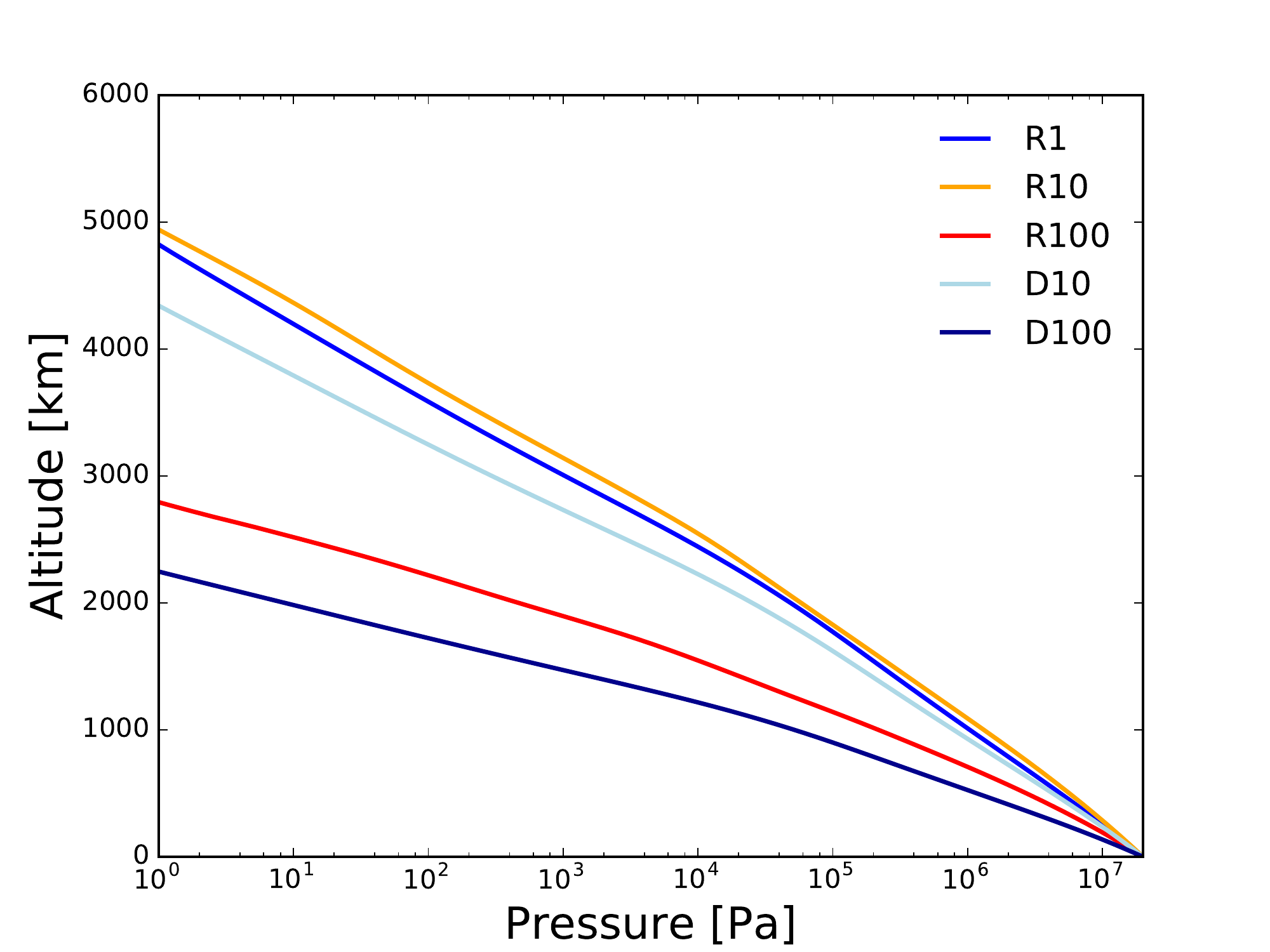}
\caption{Pressure as a function of height at the sub-stellar point for each simulation.}
\label{figure:pressure_altitude}
\end{figure}

The dynamical effect also has some impact on the emission of the atmosphere with $F_{\rm p}/F_{\rm s}$ decreasing progressively from R1 to D100. As has been demonstrated previously (\cref{section:results_temp}) the change in temperature between these simulations is negligible. However, the increase in mean molecular weight decreases the vertical scale height and the pressure decreases more rapidly with altitude. This is demonstrated in \cref{figure:pressure_altitude} where we show the pressure as a function of altitude for each simulation. Therefore, an emitting surface at a constant pressure level will correspond to a smaller altitude as the metallicity increases, due to the dynamical effect. The decrease in $F_{\rm p}/F_{\rm s}$ from R1 to D100 is caused by a reduction in the surface area of the photosphere as $z_{\rm phot}$ decreases with increasing metallicity, primarily due to the increasing mean molecular weight.

In general $F_{\rm p}/F_{\rm s}$ depends on the ratio of the radius of the planetary photosphere ($R_{\rm phot}$) and the stellar radius ($R_{\rm s}$),
\begin{equation}
\label{equation:flux_ratio}
\frac{F_{\rm p}}{F_{\rm s}}(\lambda) \propto \left(\frac{R_{\rm phot}(\lambda)}{R_{\rm s}}\right)^2.
\end{equation}
Since in our simulations $F_{\rm s}$ and $R_{\rm s}$ are constant we neglect them and rewrite \cref{equation:flux_ratio} as 
\begin{equation}
\label{equation:flux}
	F_{\rm p}(\lambda) \propto \left(R_{\rm p} + z_{\rm phot}(\lambda)\right)^2,
\end{equation}
where $R_{\rm p}$ is the wavelength-independent radius of the bulk planet, and defines the radius at the bottom boundary of the model atmosphere, and $z_{\rm phot}(\lambda)$ is the wavelength-dependent altitude of the photosphere. $R_{\rm p}$ is constant for each of our simulations, while $z_{\rm phot}$ depends on the composition and thermodynamic structure.

Using \cref{equation:flux} and the values of $z_{\rm phot}$ in \cref{table:photosphere} we can estimate the expected change in $F_{\rm p}/F_{\rm s}$ due to this mechanism. For example, the expected ratio of $F_{\rm p}/F_{\rm s}$ between the R1 and D100 simulations at 4.5 $\si{\micro\meter}$ is $F^{\rm D100}_{\rm p}/F^{\rm R1}_{\rm p} \sim 0.9$. This is consistent with the difference between the R1 and D100 phase curves in \cref{figure:phasecurve}.

The measured $R_{\rm p}$ of GJ~1214b \citep{Carter2011}, from visible wavelength transit measurements, may be affected by the presence of clouds in the atmosphere, and the actual radius of the bulk planet may therefore actually differ to the value that we assume in our models. Uncertainty in this value may lead to quantitative differences in our simulation results, however, we do not expect it to affect our qualitative conclusions.

In addition to metallicity, the emission of the atmosphere can also be affected by gas-phase chemical kinetics. Transport-induced quenching and photochemistry can drive the chemistry out of local chemical equilibrium, thereby changing the opacity, thermal structure and emission of the atmosphere. \citet[][]{LinVC2011} used a 1D chemical kinetics code to investigate the effect of vertical quenching for the atmosphere of GJ~436b. They found that the abundances of CO and CO$_2$ can be enhanced above chemical equilibrium values. The measured emission of GJ~436b has been suggested to show an enhanced abundance of CO and a reduced abundance of CH$_4$, compared to what it expected from chemical equilibrium \citep{Stevenson2010}.

Capturing the effects of the 3D atmospheric circulation on the composition requires consistency between the dynamics, radiative transfer and chemistry and is beyond the scope of this work; however, we have previously investigated this process using a 1D atmosphere code \citep{DruTB16}. In addition, the presence of clouds or hazes is likely to be important in determining the emission phase curve \citep{LeeDH16,LinMB17}.

\subsection{Comparison with previous works}

Comparing our results to previous works that have simulated the atmosphere of GJ~1214b using 3D models \citep{Men12,KatSF14,ChaML15,ZhaS17} we find similar qualitative trends in the dynamics and thermal structure as the metallicity is increased from 1$\times$ solar. There are of course quantitative differences between the results of each model due to different levels of approximation and differences in model discretisation and numerical methods. Understanding and quantifying these differences is beyond the scope of this study.

\citet{Men12}, \citet{KatSF14} and \citet{ChaML15} each perform simulations assuming a range of metallicities and account for the dynamical, radiative heat capacity and opacity effects simultaneously, which correspond to our R1, R10 and R100 series of simulations. Our results, which represent the most accurate model of GJ~1214b to date, are in agreement with the qualitative trends found by previous studies. As the metallicity increases the day-night temperature contrast increases which results in a faster equatorial jet and weaker zonal-mean zonal winds at higher latitudes.

\citet{ZhaS17} investigated the dynamical and radiative heat capacity (represented by the radiative timescale in their model) effects in isolation from the opacity effect, which closely matches our R1, C10 and C100 series of simulations. Specifically, \citet{ZhaS17} model a series of atmospheres that are dominated by a single molecule (H$_2$, He, H$_2$O etc), as opposed to a more realistic mixture of species. In terms of the mean molecular weight $\mu$ our R1 simulation is most closely matched to their H$_2$ simulation and our C100 simulation is most closely matched to their He simulation; though $c_{P,{\rm dyn}}$ and $c_{P,{\rm rad}}$ used in our C100 simulation are significantly different to value used in their He simulation.

With these differences in mind we can approximately compare our R1 and C100 simulations to the H$_2$ and He simulations of \citet{ZhaS17}. Our simulations find the day-night temperature contrast increases as the mean molecular weight increases, in agreement with \citet{ZhaS17}. Our results also show that the zonal wind velocities increase in the equatorial jet, which is opposite trend to that found by \citet{ZhaS17}, where the jet core speed decreases with increasing mean molecular weight. This difference is intriguing, but may be due to the significant differences between the two model setups. For instance, \citet{ZhaS17} assume a chemically simple atmosphere that is dominated by a single molecule, as opposed to our more realistic gas mixture. In addition, \citet{ZhaS17} use a parameterised Newtonian cooling scheme to represent the thermal evolution, as opposed to our full radiative transfer approach.

\subsection{The benefits of a coupled chemical equilibrium scheme}

In a previous application of the UM to the atmosphere of HD~209458b \citet{AmuMB16} used a mixture of methods to compute the mole fractions of the chemical species in each grid cell. The abundances of H$_2$O, CH$_4$, CO, NH$_3$ and N$_2$ were determined using the analytical solution to chemical equilibrium derived by \citet{Burrows1999}. For the alkali species a simple parameterisation was adopted based on the chemical transformation curves \citep[][their Fig. 4]{Burrows1999} between the monatomic alkali species (e.g. Na) and the alkali chlorides (e.g. NaCl); for temperatures above the transformation curve the monatomic species was assumed to be present in its solar elemental abundance, with a zero abundance for temperatures below the curve.

An important assumption behind the \citet{Burrows1999} analytical formula is that all carbon is in CO and CH$_4$, all oxygen is in H$_2$O and CO and all nitrogen is in N$_2$ and NH$_3$. This is a good assumption for near-solar compositions at high temperatures. However as the metallicity is increased other species can become important such as HCN and CO$_2$ \citep{MosLV13} and this assumption breaks down.

Coupling a Gibbs energy minimisation scheme directly to the GCM allows for the calculation of the mole fraction of a much larger number of chemical species for a general mix of elements and for a wide range of pressures and temperatures. This has obvious benefits in terms of the flexibility of the model as we are no longer restricted to the assumptions of near-solar, high temperature conditions as in the \citet{Burrows1999} method. In theory, this also allows us to more accurately calculate the abundances of the alkali species by minimising the Gibbs energy instead of using simple parameterisations based on chemical transformation curves; this is likely to be inaccurate particularly for temperatures near to or below the curves. However, in the current study we chose to neglect condensation and assume a gas-phase only composition. 

In general, the flexibility of the Gibbs energy minimisation approach is beneficial when adding new opacity species to the model, as the equilibrium chemistry of that species can easily be incorporated, provided the thermochemical data for the relevant species are available.

The main conclusion of this study is the importance of the opacities in determining the dynamical and thermal structure of the atmosphere. It is therefore crucial that the accuracy of the opacity calculation is maximised. Our model computes the opacity on-the-fly by combining the $k$-coefficients of the individual gases. This method is more flexible and can be more accurate \citep{AmuBT14,AmuTM17} than the alternative method of pre-mixed $k$-coefficients used by other GCMs applied to exoplanet atmospheres \citep[e.g.][]{KatSF14,ChaML15}. The pre-mixed $k$-coefficients method can introduce inaccuracies when interpolating over the pressure and temperature grid, particularly in the presence of rapid changes in composition \citep[see][]{AmuTM17}.

%% file: conclusions.tex

\section{Conclusions}
\label{sec:conc}

We have presented results from a series of simulations of the atmosphere of GJ~1214b that assume a variety of metallicities using a GCM consistently coupled to a Gibbs energy minimisation scheme. Combined with the radiative transfer scheme that combines individual $k$-coefficients on-the-fly \citep{AmuMB16}, by coupling the chemical equilibrium scheme we have increased the flexibility (more chemical species over a wider parameter space) and accuracy of the model. This development represents the first flexible equilibrium chemistry scheme consistently coupled to a GCM applied to exoplanets.

Our simulations show that as the metallicity is increased the zonal temperature gradient increases at low pressures, leading to a warmer dayside and cooler nightside. For higher pressures the zonal temperature gradient remains small but the equatorial regions heat up more than the high latitude regions leading to a larger meridional temperature gradient for higher metallicities.

For solar metallicities the zonal winds are characterised by a large-scale eastward flow at all latitudes. As the metallicity is increased the winds generally become faster in the equatorial region and slower at high latitudes, leading to a circulation pattern that is dominated by a fast equatorial jet.

These qualitative trends in the circulation and thermal structure with metallicity are consistent with results from previous works using different GCMs \citep{Men12,KatSF14,ChaML15}. There are quantitative differences in both the typical wind velocities and temperatures from each model. These quantitative discrepancies are likely due to a combination of different levels of approximation, differences in the model discretisation and choice of numerical methods and differences in the physical parameterisations (e.g. radiative transfer, chemistry). Understanding and quantifying these differences is beyond the scope of this study.

We investigated the different mechanisms that lead to changes in the dynamics and thermal structure as the metallicity is increased, by separating the overall effect into the dynamical effect, the radiative heat capacity effect and the opacity effect. The opacity effect was found to be the dominant mechanism that leads to changes in both the atmospheric circulation and temperature.

We calculated emission phase curves from each of the simulations. Our simulations show a significant effect on the 3.6 $\si{\micro\metre}$ {\it Spitzer}/IRAC phase curve as the metallicity is increased from 1$\times$ to 100$\times$ solar. Specifically, the amplitude of the emission phase curve increases, as the day-night temperature contrast increases, and the peak of the phase curve shifts closer to the secondary eclipse. The signature of metallicity is smaller at 4.5 $\si{\micro\metre}$ where the emission results from deeper layers of the atmosphere with an overall smaller zonal temperature gradient.

Coupling a Gibbs energy minimistion scheme directly to a GCM has benefits in both flexibility and accuracy. The mole fractions of a large number of chemical species can be computed from a general set of elemental abundances and for a wide range of thermodynamical conditions. The mole fractions can then be used to compute the total atmospheric opacity more accurately than using pre-mixed $k$-coefficients that assume a fixed chemical composition. 

Equilibrium chemistry, however, has been shown to be a likely poor assumption even for hot gas-giant atmospheres. Physical processes, such as mixing and photochemistry, have been shown to drive the chemistry out of equilibrium using 1D \citep[e.g.][]{Liang2003,LinLY2010,Moses2011,Venot2012,Zahnle2014,DruTB16,TsaLG2017}, psuedo-2D \citep{AguPV14} and simplified 3D \citep{CooS06} models. Non-equilibrium chemistry has also been suggested to explain discrepancies between observations and models that assume chemical equilibrium \citep[e.g.][]{KnuLF12,ZelLK14,WonKK16}. A GCM consistently coupled with a chemical kinetics scheme is required to relax the assumption of chemical equilibrium and to investigate the effect of these disequilibrium processes on the mole fractions, temperature and, subsequently, the observables.

\begin{acknowledgements}
We thank the anonymous referee for helping to improve this manuscript.
This work is partly supported by the European
Research Council under the European Community’s Seventh Framework Programme
(FP7/2007-2013 Grant Agreement No. 247060-PEPS and grant No.
320478-TOFU). BD acknowledges funding from the European Research Council (ERC) under the European Unions Seventh Framework Programme (FP7/2007-2013) / ERC grant agreement no. 336792 and thanks the University of Exeter for support through a PhD studentship. DSA acknowledges support from the NASA Astrobiology Program through the Nexus for Exoplanet System Science. NJM and JG's contributions were in part funded by a Leverhulme Trust Research Project Grant, and in part by a University of Exeter College of Engineering, Mathematics and Physical Sciences studentship. This work used the DiRAC Complexity system, operated by the University of Leicester IT Services, which forms part of the STFC DiRAC HPC Facility. This equipment is funded by BIS National E-Infrastructure capital grant ST/K000373/1 and STFC DiRAC Operations grant ST/K0003259/1. DiRAC is part of the National E-Infrastructure. This work also used the University of Exeter Supercomputer, a DiRAC Facility jointly funded by STFC, the Large Facilities Capital Fund of BIS and the University of Exeter. Material produced using Met Office Software.
\end{acknowledgements}